\documentclass[12pt]{article}
\usepackage{url}
\usepackage{graphicx}
\begin{document}

\begin{titlepage}

\vspace{2.truecm}

\begin{center}

\textbf{\LARGE TASI 2011: lectures on Higgs-Boson Physics}
\vspace{1,truecm}

{\large Laura Reina}
\vspace{0.5truecm}

\textit{Physics Department, Florida State University,\\
315 Keen Building, \\ 
Tallahassee, FL 32306-4350, USA\\ 
E-mail: reina@hep.fsu.edu}

\abstract{In these lectures I briefly review the Higgs mechanism of
  electroweak symmetry breaking and focus on the most relevant aspects
  of the phenomenology of the Standard Model Higgs boson at hadron
  colliders, namely the Tevatron and the Large Hadron
  Collider. Emphasis is put in particular on the Higgs physics program
  of both LHC experiments and on the theoretical activity that has
  entailed from the the need of providing accurate predictions for
  both signal and background in Higgs searches.}

\end{center}
\end{titlepage}
\tableofcontents

\section{Introduction}
\label{sec:intro}

The mechanism through which the electroweak gauge bosons, the $W^\pm$
and the $Z^0$, as well as all elementary fermions, leptons and quarks,
develop the mass properties of which we have experimental evidence is
unknown at the moment. Generically dubbed as the mechanism of
\textit{electroweak symmetry breaking} (EWSB), this problem has been
the core question that all theories proposed as extensions of the
Standard Model try to answer.

The so called \textit{Higgs mechanism} provides a very simple and
economical solution to the problem of EWSB, and since it was first
proposed in 1964 by Higgs, Kibble, Guralnik, Hagen, Englert and Brout
\cite{Englert:1964et,Higgs:1964pj,Guralnik:1964eu}, it has become
\textit{de facto} part of the Standard Model (SM). By introducing one
complex pair of scalar fields with a non trivial potential and a
suitable interaction to all matter particles, it achieves the goal of
providing mass to both the weak force carriers and the elementary
matter particles, at the expense of introducing just one new particle,
the by now famous \textit{Higgs particle} or \textit{Higgs
  boson}. Extensions of the SM often generalize the same mechanism to
adapt it to more involved symmetry patterns. This is for instance the
case of the Minimal Supersymmetric Standard Model (MSSM), where two
pairs of complex fields are introduced instead of one, resulting in a
final set of several \textit{Higgs bosons}.  While the single Higgs
boson of the SM is a neutral scalar (i.e. spinless) particle (which we
will denote by $H$), the MSSM has four Higgs bosons, two neutral
scalars ($h^0$ and $H^0$), one neutral pseudoscalar ($A^0$), and
one charged scalar ($H^\pm$).  Extensions of the SM besides the MSSM
can have even richer spectra of scalar and pseudoscalar particles
originating in the process of EWSB.

Precision studies of the SM and of the MSSM have assumed the
corresponding Higgs particle(s) as integral part of the theory and
have been able to constrain its (their) masses and couplings. Results
from the Tevatron collider have been able to exclude regions of the
parameter space of both the SM and the MSSM.  Since his inception the
LHC has validated and extended the Tevatron bounds and has recently
found strong evidence of the existence of a Higgs boson with SM-like
properties at about 125-127~GeV~\cite{:2012gk,:2012gu}. Indeed, the
discovery of a spin-0 particle compatible with the predictions for a
SM Higgs boson has been announced on July, 4$^{th}$ 2012. This comes
as one of the most exciting result that we could have ever expected at
such an early stage of the LHC and represents a milestone in the
history of particle physics.

The success of the LHC Higgs physics program hinges however
on the crucial assumption that experimental data can be compared with
very accurate theoretical predictions capable of discriminating
between signal and background at a statistically significant
level. Theorists have been meeting this challenge by modeling the
complexity of hadronic interactions in the context of Quantum
Chromodynamics (QCD).  Since at high energies QCD is a perturbative
quantum field theory (pQCD), QCD effects at colliders can be
calculated order by order in the strong coupling constant.

Lower order predictions typically have larger uncertainties associated
with them and cannot be meaningfully used to compare to experimental
measurements. In preparation for the LHC, a huge theoretical effort
has been devoted to provide accurate QCD predictions for all the most
important processes that are and will be the core of the LHC physics
program. In particular, the Higgs physics program of the LHC has
received an incredible amount of attention and most SM Higgs
production processes have been calculated to a high level of
precision, including a few orders of perturbative QCD corrections.  In
the context of the LHC Higgs Cross Sections Working Group (LHC-HXSWG)
the most up to date theoretical results have been collected for both
inclusive and exclusive production cross sections for both SM and MSSM
Higgs bosons~\cite{Dittmaier:2011ti,Dittmaier:2012vm}, and a new phase
of activity has started that will mainly focus on the identification
of the newly discovered particle.

In these lectures I would like to present a self contained
introduction to the physics of the Higgs boson(s).  In Section
\ref{sec:theory_framework}, after a brief glance at the essence of the
Higgs mechanism, I will review how it is embedded in the Standard
Model and what constraints are directly and indirectly imposed on the
mass of the single Higgs boson that is predicted in this
context. Among the extensions of the SM, I will only consider the case
of the Minimal Supersymmetric Model (MSSM), and in this context I will
mainly focus on those aspects that could be more relevant in
distinguishing the MSSM Higgs bosons. Section \ref{sec:higgs_searches}
will review the phenomenology of both the SM and the MSSM Higgs
bosons, at the Tevatron and the LHC. Sections~\ref{sec:tev_searches}
and \ref{sec:lhc_searches} deal specifically with the SM Higgs-boson
recent results from the Tevatron and the LHC.  Finally, in Section
\ref{sec:theory}, I will summarize the state of the art of existing
theoretical calculations for both decay rates and production cross
sections of a Higgs boson, and discuss the impact of QCD corrections 
in the prototype case of the $gg\rightarrow H$ production mode.

Let me conclude by pointing the reader to some selected references
available in the literature. The theoretical bases of the Higgs
mechanism are nowadays a matter for textbooks in Quantum Field
Theory. They are presented in depth in both Refs.~\cite{Peskin:1995ev}
and \cite{Weinberg:1995mt}. An excellent review of both SM and MSSM
Higgs physics, containing a very comprehensive discussion of both
theoretical and phenomenological aspects as well as a thorough
bibliography, can be found in
Refs.~\cite{Djouadi:2005gi,Djouadi:2005gj}.  The phenomenology of
Higgs physics has also been thoroughly covered in a review paper
\cite{Carena:2002es}.  Finally, series of lectures given at previous
summer schools \cite{Dawson:1994ri,Dawson:1998yi,Reina:2005ae} can
provide further references.

\section{Theoretical framework: the Higgs mechanism and its consequences.}
\label{sec:theory_framework}
In Yang-Mills theories gauge invariance forbids to have an explicit
mass term for the gauge vector bosons in the Lagrangian. If this is
acceptable for theories like QED (Quantum Electrodynamics) and QCD
(Quantum Chromodynamics), where both photons and gluons are massless,
it is unacceptable for the gauge theory of weak interactions, since
both the charged ($W^\pm$) and neutral ($Z^0$) gauge bosons have very
heavy masses ($M_W\!\simeq\!80$~GeV, $M_Z\!\simeq\!91$~GeV). A
possible solution to this problem, inspired by similar phenomena
happening in the study of spin systems, was proposed by several
physicists in 1964 \cite{Englert:1964et,Higgs:1964pj,Guralnik:1964eu},
and it is known today simply as \emph{the Higgs mechanism}. We will
review the basic idea behind it in
Section~\ref{subsec:higgs_mechanism}. In Section~\ref{subsec:higgs_sm}
we will recall how the Higgs mechanism is implemented in the Standard
Model and we will discuss which kind of theoretical constraints are
imposed on \emph{the Higgs boson}, the only physical scalar particle
predicted by the model. Finally, in Section~\ref{subsec:higgs_mssm} we
will generalize our discussion to the case of the MSSM, and use its
extended Higgs sector to illustrate how differently the Higgs
mechanism can be implemented in extensions of the SM.

\subsection{A brief introduction to the Higgs mechanism}
\label{subsec:higgs_mechanism}
\begin{figure}
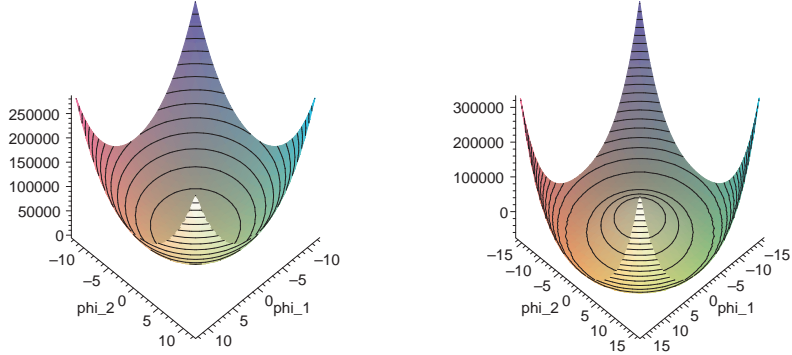

\centering
\includegraphics[scale=0.4]{./higgs_1.ps}
\hspace{1.truecm}
\includegraphics[scale=0.4]{./higgs_2.ps}
\caption[]{The potential $V(\phi)$ ($\phi\!=\phi_1+i\phi_2$) plotted
  for an arbitrary positive value of $\lambda$ and for an arbitrary
  positive (right) or negative (left) value of $\mu^2$. \label{fig:higgs_potential}
}
\end{figure}

The essence of the Higgs mechanism can be very easily illustrated
considering the case of a classical abelian Yang-Mills theory. In this
case, it is realized by adding to the Yang-Mills Lagrangian
\begin{equation}
\label{eq:L_ym_ab}
\mathcal{L}_A=-\frac{1}{4}F^{\mu\nu}F_{\mu\nu}\,\,\,\,\,\,\mbox{with}
\,\,\,\,\,\,
F^{\mu\nu}=(\partial^\mu A^\nu-\partial^\nu A^\mu)\,\,\,,
\end{equation}
a complex scalar field with Lagrangian
\begin{equation}
\label{eq:L_phi}
\mathcal{L}_\phi=(D^\mu\phi)^\ast D_\mu\phi -V(\phi)=
(D^\mu\phi)^\ast D_\mu\phi
-\mu^2\phi^\ast\phi-\lambda(\phi^\ast\phi)^2
\,\,\,,
\end{equation}
where $D^\mu\!=\!\partial^\mu +igA^\mu$, and $\lambda\!>\! 0$ for the
scalar potential to be bounded from below. The full Lagrangian
\begin{equation}
\label{eq:L_tot}
\mathcal{L}=\mathcal{L}_A+\mathcal{L}_\phi
\end{equation}
is invariant under a $U(1)$ \emph{gauge transformation} acting on
the fields as:
\begin{equation}
\label{eq:gauge_trans_ab}
\phi(x)\rightarrow e^{i\alpha(x)}\phi(x)\,\,\,,\,\,\,
A^\mu(x)\rightarrow A^\mu(x)+\frac{1}{g}\partial^\mu\alpha(x)\,\,\,,
\end{equation}
while a gauge field mass term (i.e., a term quadratic in the fields
$A^\mu$) would not be gauge invariant and cannot be added to
$\mathcal{L}$ if the $U(1)$ gauge symmetry has to be
preserved. Indeed, the Lagrangian in Eq.~(\ref{eq:L_tot}) can still
describe the physics of a massive gauge boson, provided the potential
$V(\phi)$ develops a non trivial minimum
($\phi^\ast\phi\!\neq\!0$). The occurrence of a non trivial minimum,
or, better, of a non trivial degeneracy of minima only depends on the
sign of the $\mu^2$ parameter in $V(\phi)$. For $\mu^2\!>\!0$ there is
a unique minimum at $\phi^\ast\phi\!=\!0$, while for $\mu^2\!<\!0$ the
potential develops a degeneracy of minima satisfying the equation
$\phi^\ast\phi\!=\!-\mu^2/(2\lambda)$. This is illustrated in
Fig.~\ref{fig:higgs_potential}, where the potential $V(\phi)$ is
plotted as a function of the real and imaginary parts of the field
$\phi\!=\!\phi_1+i\phi_2$.
In the case of a unique minimum at $\phi^\ast\phi\!=\!0$ the
Lagrangian in Eq.~(\ref{eq:L_tot}) describes the physics of a massless
vector boson (e.g. the photon, in electrodynamics, with $g\!=\!-e$)
interacting with a massive charged scalar particle. On the other hand,
something completely different takes place when $\mu^2\!<\!0$.
Choosing the ground state of the theory to be a particular $\phi$
among the many satisfying the equation of the minimum, and expanding
the potential in the vicinity of the chosen minimum, transforms the
Lagrangian in such a way that the original gauge symmetry is now
\emph{hidden} or \emph{spontaneously broken}, and new interesting
features emerge. To be more specific, let's pick the following
$\phi_0$ minimum (along the direction of the real part of $\phi$, as
traditional) and shift the $\phi$ field accordingly:
\begin{equation}
\label{eq:phi0}
\phi_0=\left(-\frac{\mu^2}{2\lambda}\right)^{1/2}=\frac{v}{\sqrt{2}}
\,\,\,\longrightarrow\,\,\,
\phi(x)=\phi_0+\frac{1}{\sqrt{2}}\left(\phi_1(x)+i\phi_2(x)\right)
\,\,\,. 
\end{equation}
The Lagrangian in Eq.~(\ref{eq:L_tot}) can then be rearranged as follows:
\begin{equation}
\label{eq:L_about_phi0}
{\mathcal L}=
\underbrace{
-\frac{1}{4}F^{\mu\nu}F_{\mu\nu}+\frac{1}{2}g^2v^2A^\mu A_\mu}_
{\mathrm{massive}\,\,\mathrm{vector}\,\,\mathrm{field}}
+
\underbrace{
\frac{1}{2}(\partial^\mu\phi_1)^2+\mu^2\phi_1^2}_
{\mathrm{massive}\,\,\mathrm{scalar}\,\,\mathrm{field}}+
\underbrace{
\frac{1}{2}(\partial^\mu\phi_2)^2+gvA_\mu\partial^\mu\phi_2}_{
\mathrm{Goldstone}\,\,\mathrm{boson}}+\ldots
\end{equation}
and now contains the correct terms to describe a massive vector field
$A^\mu$ with mass $m_A^2\!=\!g^2v^2$ (originating from the kinetic
term of $\mathcal{L}_\phi$), a massive real scalar field $\phi_1$ with
mass $m_{\phi_1}\!=\!-2\mu^2$, that will become a \emph{Higgs
boson}, and a massless scalar field $\phi_2$, a so called
\emph{Goldstone boson} which couples to the gauge vector boson
$A^\mu$. The terms omitted contain couplings between the $\phi_1$ and
$\phi_2$ fields irrelevant to this discussion. The gauge symmetry of
the theory allows us to make the particle content more transparent.
Indeed, if we parameterize the complex scalar field $\phi$ as:
\begin{equation}
\label{eq:phi_unit_gauge}
\phi(x)=\frac{e^{i\frac{\chi(x)}{v}}}{\sqrt{2}}(v+H(x))
\,\,\,\,\stackrel{U(1)}{\longrightarrow}\,\,\,\,
\frac{1}{\sqrt{2}}(v+H(x))\,\,\,,
\end{equation}
the $\chi$ degree of freedom can be \emph{rotated away}, as indicated
in Eq.~(\ref{eq:phi_unit_gauge}), by enforcing the $U(1)$ gauge
invariance of the original Lagrangian. With this gauge choice, known
as \emph{unitary gauge} or \emph{unitarity gauge}, the Lagrangian
becomes:
\begin{equation}
\label{eq:L_unit_gauge}
\mathcal{L}=\mathcal{L}_A+\frac{g^2v^2}{2}A^\mu A_\mu+
\frac{1}{2}\left(\partial^\mu H\partial_\mu H+2\mu^2 H^2
\right)+\ldots
\end{equation}
which unambiguously describes the dynamics of a massive vector boson
$A^\mu$ of mass $m_A^2\!=\!g^2v^2$, and a massive real scalar field of
mass $m_H^2\!=\!-2\mu^2=2\lambda v^2$, the \emph{Higgs field}. It is
interesting to note that the total counting of degrees of freedom
(d.o.f.)  before the original $U(1)$ symmetry is spontaneously broken
and after the breaking has occurred is the same. Indeed, one goes from
a theory with one massless vector field (two d.o.f.) and one complex
scalar field (two d.o.f.) to a theory with one massive vector field
(three d.o.f.) and one real scalar field (one d.o.f.), for a total of
four d.o.f. in both cases.  This is what is colorfully described by
saying that each gauge boson has \emph{eaten up} one scalar degree of
freedom, becoming massive.

We can now easily generalize the previous discussion to the case of a
non-abelian Yang-Mills theory. $\mathcal{L}_A$ in Eq.~(\ref{eq:L_tot})
now becomes:
\begin{equation}
\label{eq:L_ym_non_ab}
\mathcal{L}_A=\frac{1}{4}F^{a,\mu\nu}F^a_{\mu\nu}\,\,\,\,\,\mbox{with}
\,\,\,\,\, F^a_{\mu\nu}=\partial_\mu A^a_\nu-\partial_\nu A^a_\mu+
gf^{abc}A^b_\mu A^c_\nu\,\,\,,
\end{equation}
where the latin indices are group indices and $f^{abc}$ are the
structure constants of the Lie Algebra associated to the non abelian
gauge symmetry Lie group, defined by the commutation relations of the Lie
Algebra generators $t^a$: $[t^a,t^b]\!=\!if^{abc}t^c$. Let us also
generalize the scalar Lagrangian to include several scalar fields
$\phi_i$ which we will in full generality consider as real:
\begin{equation}
\label{eq:L_phi_multi}
\mathcal{L}_\phi=\frac{1}{2}(D^\mu\phi_i)^2-V(\phi)\,\,\,\,\,\mbox{where}\,\,\,\,\,
V(\phi)=\mu^2\phi_i^2+\frac{\lambda}{2}\phi_i^4\,\,\,,
\end {equation}
where the sum over the index $i$ is understood and
$D_\mu\!=\!\partial_\mu-igt^aA^a_\mu$. The Lagrangian of
Eq.~(\ref{eq:L_tot}) is invariant under a non-abelian gauge
transformation of the form:
\begin{eqnarray}
\label{eq:gauge_trans_nab}
\phi_i(x)&\rightarrow& (1+i\alpha^a(x)t^a)_{ij}\phi_j \,\,\,,\\
A^a_\mu(x)&\rightarrow& A^a_\mu(x)+\frac{1}{g}\partial_\mu\alpha^a(x)+
f^{abc}A^b_\mu(x)\alpha^c(x)\,\,\,.\nonumber
\end{eqnarray}
When $\mu^2\!<\!0$ the potential develops a degeneracy of minima
described by the minimum condition:
$\phi^2\!=\!\phi_{0}^2\!=\!-\mu^2/\lambda$, which only fixes the
magnitude of the vector $\phi_{0}$. By arbitrarily choosing the
direction of $\phi_{0}$, the degeneracy is removed. The Lagrangian can
be expanded in a neighborhood of the chosen minimum and mass terms for
the gauge vector bosons can be introduced as in the abelian case,
i.e.:
\begin{eqnarray}
\label{eq:gauge_boson_mass_nonab}
\frac{1}{2}(D_\mu\phi_i)^2&\longrightarrow& \ldots\,\,\,+
\frac{1}{2}g^2(t^a\phi)_i(t^b\phi)_i A_\mu^a A^{b\mu}+\ldots\\ 
&\stackrel{\phi_{min}\!=\!\phi_0}{\longrightarrow}&
\ldots\,\,\,+ \frac{1}{2}
\underbrace{g^2(t^a\phi_0)_i(t^b\phi_0)_i}_{m_{ab}^2}A_\mu^a
A^{b\mu}+\ldots \nonumber
\end{eqnarray}
Upon diagonalization of the mass matrix $m_{ab}^2$ in
Eq.~(\ref{eq:gauge_boson_mass_nonab}), all gauge vector bosons
$A_\mu^a$ for which $t^a\phi_0\ne 0$ become massive, and to each of
them corresponds a Goldstone particle, i.e. an unphysical massless
particle like the $\chi$ field of the abelian example. The remaining
scalar degrees of freedom become massive, and correspond to the Higgs
field $H$ of the abelian example.

The Higgs mechanism can be very elegantly generalized to the case of a
quantum field theory when the theory is quantized via the path
integral method\footnote{Here I assume some familiarity with path
integral quantization and the properties of various generating
functionals introduced in that context, as I did while giving these
lectures. The detailed explanation of the formalism used would take us
too far away from our main track}. In this context, the quantum analog
of the potential $V(\phi)$ is the
\emph{effective potential} $V_{eff}(\varphi_{cl})$, defined in term of
the \emph{effective action} $\Gamma[\phi_{cl}]$ (the generating
functional of the 1PI connected correlation functions) as:
\begin{equation}
\label{eq:v_eff}
V_{eff}(\varphi_{cl})=-\frac{1}{VT}\Gamma[\phi_{cl}]\,\,\,\,\,\mbox{for}
\,\,\,\,\,\phi_{cl}(x)=\mbox{constant}=\varphi_{cl}\,\,\,,
\end{equation}
where $VT$ is the space-time extent of the functional integration and
$\phi_{cl}(x)$ is the \emph{vacuum expectation value} of the field
configuration $\phi(x)$:
\begin{equation}
\label{eq:phi_cl}
\phi_{cl}(x)=\langle\Omega| \phi(x)|\Omega\rangle\,\,\,.
\end{equation}

The stable quantum states of the theory are defined by the variational
condition:
\begin{equation}
\label{eq:delta_v_eff}
\frac{\delta}{\delta\phi_{cl}}\Gamma[\phi_{cl}]\bigg|_{\phi_{cl}=\varphi_{cl}}=0
\,\,\,\,\,\,\,\,\longrightarrow\,\,\,\,\,\,\,\,
\frac{\partial}{\partial\varphi_{cl}}V_{eff}(\varphi_{cl})=0\,\,\,,
\end{equation}
which identifies in particular the states of minimum energy of the
theory, i.e. the stable vacuum states. A system with spontaneous
symmetry breaking has several minima, all with the same
energy. Specifying one of them, as in the classical case, breaks
the original symmetry on the vacuum. The relation between the
classical and quantum case is made even more transparent by the
perturbative form of the effective potential. Indeed,
$V_{eff}(\varphi_{cl})$ can be organized as a loop expansion and
calculated systematically order by order in $\hbar$:
\begin{equation}
\label{eq:veff_exp}
V_{eff}(\varphi_{cl})=V(\varphi_{cl})+\mbox{loop effects}\,\,\,,
\end{equation}
with the lowest order being the classical potential in
Eq.~(\ref{eq:L_phi}). Quantum corrections to $V_{eff}(\varphi_{cl})$
affect some of the properties of the potential and therefore
have to be taken into account in more sophisticated studies of the
Higgs mechanism for a spontaneously broken quantum gauge theory.  We
will see how this can be important in Section 
\ref{subsec:higgs_sm_constraints} when
we discuss how the mass of the SM Higgs boson is related to the
energy scale at which we expect new physics effect to become relevant
in the SM.

Finally, let us observe that at the quantum level the choice of gauge
becomes a delicate issue. For example, in the \emph{unitarity gauge}
of Eq.~(\ref{eq:phi_unit_gauge}) the particle content of the theory
becomes transparent but the propagator of a massive vector field
$A^\mu$ turns out to be:
\begin{equation}
\label{eq:prop_unit_gauge}
\Pi^{\mu\nu}(k)=-\frac{i}{k^2-m_A^2} \left(g^{\mu\nu}-\frac{k^\mu
k^\nu}{m_A^2}\right)\,\,\,,
\end{equation}
and has a problematic ultra-violet behavior, which makes more
difficult to consistently define and calculate ultraviolet-stable
scattering amplitudes and cross sections. Indeed, for the very purpose
of studying the renormalizability of quantum field theories with
spontaneous symmetry breaking, the so called \emph{renormalizable} or
\emph{renormalizability gauges} ($R_\xi$ \emph{gauges}) are
introduced. If we consider the abelian Yang-Mills theory of
Eqs.~(\ref{eq:L_ym_ab})-(\ref{eq:L_tot}), the \emph{renormalizable
gauge} choice is implemented by quantizing with a gauge condition $G$
of the form:
\begin{equation}
\label{eq:ren_gauge}
G=\frac{1}{\sqrt{\xi}}(\partial_\mu A^\mu+\xi gv\phi_2)\,\,\,,
\end{equation}
in the generating functional
\begin{equation}
\label{eq:Z_ren_gauge}
Z[J]=C\int DA\,D\phi_1\,D\phi_2\exp\left[i\int(\mathcal{L}-\frac{1}{2}G^2)\right]
\mbox{det}\left(\frac{\delta G}{\delta\alpha}\right)\,\,\,,
\end{equation}
where C is an overall factor independent of the fields, $\xi$ is an
arbitrary parameter, and $\alpha$ is the gauge transformation
parameter in Eq.~(\ref{eq:gauge_trans_ab}).  After having reduced the
determinant in Eq.~(\ref{eq:Z_ren_gauge}) to an integration over ghost
fields ($c$ and $\bar{c}$), the gauge plus scalar fields Lagrangian
looks like:
\begin{eqnarray}
\label{eq:L_ren_gauge}
\mathcal{L}-\frac{1}{2}G^2+\mathcal{L}_{ghost}&=&
-\frac{1}{2}A_\mu\left(-g^{\mu\nu}\partial^2+
\left(1-\frac{1}{\xi}\right)
\partial^\mu\partial^\nu-(gv)^2g^{\mu\nu}\right)A_\nu\nonumber\\
&+&\frac{1}{2}(\partial_\mu\phi_1)^2-\frac{1}{2}m_{\phi_1}^2\phi_1^2
+\frac{1}{2}(\partial_\mu\phi_2)^2
-\frac{\xi}{2}(gv)^2\phi_2^2+\cdots\nonumber\\
&+&\bar{c}\left[-\partial^2-
\xi(gv)^2\left(1+\frac{\phi_1}{v}\right)\right]c\,\,\,,
\end{eqnarray}
such that:
\begin{eqnarray}
\label{eq:prop_ren_gauge}
\langle A^\mu(k)A^\nu(-k)\rangle&=&
\frac{-i}{k^2-m_A^2}\left(g^{\mu\nu}-\frac{k^\mu k^\nu}{k^2}\right)+
\frac{-i\xi}{k^2-\xi m_A^2}
\left(\frac{k^\mu k^\nu}{k^2}\right)\,\,\,,
\nonumber\\
\langle \phi_1(k)\phi_1(-k)\rangle&=&\frac{-i}{k^2-m_{\phi_1}^2}\,\,\,,\\
\langle \phi_2(k)\phi_2(-k)\rangle&=&\langle c(k)\bar{c}(-k)\rangle=
\frac{-i}{k^2-\xi m_A^2}\,\,\,,
\nonumber
\end{eqnarray}
where the vector field propagator has now a safe ultraviolet behavior.
Moreover we notice that the $\phi_2$ propagator has the same
denominator of the longitudinal component of the gauge vector boson
propagator. This shows in a more formal way the relation between the
$\phi_2$ degree of freedom and the longitudinal component of the
massive vector field $A^\mu$, upon spontaneous symmetry breaking. 

\subsection{The Higgs sector of the Standard Model}
\label{subsec:higgs_sm}
The Standard Model is a spontaneously broken Yang-Mills theory based
on the $SU(2)_L\times U(1)_Y$ non-abelian symmetry
group\cite{Peskin:1995ev,Weinberg:1995mt}. The Higgs mechanism is
implemented in the Standard Model by introducing a complex scalar
field $\phi$, doublet of $SU(2)$ with hypercharge $Y_\phi=1/2$,
\begin{equation}
\label{eq:phi_sm}
\phi=
\left(
\begin{array}{c}
\phi^+\\
\phi^0
\end{array}
\right)\,\,\,,
\end{equation}
with Lagrangian
\begin{equation}
\label{eq:L_phi_sm}
\mathcal{L}_\phi=(D^\mu\phi)^\dagger D_\mu\phi-\mu^2\phi^\dagger\phi-
\lambda(\phi^\dagger\phi)^2\,\,\,,
\end{equation}
where $D_\mu\phi=(\partial_\mu-igA^a_\mu\tau^a-ig^\prime Y_\phi
B_\mu)$, and $\tau^a\!=\!\sigma^a/2$ 
(for $a\!=\!1,2,3$) are the $SU(2)$ Lie Algebra generators,
proportional to the Pauli matrix $\sigma^a$. The gauge symmetry 
of the Lagrangian is broken to $U(1)_{em}$ when a particular vacuum 
expectation value is chosen, e.g.:
\begin{equation}
\label{eq:phi_vev}
\langle\phi\rangle=\frac{1}{\sqrt{2}}
\left(
\begin{array}{c}
0\\ v
\end{array}
\right)\,\,\,\,\,\,\mbox{with}\,\,\,\,\,\,
v=\left(\frac{-\mu^2}{\lambda}\right)^{1/2}
\,\,\,\,\,(\mu^2<0,\,\lambda >0)\,\,\,.
\end{equation}
Upon spontaneous symmetry breaking the kinetic term in
Eq.~(\ref{eq:L_phi_sm}) gives origin to the SM gauge boson mass
terms. Indeed, specializing Eq.~(\ref{eq:gauge_boson_mass_nonab}) to
the present case, and using Eq.~(\ref{eq:phi_vev}), one gets:
\begin{eqnarray}
\label{eq:SM_gauge_boson_mass_terms}
(D^\mu\phi)^\dagger D_\mu\phi&\longrightarrow&\cdots +
\frac{1}{8}(0\,\,\,v)\left(gA_\mu^a\sigma^a+g^\prime B_\mu\right)
\left(gA^{b\mu}\sigma^b+g^\prime B^\mu\right)
\left(
\begin{array}{c}
0\\ v
\end{array}
\right)+\cdots
\nonumber\\
&\longrightarrow&
\cdots+\frac{1}{2}\frac{v^2}{4}\left[
g^2(A_\mu^1)^2+g^2(A_\mu^2)^2+(-gA_\mu^3+g^\prime B_\mu)^2\right] 
+\cdots\nonumber\\
\end{eqnarray}
One recognizes in Eq.~(\ref{eq:SM_gauge_boson_mass_terms}) the mass 
terms for the charged gauge bosons $W^\pm_\mu$:
\begin{equation}
\label{eq:W_mass}
W^\pm_\mu=\frac{1}{\sqrt{2}}(A_\mu^1\pm A_\mu^2)
\,\,\,\,\longrightarrow\,\,\,\,M_W=g\frac{v}{2}\,\,\,,
\end{equation}
and for the neutral gauge boson $Z^0_\mu$:
\begin{equation}
\label{eq:Z_mass}
Z^0_\mu=\frac{1}{\sqrt{g^2+g^{\prime 2}}}(gA_\mu^3-g^\prime B_\mu)
\,\,\,\,\longrightarrow\,\,\,\,
M_Z=\sqrt{g^2+g^{\prime 2}}\frac{v}{2}\,\,\,,
\end{equation} 
while the orthogonal linear combination of $A^3_\mu$ and $B_\mu$
remains massless and corresponds to the photon field ($A_\mu$):
\begin{equation}
\label{eq:photon_mass}
A_\mu=\frac{1}{\sqrt{g^2+g^{\prime 2}}}(g^\prime A_\mu^3+gB_\mu)
\,\,\,\,\longrightarrow\,\,\,\,M_A=0\,\,\,,
\end{equation}
the gauge boson of the residual $U(1)_{em}$ gauge symmetry.

The content of the scalar sector of the theory becomes more
transparent if one works in the unitary gauge and eliminate the
unphysical degrees of freedom using gauge invariance. In analogy to
what we wrote for the abelian case in Eq.~(\ref{eq:phi_unit_gauge}), 
this amounts to parameterize and rotate the $\phi(x)$ complex scalar 
field as follows:
\begin{equation}
\label{eq:phi_sm_unit_gauge}
\phi(x)=\frac{e^{\frac{i}{v}\vec\chi(x)\cdot\vec\tau}}{\sqrt{2}}
\left(
\begin{array}{c}
0\\ v+H(x)
\end{array}
\right)
\,\,\,\stackrel{SU(2)}{\longrightarrow}\,\,\,
\phi(x)=\frac{1}{\sqrt{2}}
\left(
\begin{array}{c}
0\\ v+H(x)
\end{array}
\right)\,\,\,,
\end{equation}
after which the scalar potential in Eq.~(\ref{eq:L_phi_sm}) becomes:
\begin{equation}
\label{eq:L_phi_sm_unit_gauge}
\mathcal{L}_\phi=\mu^2 H^2-\lambda v H^3-\frac{1}{4}H^4=
-\frac{1}{2}M_H^2 H^2-\sqrt{\frac{\lambda}{2}} M_H H^3
-\frac{1}{4}\lambda H^4\,\,\,.
\end{equation}
Three degrees of freedom, the $\chi^a(x)$ Goldstone bosons, have been
reabsorbed into the longitudinal components of the $W^\pm_\mu$ and
$Z^0_\mu$ weak gauge bosons. One real scalar field remains, the
\emph{Higgs boson} $H$, with mass $M_H^2\!=\!-2\mu^2=2\lambda v^2$ and
self-couplings:

\vspace{0.5truecm}
\begin{tabular}{cc}
\begin{minipage}{0.5\linewidth}
\includegraphics[scale=0.6]{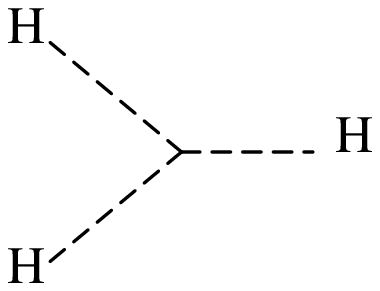}\parbox[b]{4.truecm}
{{\large $=-3i\frac{M_H^2}{v}$}\vspace{0.75truecm}}
\end{minipage}&
\begin{minipage}{0.5\linewidth}
\includegraphics[scale=0.6]{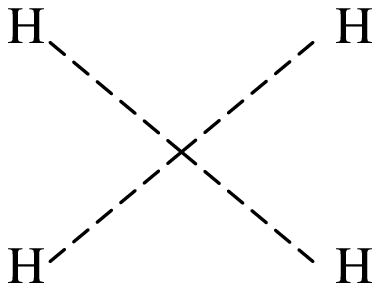}\parbox[b]{4.truecm}
{{\large $=-3i\frac{M_H^2}{v^2}$}\vspace{0.75truecm}}
\end{minipage}
\end{tabular}
\vspace{0.5truecm}

Furthermore, some of the terms that we omitted in
Eq.~(\ref{eq:SM_gauge_boson_mass_terms}), the terms linear in the
gauge bosons $W^\pm_\mu$ and $Z^0_\mu$, define the coupling
of the SM Higgs boson to the weak gauge fields:

\vspace{0.5truecm}
\begin{tabular}{cc}
\begin{minipage}{0.5\linewidth}
\includegraphics[scale=0.6]{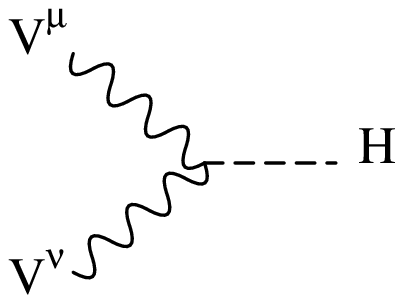}\parbox[b]{4.5truecm}
{{\large $=2i\frac{M_V^2}{v}g^{\mu\nu}$}\vspace{0.75truecm}}
\end{minipage}&
\begin{minipage}{0.5\linewidth}
\includegraphics[scale=0.6]{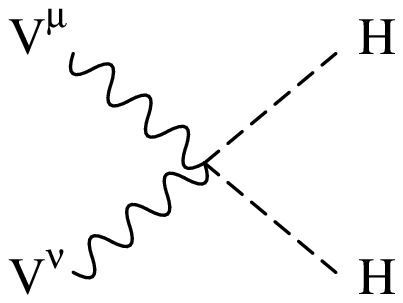}\parbox[b]{4.5truecm}
{{\large $=2i\frac{M_V^2}{v^2}g^{\mu\nu}$}\vspace{0.75truecm}}
\end{minipage}
\end{tabular}
\vspace{0.5truecm}

We notice that the couplings of the Higgs boson to the gauge fields
are proportional to their mass. Therefore $H$ does not couple to the
photon at tree level. It is important, however, to observe that
couplings that are absent at tree level may be induced at higher order
in the gauge couplings by loop corrections. Particularly relevant to
the SM Higgs-boson phenomenology that will be discussed in
Section~\ref{sec:higgs_searches} are the couplings of the SM Higgs
boson to pairs of photons, and to a photon and a $Z^0_\mu$ weak boson:

\begin{center}
\vspace{0.5truecm}
\begin{minipage}{0.8\linewidth}
\includegraphics[scale=0.6]{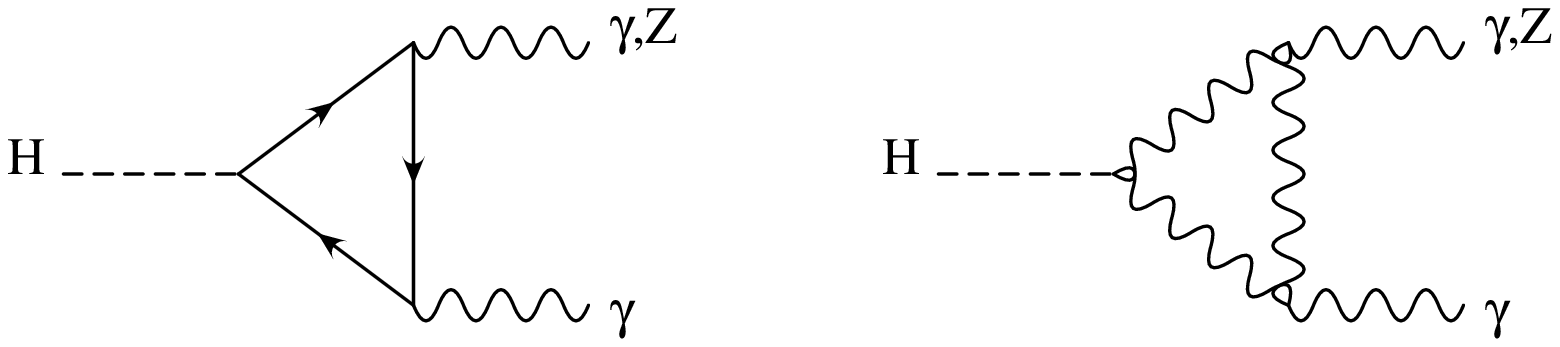}
\end{minipage}
\vspace{0.5truecm}
\end{center}

\noindent
as well as the coupling to pairs of gluons, when the SM Lagrangian is
extended through the QCD Lagrangian to include also the strong
interactions:

\begin{center}
\vspace{0.5truecm}
\begin{minipage}{0.5\linewidth}
\includegraphics[scale=0.6]{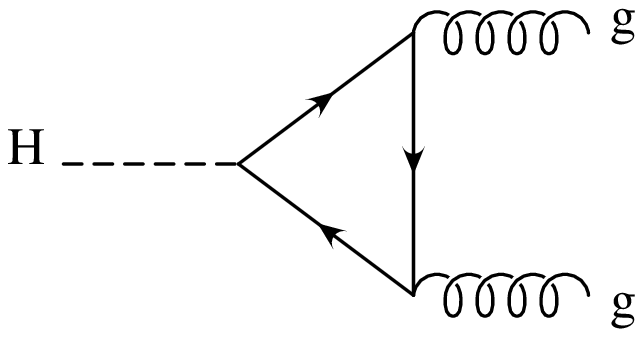}
\end{minipage}
\vspace{0.5truecm}
\end{center}

The analytical expressions for the $H\gamma\gamma$, $H\gamma Z$, and
$Hgg$ one-loop vertices are more involved and will be given in
Section~\ref{subsec:sm_higgs_branching_ratios}. As far as the Higgs
boson tree level couplings go, we observe that they are all expressed
in terms of just two parameters, either $\lambda$ and $\mu$ appearing
in the scalar potential of $\mathcal{L}_\phi$ (see Eq.~\ref{eq:L_phi_sm})) 
or, equivalently, $M_H$
and $v$, the Higgs-boson mass and the scalar-field vacuum expectation
value. Since $v$ is measured in muon decay to be
$v\!=\!(\sqrt{2}G_F)^{-1/2}\!=\!246$~GeV, the physics of the SM Higgs
boson is actually just function of its mass $M_H$.

The Standard Model gauge symmetry also forbids explicit mass terms for
the fermionic degrees of freedom of the Lagrangian. The fermion mass
terms are then generated via gauge invariant renormalizable Yukawa 
couplings to the scalar field $\phi$:
\begin{equation}
\label{eq:yukawa_lagrangian}
\mathcal{L}_\mathit{Yukawa}=-\Gamma_u^{ij}\bar{Q}^i_L\phi^c u^j_R
            -\Gamma_d^{ij}\bar{Q}^i_L\phi d^j_R
            -\Gamma_e^{ij}\bar{L}^i_L\phi l^j_R +h.c.
\end{equation}
where $\phi^c\!=\!-i\sigma^2\phi^\star$, and $\Gamma_f$ ($f=u,d,l$)
are matrices of couplings arbitrarily introduced to realize the Yukawa
coupling between the field $\phi$ and the fermionic fields of the SM.
$Q_L^i$ and $L_L^i$ (where $i=1,2,3$ is a generation index) represent
quark and lepton left handed doublets of $SU(2)_L$, while $u_R^i$,
$d_R^i$ and $l_R^i$ are the corresponding right handed singlets. When
the scalar fields $\phi$ acquires a non zero vacuum expectation value
through spontaneous symmetry breaking, each fermionic degree of
freedom coupled to $\phi$ develops a mass term with mass parameter
\begin{equation}
\label{eq:yukawa_coupling}
m_f=\Gamma_f\frac{v}{\sqrt{2}}\,\,\,,
\end{equation}
where the process of diagonalization from the current eigenstates in
Eq.~(\ref{eq:yukawa_lagrangian}) to the corresponding mass eigenstates
is understood, and $\Gamma_f$ are therefore the elements of the
diagonalized Yukawa matrices corresponding to a given fermion $f$. The
Yukawa couplings of the $f$ fermion to the Higgs boson ($y_f$) is
proportional to $\Gamma_f$:

\begin{center}
\vspace{0.5truecm}
\begin{minipage}{0.8\linewidth}
\includegraphics[scale=0.6]{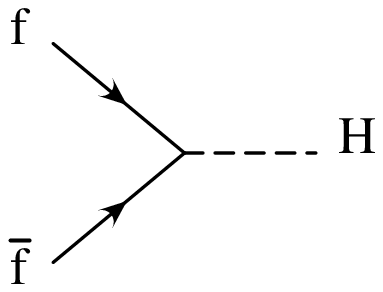}
\parbox[b]{6.truecm}
{{\large $=-i\frac{m_f}{v}=-i\frac{\Gamma_f}{\sqrt{2}}=-iy_f$}\vspace{0.75truecm}}
\end{minipage} 
\vspace{0.5truecm}
\end{center}

As long as the origin of fermion masses is not better understood in
some more general context beyond the Standard Model, the Yukawa
couplings $y_f$ represent free parameter of the SM Lagrangian. The
mechanism through which fermion masses are generated in the Standard
Model, although related to the mechanism of spontaneous symmetry
breaking, requires therefore further assumptions and involves a larger
degree of arbitrariness as compared to the gauge boson sector of the
theory.

\subsection{Theoretical constraints on the SM Higgs boson mass}
\label{subsec:higgs_sm_constraints}

Several issues arising in the scalar sector of the Standard Model
link the mass of the Higgs boson to the energy scale where the
validity of the Standard Model is expected to fail. Below that scale,
the Standard Model is the extremely successful effective field theory
that emerges from the electroweak precision tests of the last
decades. Above that scale, the Standard Model has to be embedded into
some more general theory that gives origin to a wealth of new physics
phenomena. From this point of view, the Higgs sector of the Standard
Model contains actually two parameters, the Higgs mass ($M_H$) and the
scale of new physics ($\Lambda$).

In this Section we will review the most important theoretical
constraints that are imposed on the mass of the Standard Model Higgs
boson by the consistency of the theory up to a given energy scale
$\Lambda$. In particular we will touch on issues of unitarity,
triviality, vacuum stability, fine tuning and, finally, electroweak
precision measurements.

\subsubsection{Unitarity}
\label{subsubsec:unitarity}
The scattering amplitudes for longitudinal gauge bosons
($V_LV_L\rightarrow V_LV_L$, where $V=W^\pm,Z^0$) grow as the square
of the Higgs-boson mass. This is easy to calculate using the
\emph{electroweak equivalence
theorem}~\cite{Peskin:1995ev,Weinberg:1995mt}, valid in the high
energy limit (i.e. for energies $s\!=\!Q^2\!\gg\! M_V^2$), according
to which the scattering amplitudes for longitudinal gauge bosons can
be expressed in terms of the scattering amplitudes for the
corresponding Goldstone bosons, i.e.:
\begin{equation}
\label{eq:equivalence_theorem}
\mathcal{A}(V_L^1\ldots V_L^n\rightarrow V_L^1\ldots V_L^m)=(i)^n(-i)^m
\mathcal{A}(\omega^1\dots\omega^n\rightarrow\omega^1\ldots\omega^m)+
O\left(\frac{M_V^2}{s}\right)\,\,\,,
\end{equation}
where we have indicated by $\omega^i$ the Goldstone boson associated
to the longitudinal component of the gauge boson $V^i$.
For instance, in the high energy limit, the scattering amplitude
for $W^+_LW^-_L\rightarrow W^+_LW^-_L$ satisfies:
\begin{equation}
\label{eq:ampl_wl}
\mathcal{A}(W_L^+ W_L^-\rightarrow W_L^+ W_L^-)=
\mathcal{A}(\omega^+\omega^-\rightarrow\omega^+\omega^-)+O\left(\frac{M_W^2}{s}\right)\,\,\,,
\end{equation}
where
\begin{equation}
\label{eq:ampl_omega}
\mathcal{A}(\omega^+\omega^-\rightarrow\omega^+\omega^-)=
-\frac{M_H^2}{v^2}\left(\frac{s}{s-M_H^2}+\frac{t}{t-M_H^2}\right)\,\,\,.
\end{equation}
Using a partial wave decomposition, we can also write $\mathcal A$ as:
\begin{equation}
\label{eq:ampl_partial_waves}
\mathcal{A}=16\pi\sum_{l=0}^{\infty}(2l+1)P_l(\cos\theta)a_l\,\,\,,
\end{equation}
where $a_l$ is the spin $l$ partial wave and $P_l(\cos\theta)$ are the
Legendre polynomials. In terms of partial wave amplitudes $a_l$, the
scattering cross section corresponding to $\mathcal{A}$ can be
calculated to be:
\begin{equation}
\label{eq:sigma_partial_waves}
\sigma=\frac{16\pi}{s}\sum_{l=0}^{\infty}(2l+1)|a_l|^2\,\,\,,
\end{equation}
where we have used the orthogonality of the Legendre polynomials.
Using the optical theorem, we can impose the unitarity constraint by
writing that:
\begin{equation}
\label{eq:optical_theorem}
\sigma=\frac{16\pi}{s}\sum_{l=0}^{\infty}(2l+1)|a_l|^2=
\frac{1}{s}\mathrm{Im}\left[\mathcal{A}(\theta=0)\right]\,\,\,,
\end{equation}
where $\mathcal{A}(\theta=0)$ indicates the scattering amplitude in
the forward direction. This implies that:
\begin{equation}
\label{eq:unitarity_bound}
|a_l|^2=\mathrm{Re}(a_l)^2+\mathrm{Im}(a_l)^2=\mathrm{Im}(a_l)\,\,\,
\longrightarrow\,\,\,|\mathrm{Re}(a_l)|\le\frac{1}{2}\,\,\,.
\end{equation}
Via Eq.~(\ref{eq:unitarity_bound}), different $a_l$ amplitudes can
than provide constraints on $M_H$. As an example, let us consider the
$J\!=\!0$ partial wave amplitude $a_0$ for the $W^+_LW^-_L\rightarrow
W^+_LW^-_L$ scattering we introduced above:
\begin{equation}
\label{eq:a0_wlwlwlwl}
a_0=\frac{1}{16\pi s}\int_{-s}^0\mathcal{A}\,dt = 
-\frac{M_H^2}{16\pi v^2}\left[2+\frac{M_H^2}{s-M_H^2}-
\frac{M_H^2}{s}\log\left(1+\frac{s}{M_H^2}\right)\right]\,\,\,.
\end{equation}
In the high energy limit ($M_H^2\ll s$), $a_0$ reduces to:
\begin{equation}
\label{eq:he_limit}
a_0\stackrel{M_H^2\ll s}{\longrightarrow}-\frac{M_H^2}{8\pi v^2}\,\,\,,
\end{equation}
from which, using Eq.~(\ref{eq:unitarity_bound}), one gets:
\begin{equation}
\label{eq:mh_bound_wlwlwlwl}
M_H< 870 \,\,\mbox{GeV}\,\,\,.
\end{equation}
Other more constraining relations can be obtained from different
longitudinal gauge boson scattering amplitudes. For instance,
considering the coupled channels like  $W^+_LW^-_L\rightarrow Z_LZ_L$, 
one can lower the bound to:
\begin{equation}
\label{eq:mh_bound_wlwlzlzl}
M_H<710\,\,\mbox{GeV}\,\,\,.
\end{equation}
Taking a different point of view, we can observe that if there is no Higgs
boson, or equivalently if $M_H^2\gg s$, Eq.~(\ref{eq:unitarity_bound})
gives indications on the critical scale $\sqrt{s_c}$ above which new physics 
should be expected. Indeed, considering again  
$W^+_LW^-_L\rightarrow W^+_LW^-_L$ scattering, we see that:
\begin{equation}
\label{eq:a0_no_higgs}
a_0(\omega^+\omega^-\rightarrow\omega^+\omega^-)
\stackrel{M_H^2\gg s}{\longrightarrow}-\frac{s}{32\pi v^2}\,\,\,,
\end{equation}
from which, using Eq.~(\ref{eq:unitarity_bound}), we get:
\begin{equation}
\label{eq:sc_upper_bound_wlwlwlwlw}
\sqrt{s_c}< 1.8\,\,\mbox{TeV}\,\,\,.
\end{equation}
Using more constraining channels the bound can be reduced to:
\begin{equation}
\label{eq:sc_upper_bound_best}
\sqrt{s_c}< 1.2\,\,\mbox{TeV}\,\,\,.
\end{equation}
This is very suggestive: it tells us that new physics ought to be
found around 1-2 TeV, i.e. exactly in the range of energies that will
be explored by the Tevatron and the Large Hadron Collider.

\subsubsection{Triviality and vacuum stability}
\label{subsubsec:triviality_vacuumstability}

The argument of triviality in a $\lambda\phi^4$ theory goes as
follows. The dependence of the quartic coupling $\lambda$ on the
energy scale ($Q$) is regulated by the renormalization group equation
\begin{equation}
\label{eq:phi4_lambda_rge}
\frac{d\lambda(Q)}{dQ^2}=\frac{3}{4\pi^2}\lambda^2(Q)\,\,\,.
\end{equation}
This equation states that the quartic coupling $\lambda$ decreases for
small energies and increases for large energies. Therefore, in the low
energy regime the coupling vanishes and the theory becomes trivial,
i.e. non-interactive. In the large energy regime, on the other hand,
the theory becomes non-perturbative, since $\lambda$ grows, and it can
remain perturbative only if $\lambda$ is set to zero, i.e. only if the
theory is made trivial.

The situation in the Standard Model is more complicated, since the
running of $\lambda$ is governed by more interactions. Including the
lowest orders in all the relevant couplings, we can write the equation
for the running of $\lambda(Q)$ with the energy scale as follows:
\begin{equation}
\label{eq:sm_lambda_rge}
32\pi^2\frac{d\lambda}{dt}=
24\lambda^2-(3g^{\prime 2}+9g^2-24y_t^2)\lambda
+\frac{3}{8}g^{\prime 4}+\frac{3}{4}g^{\prime 2}g^2+\frac{9}{8}g^4
-24y_t^4+\cdots
\end{equation}
where $t\!=\!\ln(Q^2/Q_0^2)$ is the logarithm of the ratio of the
energy scale and some reference scale $Q_0$ square, $y_t\!=\!m_t/v$ is
the top-quark Yukawa coupling, and the dots indicate the presence of
higher order terms that have been omitted. We see that when $M_H$
becomes large, $\lambda$ also increases (since $M_H^2\!=\!2\lambda
v^2$) and the first term in Eq.~(\ref{eq:sm_lambda_rge})
dominates. The evolution equation for $\lambda$ can then be easily
solved and gives:
\begin{equation}
\label{eq:lambda_sm_large_mh}
\lambda(Q)=\frac{\lambda(Q_0)}{1-\frac{3}{4\pi^2}\lambda(Q_0)
\ln\left(\frac{Q^2}{Q_0^2}\right)}\,\,\,.
\end{equation}
When the energy scale $Q$ grows, the
denominator in Eq.~(\ref{eq:lambda_sm_large_mh}) may vanish, in which
case $\lambda(Q)$ hits a pole, becomes infinite, and a triviality
condition needs to be imposed. This is avoided imposing that the
denominator in Eq.~(\ref{eq:lambda_sm_large_mh}) never vanishes, i.e. 
that $\lambda(Q)$ is always finite
and $1/\lambda(Q)>0$. This condition gives an explicit upper bound on
$M_H$:
\begin{equation}
M_H^2<\frac{8\pi^2v^2}{3\log\left(\frac{\Lambda^2}{v^2}\right)}\,\,\,,
\end{equation}
obtained from Eq.~(\ref{eq:lambda_sm_large_mh}) by setting
$Q\!=\!\Lambda$, the scale of new physics, and $Q_0\!=\!v$, the
electroweak scale.

On the other hand, for small $M_H$ , i.e. for small $\lambda$, the
last term in Eq.~(\ref{eq:sm_lambda_rge}) dominates and the
evolution of $\lambda(Q)$ looks like:
\begin{equation}
\lambda(\Lambda)=\lambda(v)-\frac{3}{4\pi^2}y_t^4
\log\left(\frac{\Lambda^2}{v^2}\right)\,\,\,.
\end{equation}
To assure the stability of the vacuum state of the theory we need to
require that $\lambda(\Lambda)\!>\!0$ and this gives a lower bound for
$M_H$:
\begin{equation}
\lambda(\Lambda)>0 \,\,\,\longrightarrow \,\,\,
M_H^2>\frac{3v^2}{2\pi^2}y_t^4\log\left(\frac{\Lambda^2}{v^2}\right)\,\,\,.
\end{equation}
More accurate analyses include higher order quantum correction in the
scalar potential and use a 2-loop renormalization group improved
effective potential, $V_{eff}$, whose nature and meaning has been
briefly sketched in Section~\ref{subsec:higgs_mechanism}.

\subsubsection{Indirect bounds from electroweak precision measurements}
\label{subsubsec:indirect_bound}

Once a Higgs field is introduced in the Standard Model, its virtual
excitations contribute to several physical observables, from the mass
of the $W$ boson, to various leptonic and hadronic asymmetries, to
many other electroweak observables that are usually considered in
precision tests of the Standard Model.  Since the Higgs-boson mass is
the only parameter in the Standard Model that is not directly
determined either theoretically or experimentally (previous to
discovery), it can be extracted indirectly from precision fits of all
the measured electroweak observables, within the fit uncertainty. This
is actually one of the most important results that can be obtained
from precision tests of the Standard Model and greatly illustrates the
predictivity of the Standard Model itself.  All available studies can
be found on the LEP Electroweak Working Group and on the LEP Higgs
Working Group Web pages~\cite{lepewwg:wp,lephwg:wp} as well as in
their main publications. An excellent series of lectures on the
subject of \emph{Precision Electroweak Physics} is also available from
a previous TASI school \cite{Matchev:2004yw}.
\begin{figure}[htb]
\centering
\includegraphics[scale=0.4]{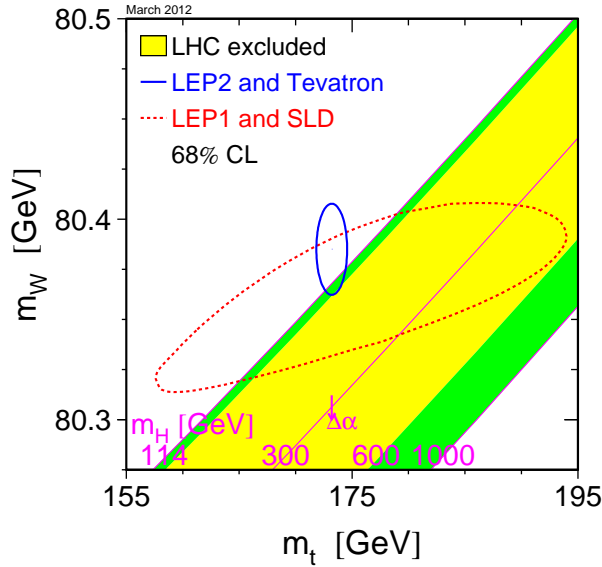}
\caption[]{Comparison of the indirect measurements of $M_W$ and $m_t$
(LEP I+SLD data) (solid contour) and the direct measurement
($p\bar{p}$ colliders and LEP II data) (dashed contour). In both cases
the 68\% CL contours are plotted. Also shown is the SM relationship
for these masses as a function of the Higgs-boson mass, $m_H$. The
arrow labeled $\Delta\alpha$ shows the variation of this relation if
$\alpha(M_Z^2)$ is varied by one standard deviation. From Ref.~\cite{lepewwg:wp}.
\label{fig:contour_mw_mt}}
\end{figure}
\begin{figure}[htb]
\begin{tabular}{lr}
\hspace{-1.truecm}
\begin{minipage}{0.5\linewidth}
{\includegraphics[scale=0.38]{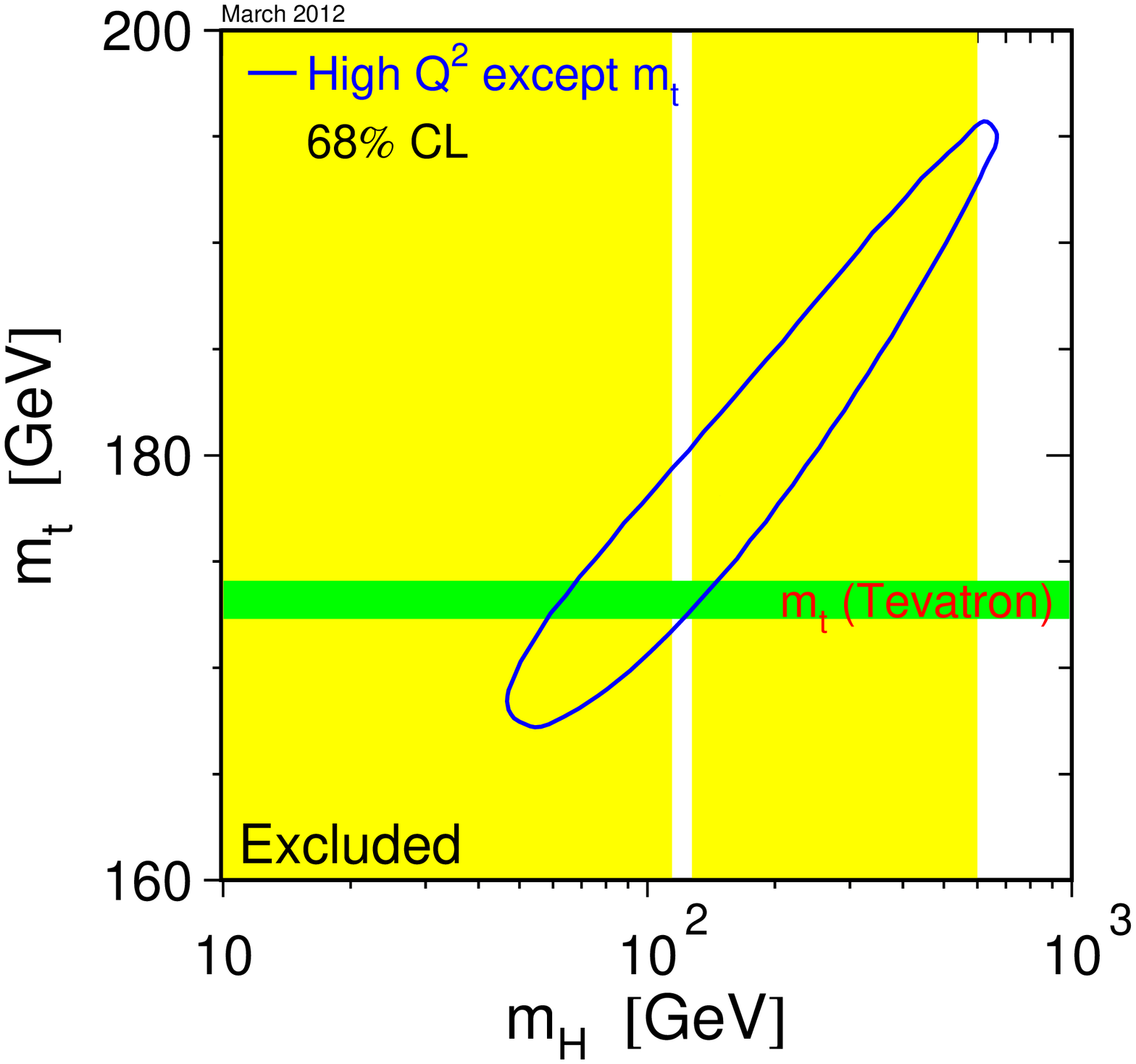}}
\end{minipage} &
\hspace{0.3truecm}
\begin{minipage}{0.5\linewidth}
{\includegraphics[scale=0.38]{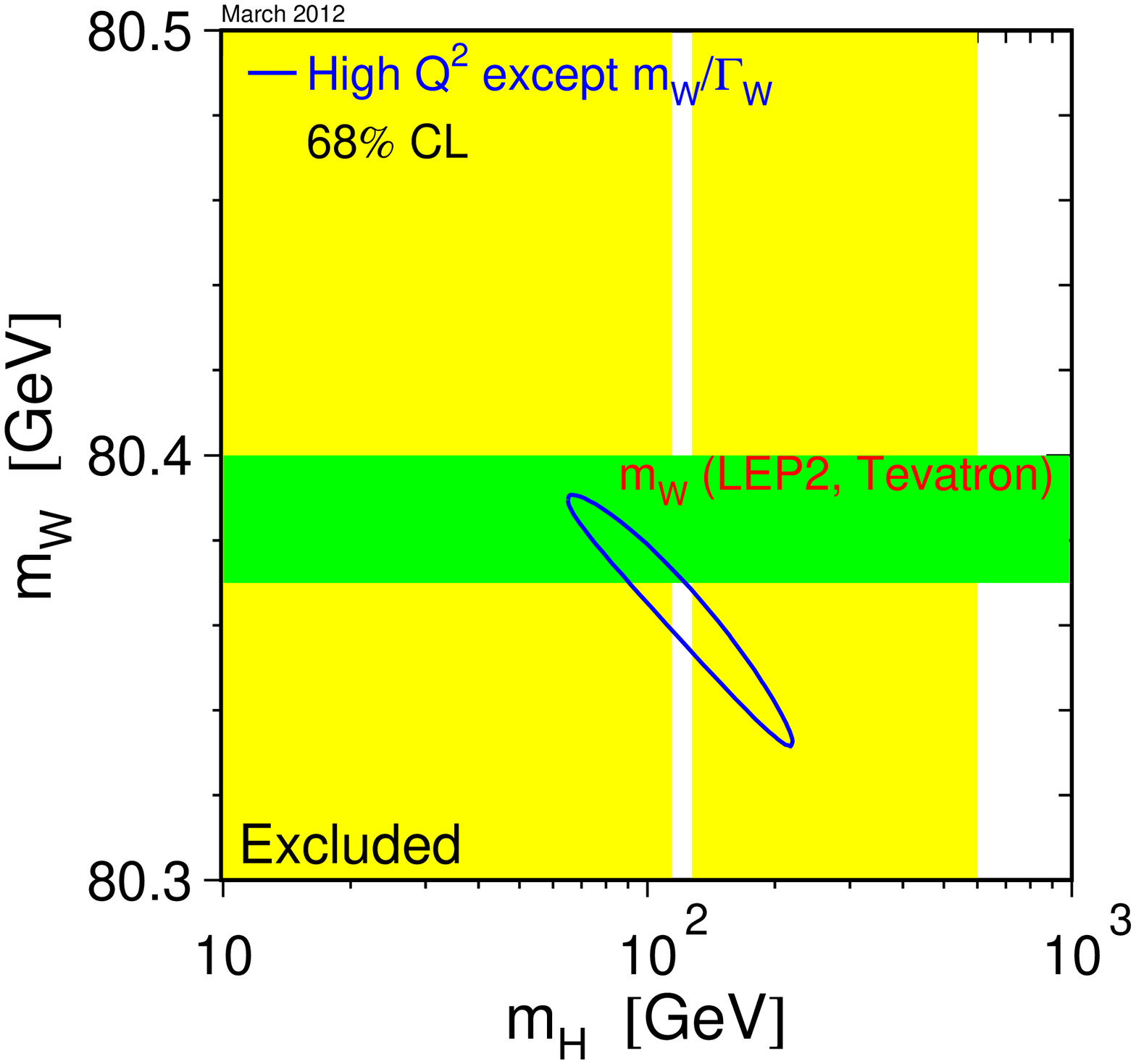}}
\end{minipage}
\end{tabular}
\caption[]{The 68\% confidence level contour in $m_t$ and $M_H$ for
the fit to all data except the direct measurement of $m_t$, indicated
by the shaded horizontal band of $\pm 1\sigma$ width. The vertical
band shows the 95\% CL exclusion limit on $M_H$ from direct
searches. From Ref.~\cite{lepewwg:wp}. \label{fig:contour_mh_mt}}
\end{figure}
The correlation between the Higgs-boson mass $M_H$, the $W$ boson mass
$M_W$, the top-quark mass $m_t$, and the precision data is illustrated
in Figs.~\ref{fig:contour_mw_mt} and \ref{fig:contour_mh_mt}. Apart
from the impressive agreement existing between the indirect
determination of $M_W$ and $m_t$ and their experimental measurements
we see in Fig.~\ref{fig:contour_mw_mt} that the 68\% CL contours from
LEP, SLD, and Tevatron measurements select a SM Higgs-boson mass
region roughly below 200~GeV. Therefore, assuming no physics beyond
the Standard Model at the weak scale, all available electroweak
precision data are consistent with a light Higgs boson.

\begin{figure}
\centering
\includegraphics[scale=0.4]{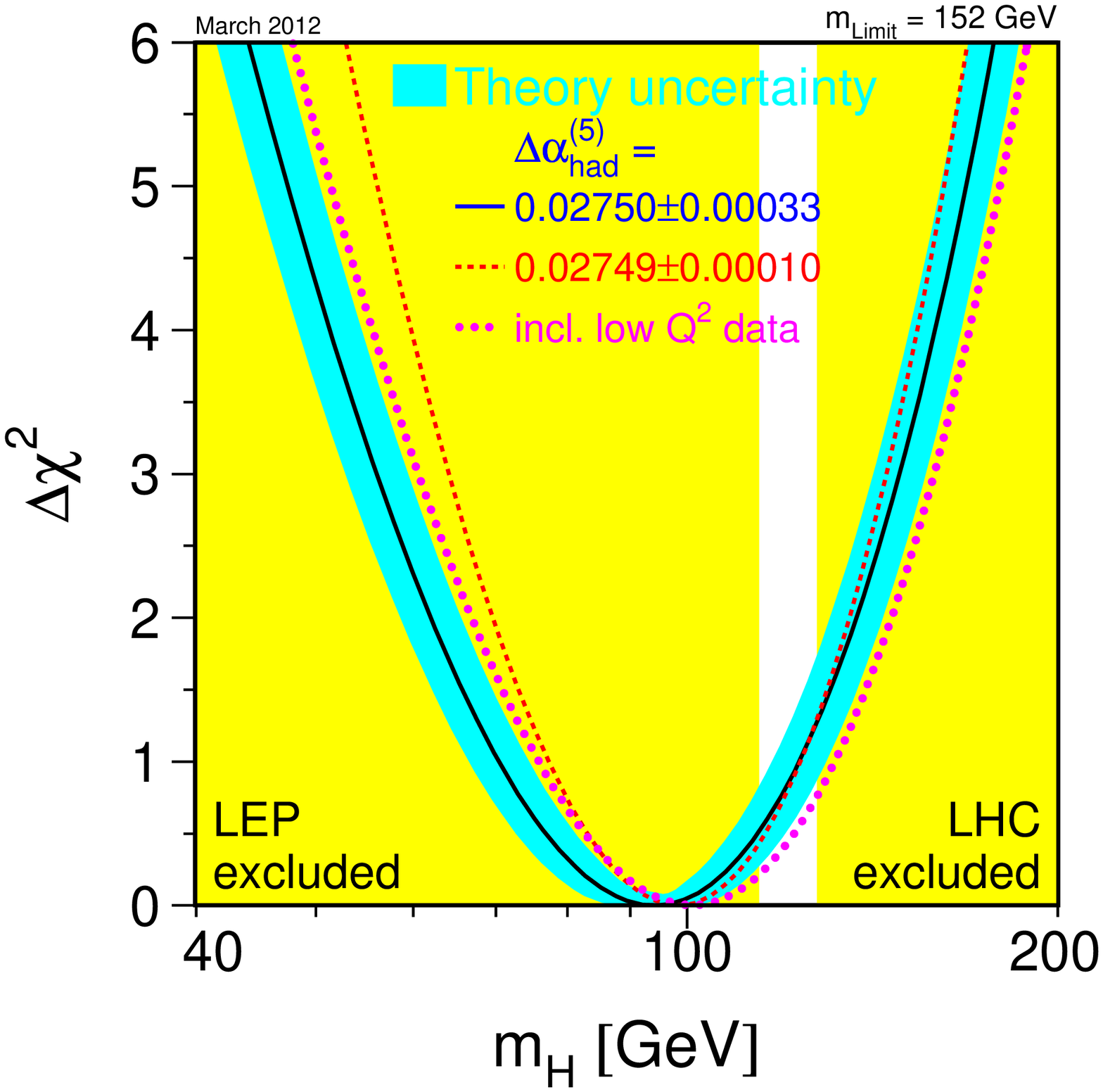}
\caption[]{$\Delta\chi^2=\chi^2-\chi^2_{min}$ \emph{vs.} $M_H$
curve. The line is the result of the fit using all electroweak data;
the band represents an estimate of the theoretical error due to
missing higher order corrections. The vertical band shows the $95\%$
CL exclusion limit on $M_H$ from direct searches. The solid and dashed
curves are derived using different evaluations of
$\Delta\alpha_{had}^{(5)}(M_Z^2)$. The dotted curve includes low $Q^2$
data. From Ref.~\cite{lepewwg:wp}. \label{fig:blue_band_mh}}
\end{figure}
The actual value of $M_H$ emerging from the electroweak precision fits
strongly depends on theoretical predictions of physical observables
that include different orders of strong and electroweak
corrections. As an example, in Fig.~\ref{fig:contour_mw_mt} the
magenta arrow shows how the yellow band would move for one standard
deviation variation in the QED fine-structure constant
$\alpha(m_Z^2)$. It also depends on the fit input parameters. As we
see in Fig.~\ref{fig:contour_mh_mt},
$M_H$ grows for larger $m_t$ and smaller $M_W$. The limits deduced from 
Fig.~\ref{fig:contour_mw_mt} and \ref{fig:blue_band_mh} is summarized as
\begin{equation}
\label{eq:mh_indirect_w12}
\left\{
\begin{array}{l}
M_H=94^{+29}_{-24}\,\,\,\mbox{GeV}\\
M_H<152 (171)\,\,\,\mbox{GeV}\,\,\,(95\%\,\,\mbox{CL})
\end{array}
\right.
\begin{array}{l}
\,\,\,\,\,\mbox{for}\,\,\,\,\,
m_t\!=\!173.2\pm 0.9\,\,\,\mbox{GeV}\,\,\,,\\
\,\,\,\,\,\mbox{and}\,\,\,\,\,
M_W\!=\!80.385\pm 0.015\,\,\,\mbox{GeV}\,\,\,.
\end{array}
\end{equation}

A large region of the $\Delta\chi^2$ band in
Fig.~\ref{fig:blue_band_mh}, in particular the region about the
minimum, is already excluded, and values of $M_H$ very close to the
experimental lower bound seem to be favored.
\begin{figure}
\centering
\includegraphics[scale=0.35]{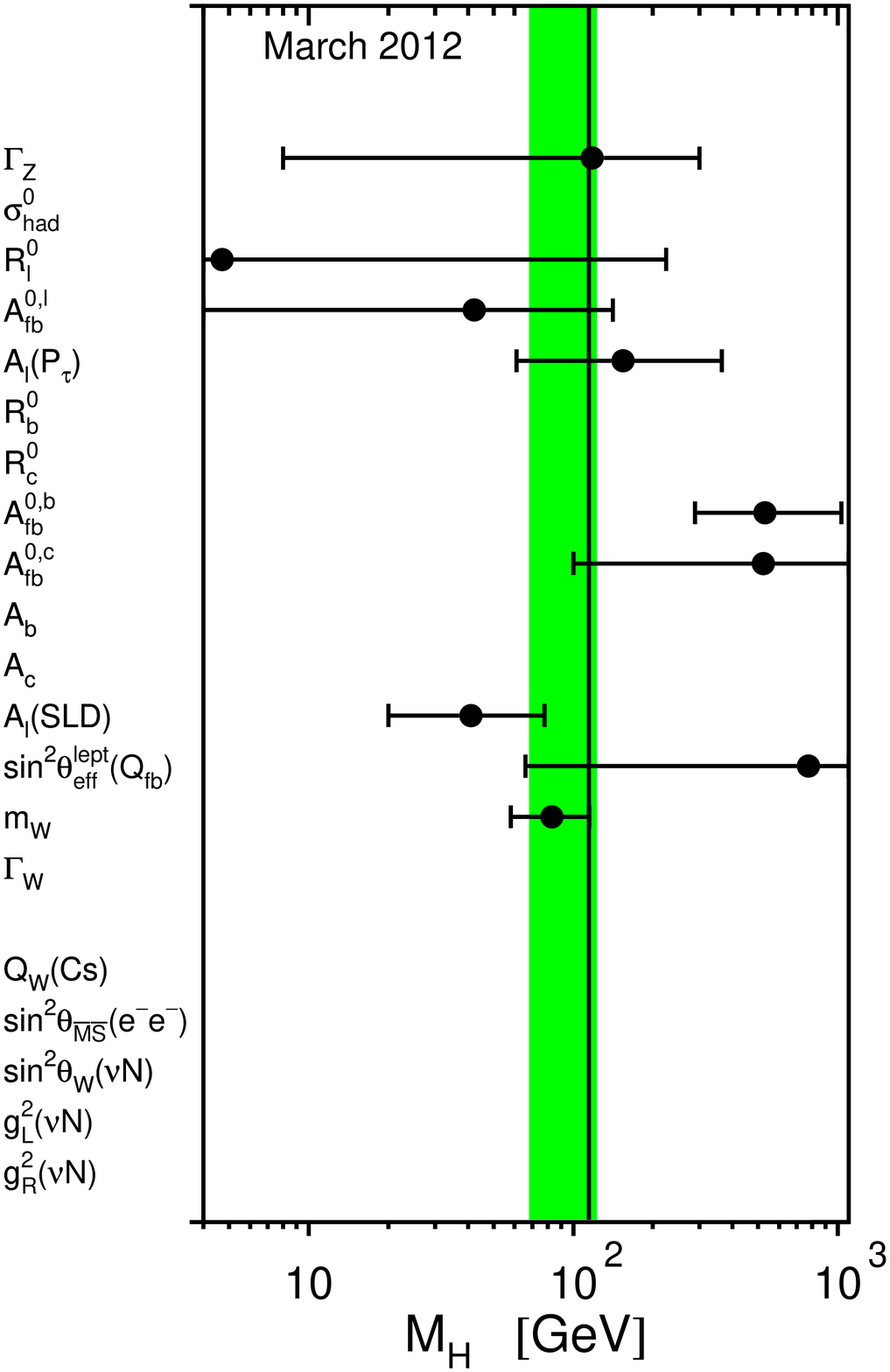}
\caption[]{Preferred range for the SM Higgs-boson mass $M_H$ as determined
from various electroweak observables. The shaded band shows the
overall constraint on the mass of the Higgs boson as derived from the
full data set. From Ref.~\cite{lepewwg:wp}. \label{fig:green_band_mh}}
\end{figure}
It is fair to conclude that the issue of constraining $M_H$ from
electroweak precision fits is open to controversies and, at a closer
look, emerges as a not clear cut statement. With this respect,
Fig.~\ref{fig:green_band_mh} illustrates the sensitivity of a few
selected electroweak observables to the Higgs-boson mass as well as
the preferred range for the SM Higgs-boson mass as determined from all
electroweak observables .  One can observe that $M_W$ and the leptonic
asymmetries prefer a lighter Higgs boson, while $A_{FB}^{b,c}$ and the
NuTeV determination of $\sin^2\theta_W$ prefer a heavier Higgs
boson. A certain
\emph{tension} is still present in the data.  We could just think that
things will progressively adjust and, after the discovery of a light
Higgs boson at the LHC, this will result in yet
another amazing success of the Standard Model. Or, one can interpret
the situation depicted in Fig.~\ref{fig:green_band_mh} as an
unavoidable indication of the presence of new physics beyond the
Standard Model and only more accurate studies of the newly discovered
spin-0 particle at the LHC will help shed some light on the puzzle.

\subsubsection{Fine-tuning}
\label{subsubsec:fine_tuning}
One aspect of the Higgs sector of the Standard Model that is
traditionally perceived as problematic is that higher order
corrections to the Higgs-boson mass parameter square contain quadratic
ultraviolet divergences. This is expected in a $\lambda\phi^4$ theory
and it does not pose a renormalizability problem, since a
$\lambda\phi^4$ theory is renormalizable. However, although per se
renormalizable, these quadratic divergences leave the \emph{inelegant}
feature that the Higgs-boson renormalized mass square has to result
from the
\emph{adjusted} or \emph{fine-tuned} balance between a bare Higgs
boson mass square and a counterterm that is proportional to the
ultraviolet cutoff square. If the physical Higgs mass has to live at
the electroweak scale, this can cause a fine-tuning of several orders
of magnitude when the scale of new physics $\Lambda$ (the ultraviolet
cutoff of the Standard Model interpreted as an effective low energy
theory) is well above the electroweak scale. Ultimately this is
related to a symmetry principle, or better to the absence of a
symmetry principle. Indeed, setting to zero the mass of the scalar
fields in the Lagrangian of the Standard Model does not restore any
symmetry to the model. Hence, the mass of the scalar fields are not
protected against large corrections.

Models of new physics beyond the Standard Model should address this
fine-tuning problem and propose a more satisfactory mechanism to
obtain the mass of the Higgs particle(s) around the electroweak
scale. Supersymmetric models, for instance, have the remarkable
feature that fermionic and bosonic degrees of freedom conspire to
cancel the Higgs mass quadratic loop divergence, when the symmetry is
exact. Other non supersymmetric models, like little Higgs models,
address the problem differently, by interpreting the Higgs boson as a
Goldstone boson of some global approximate symmetry. In both cases the
Higgs mass turns out to be proportional to some small deviation from
an exact symmetry principle, and therefore intrinsically small.

\begin{figure}
\begin{center}
\includegraphics[scale=.45]{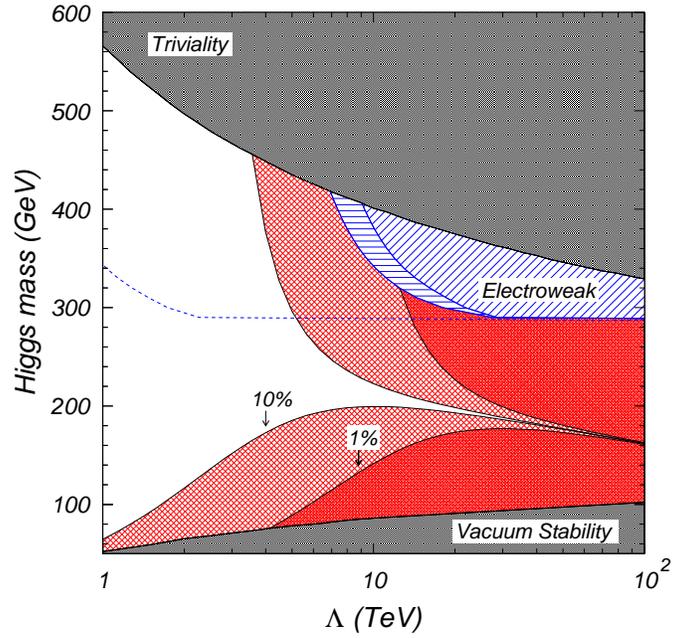}
\caption[]{The SM Higgs-boson mass $M_H$ as a function of the scale of
new physics $\Lambda$, with all the constraints derived from
unitarity, triviality, vacuum stability, electroweak precision fits,
and the requirement of a limited fine-tuning. The empty region is
consistent with all the constraints and less than 1 part in 10
fine-tuning. From Ref.~\cite{Kolda:2000wi}.\label{fig:mh_vs_lambda}}
\end{center}
\end{figure}
As suggested in Ref.~\cite{Kolda:2000wi}, the \emph{no fine-tuning}
condition in the Standard Model can be softened and translated into a
\emph{maximum amount of allowed fine-tuning}, that can be directly
related to the scale of new physics. As derived in
Section~\ref{subsec:higgs_mechanism}, upon spontaneous breaking of the
electroweak symmetry, the SM Higgs-boson mass at tree level is given
by $M_H^2\!=\!-2\mu^2$, where $\mu^2$ is the coefficient of the
quadratic term in the scalar potential. Higher order corrections to
$M_H^2$ can therefore be calculated as loop corrections to $\mu^2$,
i.e. by studying how the effective potential in
Eq.~(\ref{eq:veff_exp}) and its minimum condition are modified by loop
corrections.  If we interpret the Standard Model as the electroweak
scale effective limit of a more general theory living at a high scale
$\Lambda$, then the most general form of $\mu^2$ including all loop
corrections is:
\begin{equation}
\label{eq:mu_renormalized}
\bar\mu^2=\mu^2+\Lambda^2\sum_{n=0}^{\infty}c_n(\lambda_i)\log^n(\Lambda/Q)
\,\,\,,
\end{equation}
where $Q$ is the renormalization scale, $\lambda_i$ are a set of input
parameters (couplings) and the $c_n$ coefficients can be deduced from
the calculation of the effective potential at each loop order. As
noted originally by Veltman, there would be no fine-tuning problem if
the coefficient of $\Lambda^2$ in Eq.~(\ref{eq:mu_renormalized}) were
zero, i.e. if the loop corrections to $\mu^2$ had to vanish. This
condition, known as \emph{Veltman condition}, is usually over constraining,
since the number of independent $c_n$ (set to zero by the Veltman
condition) can be larger than the number of inputs
$\lambda_i$. However the Veltman condition can be relaxed, by
requiring that only the sum of a finite number of terms in the
coefficient of $\Lambda^2$ is zero, i.e. requiring that:
\begin{equation}
\label{eq:milder_veltman_condition}
\sum_0^{n_{max}}c_n(\lambda_i)\log^n(\Lambda/M_H)=0\,\,\,,
\end{equation}
where the renormalization scale $\mu$ has been arbitrarily set to
$M_H$ and the order $n$ has been set to $n_{max}$, fixed by the
required order of loop in the calculation of $V_{eff}$. This is based
on the fact that higher orders in $n$ come from higher loop effects
and are therefore suppressed by powers of $(16\pi^2)^{-1}$. Limiting
$n$ to $n_{max}$, Eq.~(\ref{eq:milder_veltman_condition}) can now have
a solution. Indeed, if the scale of new physics $\Lambda$ is not too
far from the electroweak scale, then the Veltman condition in
Eq.~(\ref{eq:milder_veltman_condition}) can be softened even more by
requiring that:
\begin{equation}
\label{eq:mildest_veltman_condition}
\sum_0^{n_{max}}c_n(\lambda_i)\log^n(\Lambda/M_H)<\frac{v^2}{\Lambda^2}
\,\,\,.
\end{equation}
This condition determines a value of $\Lambda_{max}$ such that for
$\Lambda\le\Lambda_{max}$ the stability of the electroweak scale does
not require any dramatic cancellation in $\bar{\mu}^2$. In other
words, for $\Lambda\le\Lambda_{max}$ the renormalization of the SM
Higgs-boson mass does not require any fine-tuning. As an example, for
$n_{max}\!=\!0$, $c_0\!=\!(32\pi^2v^2)^{-1}
3(2M_W^2+M_Z^2+M_H^2-4m_t^2)$, and the stability of the electroweak
scale is assured up to $\Lambda$ of the order of $4\pi v\simeq 2$~TeV.
For $n_{max}\!=\!1$ the maximum $\Lambda$ is pushed up to
$\Lambda\simeq 15$~TeV and for $n_{max}\!=\!2$ up to $\Lambda\simeq
50$~TeV. So, just going up to 2-loops assures us that we can consider
the SM Higgs sector free of fine-tuning up to scales that are well
beyond where we would hope to soon discover new physics.

For each value of $n_{max}$, and for each corresponding
$\Lambda_{max}$, $M_H$ becomes a function of the cutoff $\Lambda$, and
the amount of fine-tuning allowed in the theory limits the region in
the $(\Lambda,M_H)$ plane allowed to $M_H(\Lambda)$. This is well
represented in Fig.~\ref{fig:mh_vs_lambda}, where also the constraint
from the conditions of unitarity (see
Section~\ref{subsubsec:unitarity}), triviality (see
Section~\ref{subsubsec:triviality_vacuumstability}), vacuum stability
(see Section~\ref{subsubsec:triviality_vacuumstability}) and
electroweak precision fits (see
Section~\ref{subsubsec:indirect_bound}) are summarized. Finally, the
main lesson we take away from this plot is that if a Higgs boson is
discovered new physics is just around the corner and should manifest
itself at the LHC.

\subsection{The Higgs sector of the Minimal Supersymmetric Standard Model}
\label{subsec:higgs_mssm}

In the supersymmetric extension of the Standard Model, the electroweak
symmetry is spontaneously broken via the Higgs mechanism introducing
two complex scalar $SU(2)_L$ doublets. The dynamics of the Higgs
mechanism goes pretty much unchanged with respect to the Standard
Model case, although the form of the scalar potential is more complex
and its minimization more involved. As a result, the $W^\pm$ and $Z^0$
weak gauge bosons acquire masses that depend on the parameterization
of the supersymmetric model at hand. At the same time, fermion masses
are generated by coupling the two scalar doublets to the fermions via
Yukawa interactions. A supersymmetric model is therefore a natural
reference to compare the Standard Model to, since it is a
theoretically sound extension of the Standard Model, still
fundamentally based on the same electroweak symmetry breaking
mechanism.

Far from being a simple generalization of the SM Higgs sector, the
scalar sector of a supersymmetric model can be theoretically more
satisfactory because: \emph{(i)} spontaneous symmetry breaking is
radiatively induced (\emph{i.e.} the sign of the quadratic term in the
Higgs potential is driven from positive to negative) mainly by the
evolution of the top-quark Yukawa coupling from the scale of
supersymmetry-breaking to the electroweak scale, and \emph{(ii)}
higher order corrections to the Higgs mass do not contain quadratic
divergences, since they cancel when the contribution of both scalars
and their super-partners is considered (see
Section~\ref{subsubsec:fine_tuning}).

At the same time, the fact of having a supersymmetric theory and two
scalar doublets modifies the phenomenological properties of the
supersymmetric physical scalar fields dramatically. In this Section we
will review only the most important properties of the Higgs sector of
the MSSM, so that in Section~\ref{sec:higgs_searches} we can compare
the physics of the SM Higgs boson to that of the MSSM Higgs bosons.

I will start by recalling some general properties of a Two Higgs
Doublet Model in Section~\ref{subsubsec:2hdm}, and I will then specify
the discussion to the case of the MSSM in
Section~\ref{subsubsec:higgs_mssm}. In
Sections~\ref{subsubsec:mssm_higgs_couplings_bosons} and
\ref{subsubsec:mssm_higgs_couplings_fermions} I will review the form
of the couplings of the MSSM Higgs bosons to the SM gauge bosons and
fermions, including the impact of the most important supersymmetric
higher order corrections. A thorough introduction to Supersymmetry and
the Minimal Supersymmetric Standard Model has been given during this
school by Prof.~H.~Haber to whose lectures I
refer~\cite{Haber:tasi04}.

\subsubsection{About Two Higgs Doublet Models}
\label{subsubsec:2hdm}
The most popular and simplest extension of the Standard Model is
obtained by considering a scalar sector made of two instead of one
complex scalar doublets. These models, dubbed \emph{Two Higgs Doublet
Models} (2HDM), have a richer spectrum of physical scalar
fields. Indeed, after spontaneous symmetry breaking, only three of the
eight original scalar degrees of freedom (corresponding to two complex
doublet) are reabsorbed in transforming the originally massless vector
bosons into massive ones. The remaining five degrees of freedom
correspond to physical degrees of freedom in the form of: two neutral
scalar, one neutral pseudoscalar, and two charged scalar fields.

At the same time, having multiple scalar doublets in the Yukawa
Lagrangian (see Eq.~(\ref{eq:yukawa_lagrangian})) allows for scalar
flavor changing neutral current. Indeed, when generalized to the case 
of two scalar doublet $\phi^1$ and $\phi^2$,
Eq.~(\ref{eq:yukawa_lagrangian}) becomes (quark case only):
\begin{equation}
\label{eq:yukawa_lagrangian_2hdm}
\mathcal{L}_{Yukawa}=
-\sum_{k=1,2}\Gamma_{ij,k}^u\bar{Q}^i_L\Phi^{k,c} u^j_R
-\sum_{k=1,2}\Gamma_{ij,k}^d\bar{Q}^i_L\Phi^k d^j_R +
\mathrm{h.c.}\,\,\,,
\end{equation}
where each pair of fermions $(i,j)$ couple to a linear combination of
the scalar fields $\phi^1$ and $\phi^2$.
When, upon spontaneous symmetry
breaking, the fields $\phi^1$ and $\phi^2$ acquire vacuum
expectation values
\begin{equation}
\langle\Phi^k\rangle=\frac{v^k}{\sqrt{2}}\,\,\,\,\,\mbox{for}\,\,\,\,\,
k=1,2\,\,\,,
\end{equation}
the parameterization of $\mathcal{L}_{Yukawa}$ of
Eq.~(\ref{eq:yukawa_lagrangian_2hdm}) in the vicinity of the minimum
of the scalar potential, with $
\Phi^k=\Phi^{\prime k}+v^k$ (for $k=1,2$), gives:
\begin{equation}
\mathcal{L}_{Yukawa}=
-\bar{u}^i_L
\underbrace{\sum_k\Gamma_{ij,k}^u\frac{v^k}{\sqrt{2}}}_{M^u_{ij}}u^j_R
-\bar{d}^i_L
\underbrace{\sum_k\Gamma_{ij,k}^d\frac{v^k}{\sqrt{2}}}_{M^d_{ij}}d^j_R+ 
\mathrm{h.c.} + \mbox{\small{FC couplings}}\,\,\,,
\end{equation}
where the fermion mass matrices $M^u_{ij}$ and $M^d_{ij}$ are now
proportional to a linear combination of the vacuum expectation values
of $\phi^1$ and $\phi^2$. The diagonalization of $M^u_{ij}$ and
$M^d_{ij}$ does not imply the diagonalization of the couplings of the
$\phi^{\prime k}$ fields to the fermions, and Flavor Changing (FC)
couplings arise.  This is perceived as a problem in view of the
absence of experimental evidence to support neutral flavor changing
effects. If present, these effects have to be tiny in most processes
involving in particular the first two generations of quarks, and a
safer way to build a 2HDM is to forbid them all together at the
Lagrangian level.  This is traditionally done by requiring either that
$u$-type and $d$-type quarks couple to the same doublet (Model I) or
that $u$-type quarks couple to one scalar doublet while $d$-type
quarks to the other (Model II). Indeed, these two different
realization of a 2HDM can be justified by enforcing on
$\mathcal{L}_{Yukawa}$ the following \emph{ad hoc} discrete symmetry:
\begin{equation}
\left\{
\begin{array}{c}
\Phi^1\rightarrow -\Phi^1\,\,\,\,\mathrm{and}\,\,\,\,\Phi^2\rightarrow\Phi^2 \\
d^i\rightarrow -d^i\,\,\,\,\mathrm{and}\,\,\,\, u^j\rightarrow\pm u^j
\end{array}
\right.
\end{equation}
The case in which FC scalar neutral current are not forbidden (Model
III) has also been studied in detail. In this case both up and
down-type quarks can couple to both scalar doublets, and strict
constraints have to be imposed on the FC scalar couplings in particular
between the first two generations of quarks.

2HDMs have indeed a very rich phenomenology that has been extensively
studied. In these lectures, however, we will only compare the SM Higgs
boson phenomenology to the phenomenology of the Higgs bosons of the
MSSM, a particular kind of 2HDM that we will illustrate in the
following Sections.

\subsubsection{The MSSM Higgs sector: introduction}
\label{subsubsec:higgs_mssm}
The Higgs sector of the MSSM is actually a Model II 2HDM. It contains
two complex $SU(2)_L$ scalar doublets:
\begin{equation}
\label{eq:phiu_phid}
\Phi_1=\left(\begin{array}{c}\phi_1^+\\\phi_1^0\end{array}\right)
\,\,\,\,\,\,,\,\,\,\,\,\,
\Phi_2=\left(\begin{array}{c}\phi_2^0\\\phi_2^-\end{array}\right)
\,\,\,,
\end{equation}
with opposite hypercharge ($Y\!=\!\pm 1$), as needed to make the
theory anomaly-free\footnote{Another reason for the choice of a 2HDM
is that in a supersymmetric model the superpotential should be
expressed just in terms of superfields, not their conjugates. So, one
needs to introduce two doublets to give mass to fermion fields of
opposite weak isospin. The second doublet plays the role of $\phi^c$
in the Standard Model (see Eq.~(\ref{eq:yukawa_lagrangian})), where
$\phi^c$ has opposite hypercharge and weak isospin with respect to
$\phi$.}. $\Phi_1$ couples to the up-type and $\Phi_2$ to the
down-type quarks respectively.  Correspondingly, the Higgs part of the
superpotential can be written as:
\begin{eqnarray}
\label{eq:higgs_superpotential}
V_H&=&(|\mu|^2+m_1^2)|\Phi_1|^2+(|\mu|^2+m_2^2)|\Phi_2|^2
-\mu B\epsilon_{ij}(\Phi_1^i\Phi_2^j+h.c.)\nonumber\\
&+&\frac{g^2+g^{\prime 2}}{8}\left(|\Phi_1|^2-|\Phi_2|^2\right)^2
+\frac{g^2}{2}|\Phi_1^\dagger\Phi_2|^2\,\,\,,
\end{eqnarray}
in which we can identify three different contributions
\cite{Haber:tasi04,Djouadi:2005gj}:
\begin{itemize}
\item[\emph{(i)}] the so
called $D$ terms, containing the quartic scalar interactions, which
for the Higgs fields $\Phi_1$ and $\Phi_2$ correspond to:
\begin{equation}
\label{eq:higgs_superpotential_d_terms}
\frac{g^2+g^{\prime 2}}{8}\left(|\Phi_1|^2-|\Phi_2|^2\right)^2
+\frac{g^2}{2}|\Phi_1^\dagger\Phi_2|^2\,\,\,,
\end{equation}
with $g$ and $g^\prime$ the gauge couplings of $SU(2)_L$ and $U(1)_Y$
respectively;
\item[\emph{(ii)}] the so called $F$ terms, corresponding to:
\begin{equation}
\label{eq:higgs_superpotential_f_terms}
|\mu|^2(|\Phi_1|^2+|\Phi_2|^2)\,\,\,;
\end{equation}
\item[\emph{(iii)}] the soft SUSY-breaking scalar Higgs mass and bilinear
terms, corresponding to:
\begin{equation}
\label{eq:higgs_superpotential_soft_terms}
m_1^2|\Phi_1|^2+m_2^2|\Phi_2|^2
-\mu B\epsilon_{ij}(\Phi_1^i\Phi_2^j+h.c.)\,\,\,.
\end{equation}
\end{itemize}
Overall, the scalar potential in Eq.~(\ref{eq:higgs_superpotential})
depends on three independent combinations of parameters,
$|\mu|^2+m_1^2$, $|\mu|^2+m_2^2$, and $\mu B$. One basic difference
with respect to the SM case is that the quartic coupling has been
replaced by gauge couplings. This reduced arbitrariness will play an
important role in the following.

Upon spontaneous symmetry breaking, the neutral components of
$\Phi_1$ and $\Phi_2$ acquire vacuum expectation values
\begin{equation}
\label{eq:phiu_phid_vev}
\langle\Phi_1\rangle=
\frac{1}{\sqrt{2}}\left(\begin{array}{c}0\\v_1\end{array}\right)
\,\,\,\,\,\,,\,\,\,\,\,\,
\langle\Phi_2\rangle=
\frac{1}{\sqrt{2}}\left(\begin{array}{c}v_2\\0\end{array}\right)\,\,\,,
\end{equation}
and the Higgs mechanism proceed as in the Standard Model except that
now one starts with eight degrees of freedom, corresponding to the
two complex doublets $\Phi_1$ and $\Phi_2$. Three degrees of freedom
are absorbed in making the $W^\pm$ and the $Z^0$ massive. The $W$ mass
is chosen to be: $M_W^2=g^2(v_1^2+v_2^2)/4=g^2v^2/4$, and this fixes
the normalization of $v_1$ and $v_2$, leaving only two independent
parameters to describe the entire MSSM Higgs sector.  The remaining
five degrees of freedom are physical and correspond to two neutral
scalar fields
\begin{eqnarray}
\label{eq:mssm_scalar_higgses}
h^0&=&-(\sqrt{2}\mbox{Re}\phi_2^0-v_2)\sin\alpha+
 (\sqrt{2}\mbox{Re}\phi_1^0-v_1)\cos\alpha\\
H^0&=&
(\sqrt{2}\mbox{Re}\phi_2^0-v_2)\cos\alpha+
(\sqrt{2}\mbox{Re}\phi_1^0-v_1)\sin\alpha\,\,\,,\nonumber
\end{eqnarray}
one neutral pseudoscalar field
\begin{equation}
\label{eq:mssm_pseudoscalar_higgs}
A^0=
\sqrt{2}\left(\mbox{Im}\phi_2^0\sin\beta+\mbox{Im}\phi_1^0\cos\beta\right)\,\,\,,
\end{equation}
and two charged scalar fields
\begin{equation}
\label{eq:mssm_charged_higgs}
H^\pm=\phi_2^\pm\sin\beta+\phi_1^\pm\cos\beta\,\,\,,
\end{equation}
where $\alpha$ and $\beta$ are mixing angles, and
$\tan\beta\!=\!v_1/v_2$. At tree level, the masses of the scalar and
pseudoscalar degrees of freedom satisfy the following relations:
\begin{eqnarray}
\label{eq:mssm_higgs_masses}
M_{H^\pm}^2&=&M_A^2+M_W^2\,\,\,,\\
M_{H,h}^2&=&\frac{1}{2}\left(
M_A^2+M_Z^2\pm((M_A^2+M_Z^2)^2-4M_Z^2M_A^2\cos^2 2\beta)^{1/2}\right)
\,\,\,,\nonumber
\end{eqnarray}
making it natural to pick $M_A$ and $\tan\beta$ as the two independent
parameters of the Higgs sector.

\begin{figure}
\centering
\includegraphics[scale=0.5]{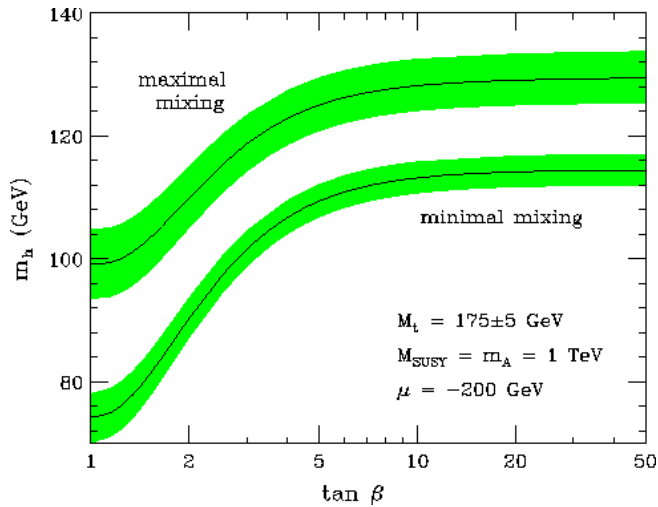}
\caption[]{The mass of the light neutral scalar Higgs boson, $h^0$, as
a function of $\tan\beta$, in the \emph{minimal mixing} and
\emph{maximal mixing} scenario. The green bands are obtained by varying
the top-quark mass in the $m_t\!=\!175\pm 5$~GeV range. 
The plot is built by fixing $M_A\!=\!1$~TeV and
$M_{SUSY}\!\equiv\!M_Q\!=\!\!M_U\!=\!M_D\!=\!1$~TeV. From
Ref.~\cite{Carena:2002es}.\label{fig:mh_upper_bound}}
\end{figure}

\begin{figure}
\centering
\includegraphics[scale=0.5]{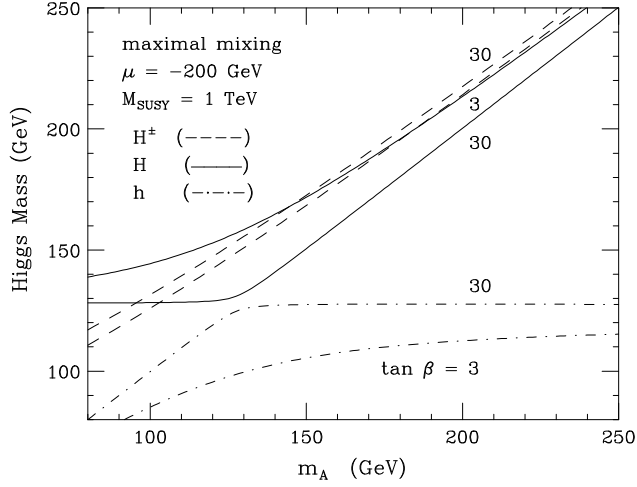}
\caption[]{The mass of the light ($h^0$) and heavy ($H^0$) neutral
scalar Higgs bosons, and of the charged scalar Higgs boson ($H^\pm$)
as a function of the neutral pseudoscalar mass $M_A$, for two
different values of $\tan\beta$ ($\tan\beta\!=\!3,30$). The top-quark
mass is fixed to $m_t\!=\!174.3$~GeV and
$M_{SUSY}\!\equiv\!M_Q\!=\!\!M_U\!=\!M_D\!=\!1$~TeV. The maximal
mixing scenario is chosen. From
Ref.~\cite{Carena:2002es}.\label{fig:mass_higgs_ma_tanb}}
\end{figure}

Eq.~(\ref{eq:mssm_higgs_masses}) provides the famous tree level upper
bound on the mass of one of the neutral scalar Higgs bosons, $h^0$:
\begin{equation}
\label{eq:mh0_upper_bound_tree_level}
M_{h}^2\le M_Z^2\cos 2\beta\le M_Z^2\,\,\,,
\end{equation}
which already contradicts the current experimental lower bound set by
LEP II: $M_{h}>93.0$~GeV~\cite{lephwg:2004mssm}. The contradiction is
lifted by including higher order radiative corrections to the Higgs
spectrum, in particular by calculating higher order corrections to the
neutral scalar mass matrix. Over the past few years a huge effort has
been dedicated to the calculation of the full one-loop corrections and
of several leading and sub-leading sets of two-loop corrections,
including resummation of leading and sub-leading logarithms via
appropriate renormalization group equation (RGE) methods. A detailed
discussion of this topic can be found in some recent
reviews~\cite{Carena:2002es,Heinemeyer:2004ms,Heinemeyer:2004gx} and
in the original literature referenced therein. For the purpose of
these lectures, let us just observe that, qualitatively, the impact of
radiative corrections on $M_{h}^{max}$ can be seen by just including
the leading two-loop corrections proportional to $y_t^2$, the square
of the top-quark Yukawa coupling, and applying RGE techniques to resum
the leading orders of logarithms. In this case, the upper bound on the
light neutral scalar in Eq.~(\ref{eq:mh0_upper_bound_tree_level}) is
modified as follows:
\begin{equation}
\label{eq:mh0_upper_bound_loop_level}
M_{h}^2\le M_Z^2+\frac{3g^2m_t^2}{8\pi^2M_W^2}
\left[\log\left(\frac{M_S^2}{m_t^2}\right)+
\frac{X_t^2}{M_S^2}\left(1-\frac{X_t^2}{12 M_S^2}\right)\right]\,\,\,,
\end{equation}
where $M_S^2=(M_{\tilde t_1}^2+M_{\tilde t_2}^2)/2$ is the average of
the two top-squark masses, $m_t$ is the running top-quark mass (to
account for the leading two-loop QCD corrections), and $X_t$ is the
top-squark mixing parameter defined by the top-squark mass matrix:
\begin{equation}
\label{eq:stop_mass_matrix}
\left(
\begin{array}{cc}
M_{Q_t}^{2}+m_t^2+D_L^t & m_t X_t\\
m_t X_t & M_{R_t}^2+m_t^2+D_R^t
\end{array}
\right)\,\,\,,
\end{equation}
with $X_t\equiv A_t-\mu\cot\beta$ ($A_t$ being one of the top-squark soft
SUSY breaking trilinear coupling),
$D_L^t=(1/2-2/3\sin\theta_W)M_Z^2\cos 2\beta$, and
$D_R^t=2/3\sin^2\theta_W M_Z^2\cos 2\beta$.
Fig.~\ref{fig:mh_upper_bound} illustrates the behavior of $M_h$
as a function of $\tan\beta$, in the case of minimal and maximal
mixing. For large $\tan\beta$ a plateau (i.e. an upper bound) is
clearly reached. The green bands represent the variation of $M_h$
as a function of $m_t$ when $m_t\!=\!175\pm 5$~GeV.
If top-squark mixing is maximal, the upper bound on $M_{h}$ is
approximately $M_h^{max}\!\simeq\! 135$~GeV\footnote{This limit is
obtained for $m_t\!=\!175$ GeV, and it can go up to
$M_h^{max}\!\simeq\!144$~GeV for $m_t\!=\!178$~GeV.}.  The behavior of
both $M_{h,H}$ and $M_{H^\pm}$ as a function of $M_A$ and $\tan\beta$
is summarized in Fig.~\ref{fig:mass_higgs_ma_tanb}, always for the
case of maximal mixing. It is interesting to notice that for all
values of $M_A$ and $\tan\beta$ the $M_H\!>\!M_h^{max}$. Also we
observe that, in the limit of large $\tan\beta$, \emph{i)} for
$M_A\!<\!M_h^{max}$: $M_h\simeq M_A$ and $M_H\simeq M_h^{max}$, while
\emph{ii)} for $M_A\!>\!M_h^{max}$: $M_H\simeq
M_A$ and $M_h\simeq M_h^{max}$.

\subsubsection{MSSM Higgs-boson couplings to electroweak gauge bosons}
\label{subsubsec:mssm_higgs_couplings_bosons}
The Higgs-boson couplings to the electroweak gauge bosons are obtained
from the kinetic term of the scalar Lagrangian, in strict analogy to
what we have explicitly seen in the case of the SM Higgs boson.
Here, we would like to recall the form of the $H_iVV$ and $H_iH_jV$
couplings (for $H_i\!=\!h^0,H^0,A^0,H^\pm$, and $V\!=\!W^\pm,Z^0$) that
are most important in order to understand the main features of the
MSSM plots that will be shown in Section~\ref{sec:higgs_searches}.  

First of all, the couplings of the neutral scalar Higgs bosons to both
$W^\pm$ and $Z^0$ can be written as:
\begin{equation}
\label{eq:couplings_hVV}
g_{hVV}=g_V M_V\sin(\beta-\alpha)g^{\mu\nu}\,\,\,\,\,,\,\,\,\,\,
g_{HVV}=g_V M_V\cos(\beta-\alpha)g^{\mu\nu}\,\,\,,
\end{equation}
where $g_V\!=\!2M_V/v$, while the $A^0VV$ and $H^\pm VV$ couplings
vanish because of CP-invariance. As in the SM case, since the photon
is massless, there are no tree level $\gamma\gamma H_i$ and $\gamma
Z^0 H_i$ couplings.

Moreover, in the neutral Higgs sector, only the $h^0A^0Z^0$ and
$H^0A^0Z^0$ couplings are allowed and given by:
\begin{equation}
\label{eq:coupling_hAZ_HAZ}
g_{hAZ}=\frac{g\cos(\beta-\alpha)}{2\cos\theta_W}(p_h-p_A)^\mu
\,\,\,\,\,,\,\,\,\,\,
g_{HAZ}=-\frac{g\sin(\beta-\alpha)}{2\cos\theta_W}(p_H-p_A)^\mu\,\,\,,
\end{equation}
where all momenta are incoming. We also have several $H_iH_jV$
couplings involving the charge Higgs boson, namely:
\begin{eqnarray}
\label{eq:coupling_H+}
g_{H^+H^-Z}&=&
-\frac{g}{2\cos\theta_W}\cos 2\theta_W(p_{H^+}-p_{H^-})^\mu\,\,\,,\\
g_{H^+H^-\gamma}&=&-ie(p_{H^+}-p_{H^-})^\mu\,\,\,,\nonumber\\
g_{H^\mp hW^\pm}&=&
\mp i\frac{g}{2}\cos(\beta-\alpha)(p_{h}-p_{H^\mp})^\mu\,\,\,,
\nonumber\\
g_{H^\mp HW^\pm}&=&\pm i\frac{g}{2}\sin(\beta-\alpha)(p_{H}-p_{H^\mp})^\mu
\,\,\,,\nonumber\\
g_{H^\mp AW^\pm}&=&
\frac{g}{2}(p_{A}-p_{H^\pm})^\mu\,\,\,.\nonumber
\end{eqnarray}

At this stage it is interesting to introduce the so called
\emph{decoupling limit}, i.e. the limit of $M_A\gg M_Z$, and to
analyze how masses and couplings behave in this particular limit.
$M_{H^\pm}$ in Eq.~(\ref{eq:mssm_higgs_masses}) is unchanged, while
$M_{h,H}$ become:
\begin{equation}
\label{eq:M_H_decoupling_limit}
M_{h}\simeq M_{h}^{max}\,\,\,\,\,\mbox{and}\,\,\,\,\,
M_{H}\simeq M_A^2+M_Z^2\sin^2 2\beta\,\,\,.
\end{equation}
Moreover, as one can derive from the diagonalization of the
neutral scalar Higgs-boson mass matrix:
\begin{equation}
\label{eq:cos_betamalpha_decoupling_limit}
\cos^2(\beta-\alpha)=\frac{M_h^2(M_Z^2-M_h^2)}
{M_A^2(M_H^2-M_h^2)}\,\,\,\stackrel{M_A^2\gg
M_Z^2}{\longrightarrow}\,\,\,
\frac{M_Z^4\sin^2 4\beta}{4M_A^4}\,\,\,.
\end{equation}
From the previous equations we then deduce that, in the decoupling
limit, the only light Higgs boson is $h^0$ with mass $M_{h}\simeq
M_{h}^{max}$, while $M_{H}\simeq M_{H^\pm}\simeq M_A\gg M_Z$, and
because $\cos(\beta-\alpha)\rightarrow 0$
($\sin(\beta-\alpha)\rightarrow 1)$), the couplings of $h^0$ to the
gauge bosons tend to the SM Higgs-boson limit. This is to say that, in the
decoupling limit, the light MSSM Higgs boson will be hardly
distinguishable from the SM Higgs boson.

Finally, we need to remember that the tree level couplings may be
modified by radiative corrections involving both loops of SM and MSSM
particles, among which loops of third generation quarks and squarks
dominate. The very same radiative corrections that modify the Higgs
boson mass matrix, thereby changing the definition of the mass
eigenstates, also affect the couplings of the corrected mass
eigenstates to the gauge bosons. This can be reabsorbed into the
definition of a \emph{renormalized} mixing angle $\alpha$ or a
\emph{radiatively corrected} value for $\cos(\beta-\alpha)$
($\sin(\beta-\alpha)$). Using the notation of
Ref.~\cite{Carena:2002es}, the radiatively corrected
$\cos(\beta-\alpha)$ can be written as:
\begin{equation}
\label{eq:cos_betamalpha_rad_corrected}
\cos(\beta-\alpha)=K\left[\frac{M_Z^2\sin 4\beta}{2M_A^2}+
\mathcal{O}\left(\frac{M_Z^4}{M_A^4}\right)\right]\,\,\,,
\end{equation}
where
\begin{equation}
\label{eq:cos_betamalpha_K_factor}
K\equiv 1+
\frac{\delta\mathcal{M}_{11}^2-\delta\mathcal{M}_{22}^2}{2M_Z^2\cos 2\beta}-
\frac{\delta\mathcal{M}_{12}^2}{M_Z^2\sin 2\beta}\,\,\,,
\end{equation}
and $\delta\mathcal{M}_{ij}$ are the radiative corrections to the
corresponding elements of the CP-even Higgs squared-mass matrix (see
Ref.~\cite{Carena:2002es}).  It is interesting to notice that on top
of the traditional decoupling limit introduced above ($M_A\gg M_Z$),
there is now also the possibility that $\cos(\beta-\alpha)\rightarrow
0$ if $K\rightarrow 0$, and this happens independently of the value of
$M_A$.

\subsubsection{MSSM Higgs-boson couplings to fermions}
\label{subsubsec:mssm_higgs_couplings_fermions}
As anticipated, $\Phi_1$ and $\Phi_2$ have Yukawa-type couplings to the
up-type and down-type components of all $SU(2)_L$ fermion
doublets. For example, the Yukawa Lagrangian corresponding to the
third generation of quarks reads:
\begin{equation}
\label{eq:yukawa_lagrangian_mssm}
\mathcal{L}_{Yukawa}=
-h_t\left[\bar{t}_R\phi_1^0 t_L-\bar{t}_R\phi_1^+ b_L\right]
-h_b\left[\bar{b}_R\phi_2^0 b_L-\bar{b}_R\phi_2^- t_L\right]+\mathrm{h.c.}
\end{equation}
Upon spontaneous symmetry breaking $\mathcal{L}_{Yukawa}$ provides both the
corresponding quark masses:
\begin{equation}
m_t=h_t\frac{v_1}{\sqrt{2}}=h_t\frac{v\sin\beta}{\sqrt{2}}
\,\,\,\,\,\mbox{and}\,\,\,\,\,
m_b=h_b\frac{v_2}{\sqrt{2}}=h_b\frac{v\cos\beta}{\sqrt{2}}\,\,\,,
\end{equation}
and the corresponding Higgs-quark couplings:
\begin{eqnarray}
\label{eq:yukawa_couplings_3rd_generation}
g_{ht\bar{t}}&=&\frac{\cos\alpha}{\sin\beta}y_t=
   \left[\sin(\beta-\alpha)+\cot\beta\cos(\beta-\alpha)\right]y_t\,\,\,,\\
g_{hb\bar{b}}&=&-\frac{\sin\alpha}{\cos\beta}y_b=
   \left[\sin(\beta-\alpha)-\tan\beta\cos(\beta-\alpha)\right]y_b\,\,\,,\nonumber\\
g_{Ht\bar{t}}&=&\frac{\sin\alpha}{\sin\beta}y_t=
   \left[\cos(\beta-\alpha)-\cot\beta\sin(\beta-\alpha)\right]y_t\,\,\,,\nonumber\\
g_{Hb\bar{b}}&=&\frac{\cos\alpha}{\cos\beta}y_b=
   \left[\cos(\beta-\alpha)+\tan\beta\sin(\beta-\alpha)\right]y_b\,\,\,,\nonumber\\
g_{At\bar{t}}&=&\cot\beta\, y_t\,\,\,\,,\,\,\,\,
g_{Ab\bar{b}}=\tan\beta\, y_b\,\,\,,\nonumber\\
g_{H^\pm t\bar{b}}&=&\frac{g}{2\sqrt{2}M_W}
\left[m_t\cot\beta(1-\gamma_5)+m_b\tan\beta(1+\gamma_5)\right]\,\,\,,\nonumber
\end{eqnarray}
where $y_q\!=\!m_q/v$ (for $q\!=\!t,b$) are the SM couplings.  It is
interesting to notice that in the $M_A\gg M_Z$ decoupling limit, as
expected, all the couplings in
Eq.~(\ref{eq:yukawa_couplings_3rd_generation}) reduce to the SM limit,
i.e. all $H^0$, $A^0$, and $H^\pm$ couplings vanish, while the
couplings of the light neutral Higgs boson, $h^0$, reduce to the
corresponding SM Higgs-boson couplings.

The Higgs boson-fermion couplings are also modified directly by
one-loop radiative corrections (squarks-gluino loops for quarks
couplings and slepton-neutralino loops for lepton couplings). A
detailed discussion can be found in
Ref.~\cite{Carena:2002es,Djouadi:2005gj} and in the
literature referenced therein. Of particular relevance are the
corrections to the couplings of the third quark generation. These
can be parameterized at the Lagrangian level by writing the
radiatively corrected \emph{effective} Yukawa Lagrangian as:
\begin{eqnarray}
\label{eq:yukawa_lagrangian_mssm_rad_corrected}
\mathcal{L}_{Yukawa}^{eff}&=&
-\epsilon_{ij}\left[
(h_b+\delta h_b)\bar{b}_R Q^j_L\Phi^i_2+(h_t+\delta h_t)\bar{t}_R Q^i_L\Phi^j_1
\right]\\
&-&\Delta h_t\bar{t}_RQ^k_L\Phi^{k\ast}_2-\Delta h_b\bar{b}_RQ^k_L\Phi^{k\ast}_1+
\mathrm{h.c.}\,\,\,,\nonumber
\end{eqnarray}
where we notice that radiative corrections induce a small coupling
between $\Phi_1$ and down-type fields and between $\Phi_2$ and
up-type fields. Moreover the tree level relation between $h_b$, $h_t$,
$m_b$ and $m_t$ are modified as follows:
\begin{eqnarray}
\label{eq:mb_mt_rad_corrected}
m_b&=&\frac{h_b v}{\sqrt{2}}\cos\beta\left(
1+\frac{\delta h_b}{h_b}+\frac{\Delta h_b\tan\beta}{h_b}\right)\equiv
\frac{h_b v}{\sqrt{2}}\cos\beta(1+\Delta_b)\,\,\,,\\
m_t&=&\frac{h_t v}{\sqrt{2}}\sin\beta\left(
1+\frac{\delta h_t}{h_t}+\frac{\Delta h_t\tan\beta}{h_t}\right)\equiv
\frac{h_t v}{\sqrt{2}}\sin\beta(1+\Delta_t)\,\,\,,\nonumber
\end{eqnarray}
where the leading corrections are proportional to $\Delta h_b$ and
turn out to also be $\tan\beta$ enhanced.  On the other hand, the
couplings between Higgs mass eigenstates and third generation quarks given in
Eq.~(\ref{eq:yukawa_couplings_3rd_generation}) are corrected as
follows:
\begin{eqnarray}
\label{eq:yukawa_couplings_3rd_generation_rad_corrected}
g_{ht\bar{t}}&=&\frac{\cos\alpha}{\sin\beta}y_t\,
  \left[1-\frac{1}{1+\Delta_t}\frac{\Delta h_t}{h_t}\left(\cot\beta+\tan\alpha\right)
  \right]\,\,\,,\\
g_{hb\bar{b}}&=&-\frac{\sin\alpha}{\cos\beta}y_b\,
   \left[1+\frac{1}{1+\Delta_b}\left(\frac{\delta h_b}{h_b}-\Delta_b\right)
   \left(1+\cot\alpha\cot\beta\right)\right]\,\,\,,\nonumber\\
g_{Ht\bar{t}}&=&\frac{\sin\alpha}{\sin\beta}y_t\,
   \left[1-\frac{1}{1+\Delta_t}\frac{\Delta h_t}{h_t}\left(\cot\beta-\cot\alpha\right)
   \right]\nonumber\,\,\,,\\
g_{Hb\bar{b}}&=&\frac{\cos\alpha}{\cos\beta}y_b\,
   \left[1+\frac{1}{1+\Delta_b}\left(\frac{\delta h_b}{h_b}-\Delta_b\right)
   \left(1-\tan\alpha\cot\beta\right)\right]\,\,\,,\nonumber\\
g_{At\bar{t}}&=&\cot\beta\,y_t\,
   \left[1-\frac{1}{1+\Delta_t}\frac{\Delta h_t}{h_t}\left(\cot\beta+\tan\beta\right)
   \right]\nonumber\,\,\,,\\
g_{Ab\bar{b}}&=&\tan\beta\,y_b\,
   \left[1+\frac{1}{(1+\Delta_b)\sin^2\beta}\left(\frac{\delta h_b}{h_b}-\Delta_b\right)
   \right]\,\,\,,\nonumber\\
g_{H^\pm t\bar{b}}&\simeq&\frac{g}{2\sqrt{2}M_W}
 \left\{ m_t\cot\beta\left[1-\frac{1}{1+\Delta_t}\frac{\Delta h_t}{h_t}
        \left(\cot\beta+\tan\beta\right)\right](1+\gamma_5)\right.\nonumber\\
&+&\left.m_b\tan\beta\left[1+\frac{1}{(1+\Delta_b)\sin^2\beta}
        \left(\frac{\delta h_b}{h_b}-\Delta_b\right)\right](1-\gamma_5)
\right\}\,\,\,,\nonumber
\end{eqnarray}
where the last coupling is given in the approximation of small isospin
breaking effects, since interactions of this kind have been neglected
in the Lagrangian of
Eq.~(\ref{eq:yukawa_lagrangian_mssm_rad_corrected}).

\section{Higgs searches}
\label{sec:higgs_searches}

The search for a SM-like Higgs boson and for more exotic Higgs bosons
of the kind predicted by multi-Higgs models is one of the most
important goals of the physics programme of both the Tevatron and the
LHC. After LEP ended its lifetime by setting a lower bound on the mass
of a SM-like Higgs at 114.4~GeV, the Tevatron has excluded larger
windows of the SM-Higgs (mass) parameter space and is still actively
analyzing data collected during Run II, while the LHC is breaking new
ground with an unprecedented amount of high energy data and has
confirmed and extended the Tevatron exclusion bounds, promising to
confirm or exclude the existence of a SM Higgs boson by the end of
2012, i.e. before the 2013 shutdown.  Indeed on July, $4^{th}$ 2012
both the ATLAS and CMS experiments at CERN announced the discovery of
a spin-0 particle with SM-Higgs-like properties and mass around
125-127~GeV. Two days earlier, on July 2$^{nd}$ 2012, the Tevatron
experiments, CDF and D0, presented their Summer 2012 results and
showed how their data would confirm or at least not contradict a
discovery of a Higgs boson at about 126~GeV.  Searches for more exotic
Higgs bosons are meanwhile providing more and more stringent bounds on
supersymmetric as well as non-supersymmetric extensions of the SM.

Having discussed the nature and implications of EWSB via the Higgs
mechanism in Section~\ref{sec:theory_framework}, we can now turn to
investigate the more phenomenological aspects of Higgs searches. In
this Section I would like to build some background to understand the
main properties of Higgs searches at hadron colliders and then
specialize the discussion to Higgs searches at both the Tevatron and
the LHC. For the sake of clarity, I will focus on the case of a SM
Higgs boson and take the opportunity to go in detail on some important
aspects. The possibility of interpreting the ATLAS and CMS signals in
the context of various extensions of the SM is being thoroughly
investigated by the theory community. Since the discussion was not
part of these lectures when they were originally delivered, and since
including it would go far beyond the scope of this lectures, I would
rather not touch on it and refer the interested reader to the
rapidly growing literature on the subject.

\subsection{SM Higgs-boson decay branching ratios}
\label{subsec:sm_higgs_branching_ratios}

Different search channels are at the moment distinguished by the
corresponding Higgs-boson decay channels. A precise calculation of
both production cross sections and decay widths with their respective
uncertainties is therefore essential to a correct interpretation of
the data. In this section we will review the main decay properties of
a SM Higgs boson and the major sources of uncertainties in the
theoretical calculation of the corresponding decay widths.

\begin{figure}
\begin{tabular}{lr}
\hspace{-0.5truecm}
\begin{minipage}{0.5\linewidth}
{\includegraphics[scale=0.35]{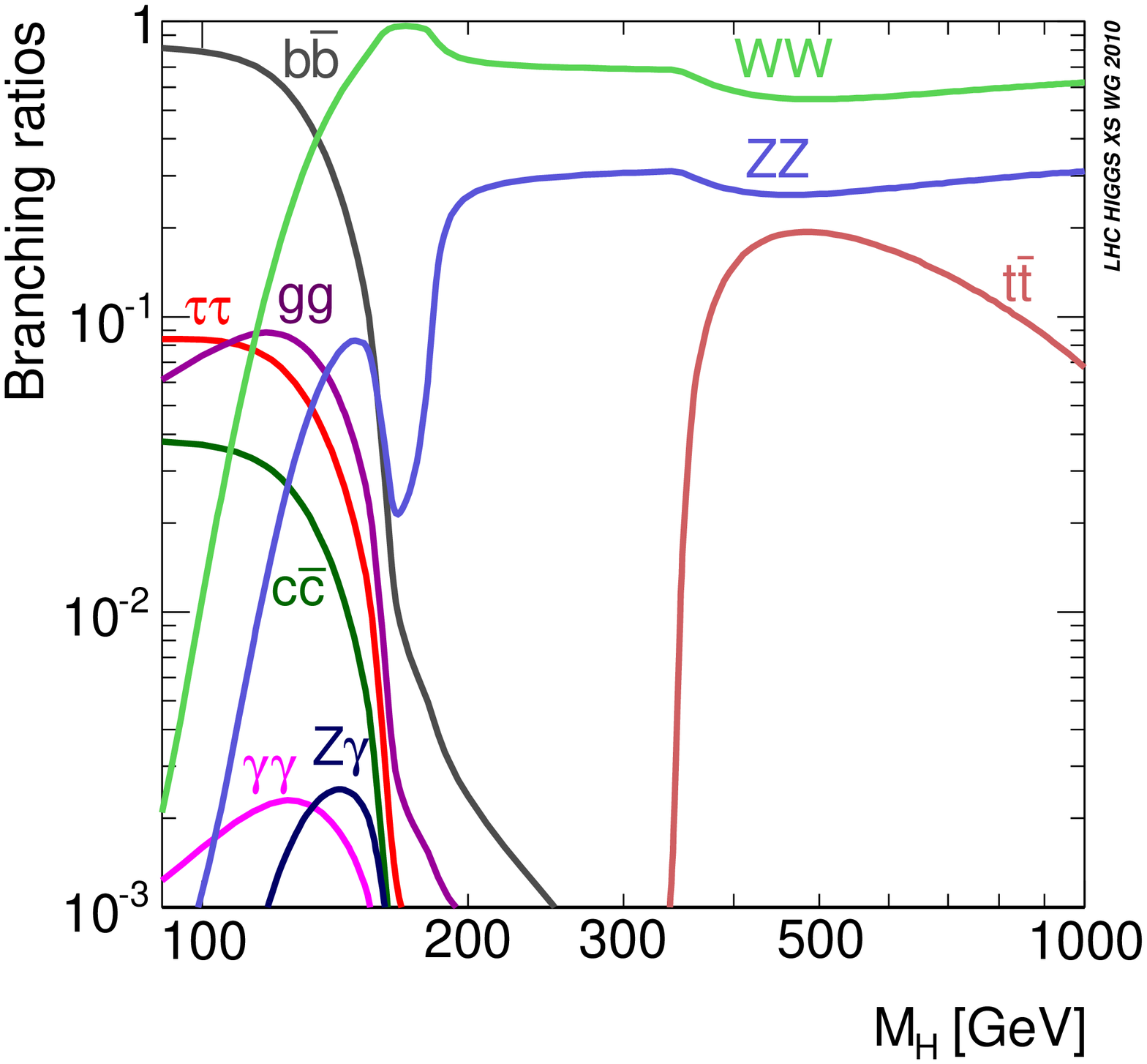}}
\end{minipage} &
\hspace{-0.3truecm}
\begin{minipage}{0.5\linewidth}
{\includegraphics[scale=0.35]{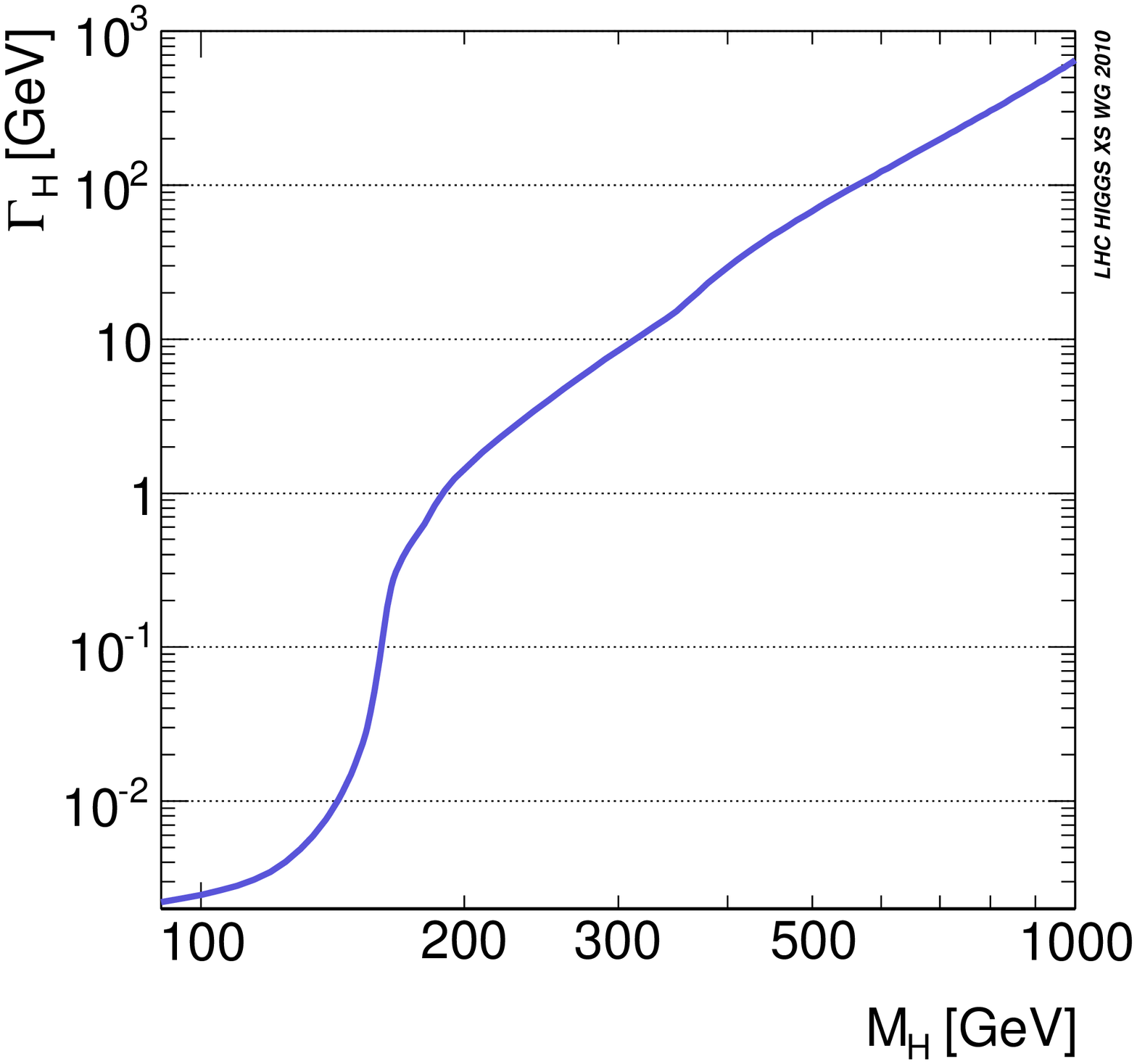}}
\end{minipage} 
\end{tabular}
\caption[]{SM Higgs decay branching ratios (left) and width (right) as
  a function of $M_H$. From Ref.~\cite{Dittmaier:2011ti}.
\label{fig:sm_higgs_br}}
\end{figure}
In Section~\ref{subsec:higgs_sm} we have derived the SM Higgs
couplings to gauge bosons and fermions. In particular we have seen
that, at the tree level the SM Higgs boson can decay into pairs of
electroweak gauge bosons ($H\rightarrow W^+W^-,ZZ$), and into pairs of
quarks and leptons ($H\rightarrow Q\bar{Q},l^+l^-$), while at one-loop
it can also decay into two photons ($H\rightarrow\gamma\gamma$), two
gluons ($H\rightarrow gg$), or a $\gamma Z$ pair ($H\rightarrow\gamma
Z$).  Fig.~\ref{fig:sm_higgs_br} represents all the decay branching
ratios of the SM Higgs boson as functions of its mass $M_H$. The SM
Higgs-boson total width, sum of all the partial widths
$\Gamma(H\rightarrow XX)$, is represented in
Fig.~\ref{fig:sm_higgs_br}.

In particular, Fig.~\ref{fig:sm_higgs_br} shows that a light Higgs
boson ($M_H\le 130-140$~GeV) behaves very differently from a heavy
Higgs boson ($M_H\ge 130-140$~GeV). Indeed, a light SM Higgs boson
mainly decays into a $b\bar{b}$ pair, followed hierarchically by all
other pairs of lighter fermions. Loop-induced decays also play a role
in this region. $H\rightarrow gg$ is dominant among them, and it is
actually larger than many tree level decays.  Unfortunately, this
decay mode is almost useless, in particular at hadron colliders,
because of background limitations. Among radiative decays,
$H\rightarrow\gamma\gamma$ is tiny, but it is actually
phenomenologically very important because the two photon signal can be
seen over large hadronic backgrounds. On the other hand, for larger
Higgs masses, the decays to $W^+W^-$ and $ZZ$ dominates. All decays
into fermions or loop-induced decays are suppressed, except
$H\rightarrow t\bar{t}$ for Higgs masses above the $t\bar{t}$
production threshold. There is an intermediate region, around
$M_H\simeq 160$~GeV, i.e. below the $W^+W^-$ and $ZZ$ threshold, where
the decays into $WW^*$ and $ZZ^*$ (when one of the two gauge bosons is
off-shell) become important. These are indeed three-body decays of the
Higgs boson that start to dominate over the $H\rightarrow b\bar{b}$
two-body decay mode when the largeness of the $HWW$ or $HZZ$ couplings
compensate for their phase space suppression\footnote{Actually, even
  four-body decays, corresponding to $H\rightarrow W^*W^*,Z^*Z^*$ may
  become important in the intermediate mass region and are indeed
  accounted for in Fig.~\ref{fig:sm_higgs_br}.}.  The different decay
pattern of a light vs a heavy Higgs boson influences the role played,
in each mass region, by different Higgs production processes at hadron
and lepton colliders.

\begin{figure}
\begin{tabular}{lr}
\hspace{-0.5truecm}
\begin{minipage}{0.5\linewidth}
{\includegraphics[scale=0.35]{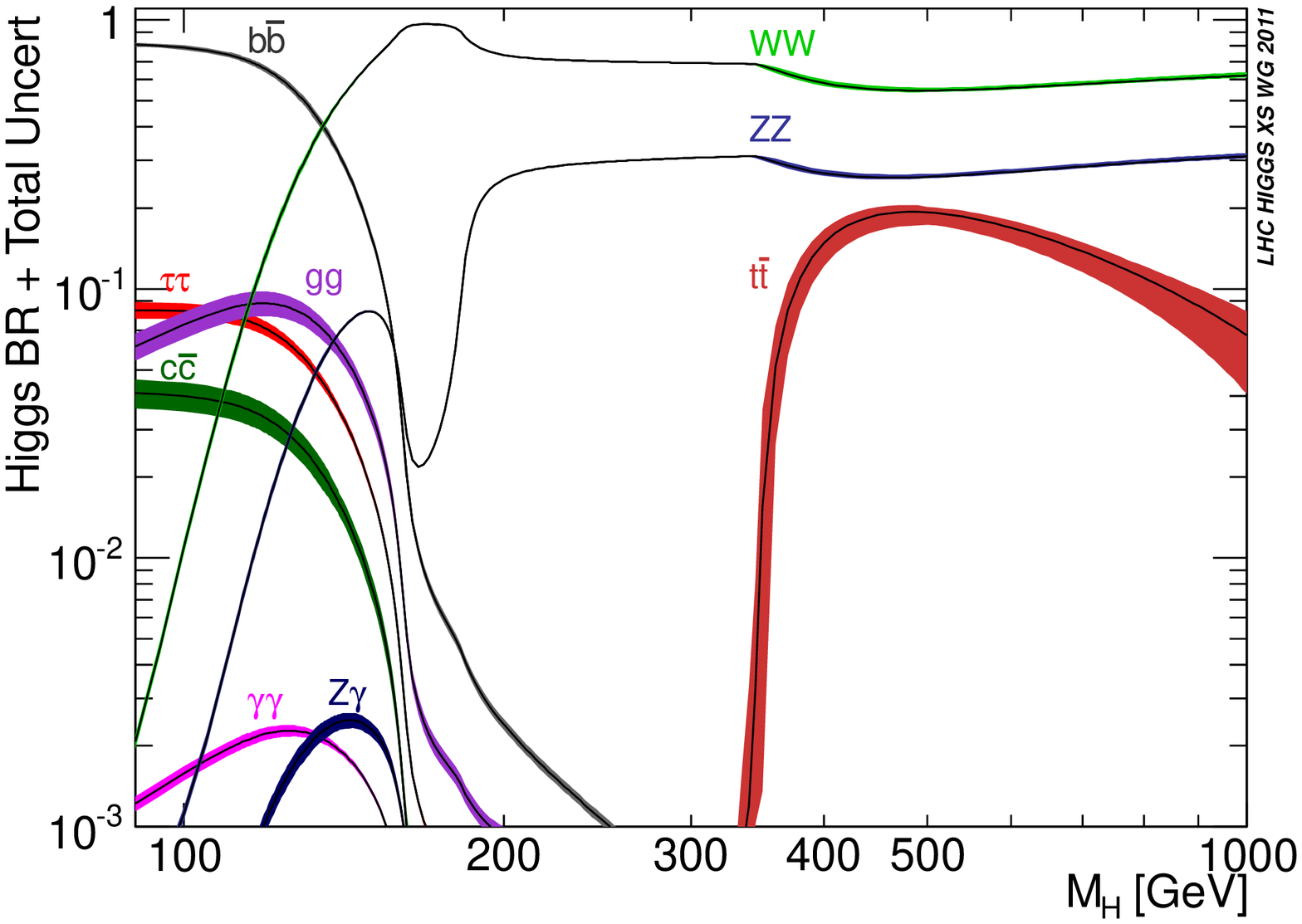}}
\end{minipage} &
\begin{minipage}{0.5\linewidth}
{\includegraphics[scale=0.35]{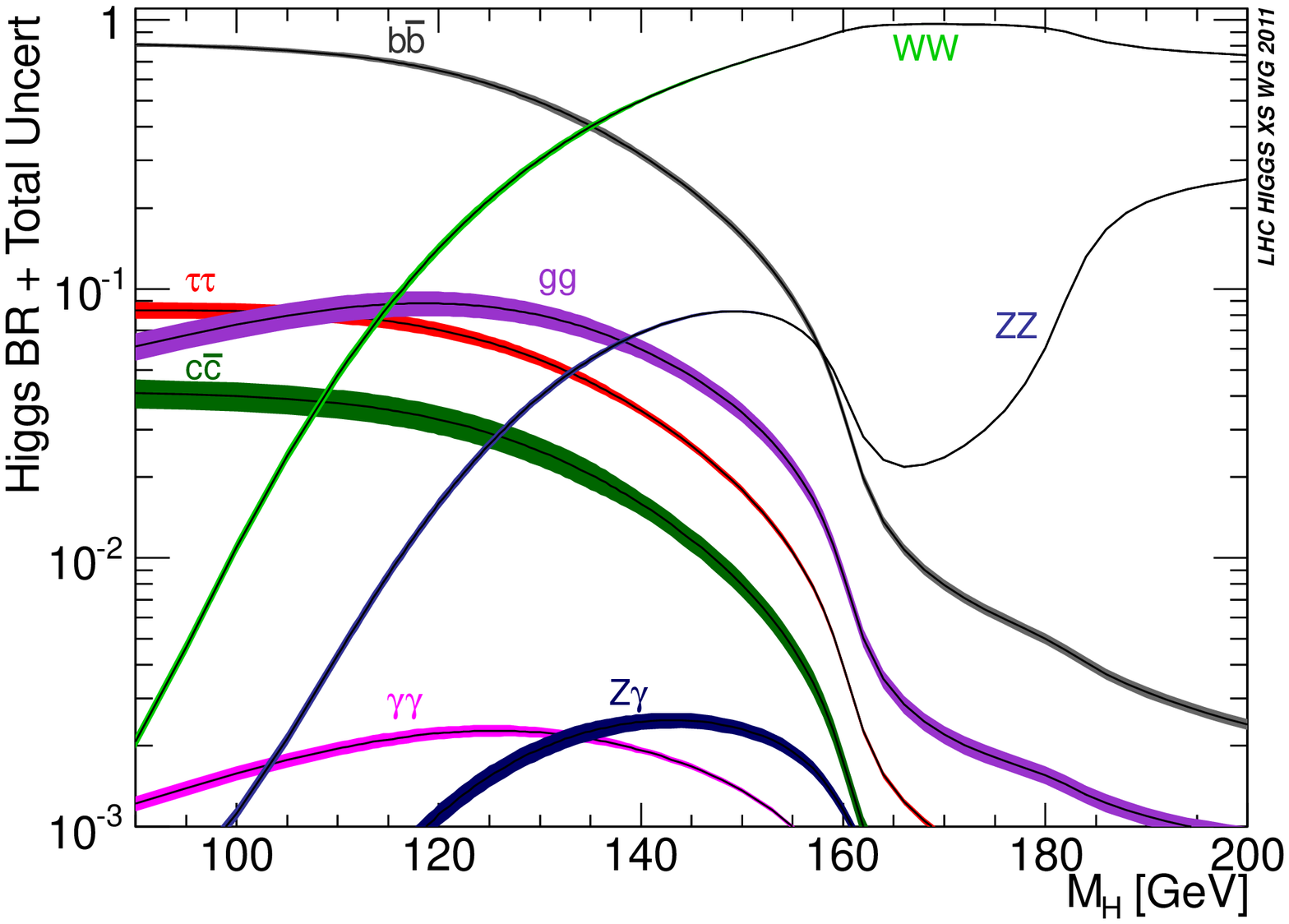}}
\end{minipage} 
\end{tabular}
\caption[]{SM Higgs decay branching ratios (left) and width (right) as
  a function of $M_H$. From Ref.~\cite{Dittmaier:2011ti}.
\label{fig:sm_higgs_br_band}}
\end{figure}
The curves in Fig.~\ref{fig:sm_higgs_br} are obtained by including all
available QCD and electroweak (EW) radiative corrections. Indeed, the
problem of computing the relevant orders of QCD and EW corrections for
Higgs decays has been thoroughly explored and the results are nowadays
available in public codes like HDECAY~\cite{Djouadi:1997yw}, which has
been used to produce Fig.~\ref{fig:sm_higgs_br}. Indeed it would be
more accurate to represent each curve as a band, including both
parametric (due to the variation of the input parameters $\alpha_s$,
$m_c$, $m_b$, and $m_t$) and theoretical uncertainties (resulting from
approximations in the theoretical calculations, the dominant effects
being due to missing higher orders). The effect of including both
kinds of uncertainties have been studied in detail in
Ref.~\cite{Dittmaier:2011ti} and \cite{Dittmaier:2012vm} and is
illustrated in Fig.~\ref{fig:sm_higgs_br_band} where the
right-hand-side plot gives an expanded view of the low mass region.
Furthermore, for $H\rightarrow WW$ and $H\rightarrow ZZ$ the full
decay chains into all possible 4-fermion final states have been
calculated including NLO QCD and EW corrections, and has been included
in the PROPHECY4f Monte Carlo event
generator~\cite{Bredenstein:2006rh,Bredenstein:2006ha}, which also
takes into account all possible interferences between common final
states as well as leading two-loop heavy-Higgs corrections. This has
been used in estimating the overall uncertainties of
Fig.~\ref{fig:sm_higgs_br_band}.

The theoretical uncertainties are most relevant for the $H\rightarrow
gg$, $H\rightarrow Z\gamma$, and $H\rightarrow t\bar{t}$ branching
ratios, reaching $O$($10\%$). For the $H\rightarrow b\bar{b}$,
$H\rightarrow c\bar{c}$, and $H\rightarrow\tau\tau$ branching ratios
they remain below a few per cent.  Parametric uncertainties are
relevant mostly for the $H\rightarrow c\bar{c}$ and $H\rightarrow gg$
branching ratios, reaching up to $O$($10\%$) and $O$($5\%$),
respectively.  They are mainly induced by the parametric uncertainties
in $\alpha_s$ and $m_c$.  The parametric uncertainties resulting from
$m_b$ affect the $\mathrm{Br}(H\rightarrow b\bar{b})$ at the level of
$3\%$, and the parametric uncertainty from $m_t$ influences in
particular the $\mathrm{Br}(H\rightarrow t\bar{t})$ near the
$t\bar{t}$ threshold.  For the $H\rightarrow\gamma\gamma$ channel the
total uncertainty can reach up to about $5\%$ in the relevant mass
range.  Both theoretical and parametric uncertainties on the
$H\rightarrow ZZ$ and $H\rightarrow WW$ channels remain at the level
of $1\%$ over the full mass range, giving rise to a total uncertainty
below $3\%$ for $m_H > 135$~GeV.

\subsubsection{General properties of radiative corrections to Higgs decays}
\label{subsubsection:sm_higgs_decays_rad_corr_general}
All Higgs-boson decay rates are modified by both EW and QCD radiative
corrections. QCD corrections are particularly important for $H\rightarrow
Q\bar{Q}$ decays, where they mainly amount to a redefinition of the
Yukawa coupling by shifting the mass parameter in it from the pole
mass value to the running mass value, and for $H\rightarrow gg$.  EW
corrections can be further separated into: \emph{i)} corrections due
to fermion loops, \emph{ii)} corrections due to the Higgs-boson
self-interaction, and \emph{iii)} other EW corrections. Both
corrections of type \emph{(ii)} and
\emph{(iii)} are in general very small if not for large Higgs-boson
masses, i.e. for $M_H\gg M_W$. On the other hand, corrections of type
\emph{(i)} are very important over the entire Higgs mass range, and
are particularly relevant for $M_H\ll 2m_t$, where the top-quark loop
corrections play a leading role.  Indeed, for $M_H\ll 2m_t$, the
dominant corrections for both Higgs decays into fermion and gauge
bosons come from the top-quark contribution to the renormalization of
the Higgs wave function and vacuum expectation value.

Several higher order radiative corrections to Higgs decays have been
calculated in the large $m_t$ limit, specifically in the limit when
$M_H\ll 2m_t$. Results can then be derived applying some very powerful
\emph{low energy theorems}. The idea
is that, for an on-shell Higgs field ($p_H^2\!=\!M_H^2$), the limit of
small masses ($M_H\ll 2m_t$) is equivalent to a $p_H\rightarrow 0$
limit, in which case the Higgs couplings to the fermion fields can be
simply obtained by substituting
\begin{equation}
\label{eq:mi_subs_let}
m_i^0\rightarrow m_i^0\left(1+\frac{H^0}{v^0}\right)\,\,\,,
\end{equation}
in the (bare) Yukawa Lagrangian, for each massive particle $i$.  
In Eq.~(\ref{eq:mi_subs_let}) $H^0$
is a constant field and the upper zero indices indicate that all
formal manipulations are done on bare quantities. This induces a
simple relation between the bare matrix element for a process with
($X\rightarrow Y+H$) and without ($X\rightarrow Y$) a Higgs field,
namely
\begin{equation}
\label{eq:matrix_element_relation_let}
\lim_{p_H\rightarrow 0}\mathcal{A}(X\rightarrow Y+H)=
\frac{1}{v^0}\sum_im_i^0\frac{\partial}{\partial m_i^0}\mathcal{A}(X\rightarrow Y)\,\,\,.
\end{equation}
When the theory is renormalized, the only actual difference is that
the derivative operation in Eq.~(\ref{eq:matrix_element_relation_let})
needs to be modified as follows
\begin{equation}
\label{eq:derivative_renormalized_let}
m_i^0\frac{\partial}{\partial m_i^0}\longrightarrow
\frac{m_i}{1+\gamma_{m_i}}\frac{\partial}{\partial m_i}
\end{equation}
where $\gamma_{m_i}$ is the mass anomalous dimension of fermion $f_i$.
This accounts for the fact that the renormalized Higgs-fermion Yukawa
coupling is determined through the $Z_2$ and $Z_m$ counterterms, and
not via the $Hf\bar{f}$ vertex function at zero momentum transfer (as
used in the $p_H\to 0$ limit above).

The theorem summarized by Eq.~(\ref{eq:matrix_element_relation_let})
is valid also when higher order radiative corrections are included.
Therefore, outstanding applications of Eq.~(\ref{eq:matrix_element_relation_let})
include the determination of the one-loop $Hgg$ and $H\gamma\gamma$
vertices from the gluon or photon self-energies, as well as the
calculation of several orders of their QCD and EW radiative
corrections. Indeed, in the $m_t\rightarrow\infty$ limit, the loop-induced
$H\gamma\gamma$ and $Hgg$ interactions can be seen as effective
vertices derived from an effective Lagrangian of the form:
\begin{equation}
\label{eq:effective_lagrangian_let}
\mathcal{L}_{eff}=\frac{\alpha_s}{12\pi}F^{(a)\mu\nu}F^(a)_{\mu\nu}\frac{H}{v}
(1+O(\alpha_s))\,\,\,,
\end{equation}
where $F^(a)_{\mu\nu}$ is the field strength tensor of QED (for the
$H\gamma\gamma$ vertex) or QCD (for the $Hgg$ vertex).  The
calculation of higher order corrections to the
$H\rightarrow\gamma\gamma$ and $H\rightarrow gg$ decays is then
reduced by one order of loops! Since these vertices start as one-loop
effects, the calculation of the first order of corrections would
already be a strenuous task, and any higher order effect would be a
formidable challenge. Thanks to the low energy theorem results
sketched above, QCD NNLO corrections have indeed been calculated.

\subsubsection{Higgs-boson decays to gauge bosons: 
$H\rightarrow W^+W^-,ZZ$}
\label{subsubsec:sm_higgs_to_gaugebosons}
The tree level decay rate for $H\rightarrow VV$ ($V\!=\!W^\pm,Z$) 
can be written as: 
\begin{equation}
\Gamma(H\rightarrow VV)=
\frac{G_FM_H^3}{16\sqrt{2}\pi}\delta_V\left(1-\tau_V+\frac{3}{4}\tau_V^2\right)
\beta_V\,\,\,,
\end{equation}
where $\beta_V=\sqrt{1-\tau_V}$, $\tau_V=4M_V^2/M_H^2$, and
$\delta_{W,Z}\!=\!2,1$. 

Below the $W^+W^-$ and $ZZ$ threshold, the SM Higgs-boson can still
decay via three (or four) body decays mediated by $WW^*$ ($W^*W^*$)
or $ZZ^*$ ($Z^*Z^*$) intermediate states. As we can see from
Fig.~\ref{fig:sm_higgs_br}, the off-shell decays $H\rightarrow WW^*$
and $H\rightarrow ZZ^*$ are relevant in the intermediate mass region
around $M_H\simeq 160$~GeV, where they compete and overcome the
$H\rightarrow b\bar{b}$ decay mode. The decay rates for $H\rightarrow
VV^*\rightarrow Vf_i\bar{f_j}$ ($V\!=\!W^\pm,Z$) are given by:
\begin{eqnarray}
\Gamma(H\rightarrow WW^*)&=&\frac{3g^4M_H}{512\pi^3}
F\left(\frac{M_W}{M_H}\right)\,\,\,,\\
\Gamma(H\rightarrow ZZ^*)&=&\frac{g^4M_H}{2048(1-s_W^2)^2\pi^3}
\left(7-\frac{40}{3}s_W^2+\frac{160}{9}s_W^4\right)
F\left(\frac{M_Z}{M_H}\right)\,\,\,,\nonumber
\end{eqnarray}
where $s_W\!=\!\sin\theta_W$ is the sine of the Weinberg angle and the
function $F(x)$ is given by
\begin{eqnarray}
F(x)&=&-(1-x^2)\left(\frac{47}{2}x^2-\frac{13}{2}+\frac{1}{x^2}\right)
-3\left(1-6x^2+4x^4\right)\ln(x)\nonumber\\
&+&3\,\frac{1-8x^2+20x^4}{\sqrt{4x^2-1}}
\arccos\left(\frac{3x^2-1}{2x^3}\right)\,\,\,.
\end{eqnarray}

\subsubsection{Higgs-boson decays to fermions:
$H\rightarrow Q\bar{Q},l^+l^-$}
\label{subsubsec:sm_higgs_to_fermions}
The tree level decay rate for $H\rightarrow f\bar{f}$ ($f\!=\!Q,l$,
$Q=$quark, $l=$lepton) can be written as: 
\begin{equation}
\Gamma(H\rightarrow f\bar f)=
\frac{G_FM_H}{4\sqrt{2}\pi}N_c^f m_f^2\beta_f^3\,\,\,,
\end{equation}
where $\beta_f=\sqrt{1-\tau_f}$, $\tau_f=4m_f^2/M_H^2$, and
$(N_c)^{l,Q}\!=\!1,3$.
QCD corrections dominate over other radiative corrections and they
modify the rate as follows:
\begin{equation}
\label{eq:gamma_hqq}
\Gamma(H\rightarrow Q\bar{Q})_{QCD}=
\frac{3G_FM_H}{4\sqrt{2}\pi}\bar{m}_Q^2(M_H)\beta_q^3
\left[\Delta_{QCD}+\Delta_t\right]\,\,\,,
\end{equation}
where $\Delta_t$ represents specifically QCD corrections
involving a top-quark loop. $\Delta_{QCD}$ and $\Delta_t$ have been
calculated up to three loops and are given by:
\begin{eqnarray}
\Delta_{QCD}&=&1+5.67\frac{\alpha_s(M_H)}{\pi}+(35.94-1.36N_F)
\left(\frac{\alpha_s(M_H)}{\pi}\right)^2+\\
&&(164.14-25.77N_F+0.26N_F^2)\left(\frac{\alpha_s(M_H)}{\pi}\right)^3
\,\,\,,\nonumber\\
\Delta_t&=&\left(\frac{\alpha_s(M_H)}{\pi}\right)^2
\left[1.57-\frac{2}{3}\ln\frac{M_H^2}{m_t^2}+
\frac{1}{9}\ln^2\frac{\bar{m}_Q^2(M_H)}{M_H^2}\right]\,\,\,,\nonumber
\end{eqnarray}
where $\alpha_s(M_H)$ and $\bar{m}_Q(M_H)$ are the renormalized
running QCD coupling and quark mass in the $\overline{MS}$ scheme.  It
is important to notice that using the $\overline{MS}$ running mass in
the overall Yukawa coupling square of Eq.~(\ref{eq:gamma_hqq})is very
important in Higgs decays, since it reabsorbs most of the QCD
corrections, including large logarithms of the form
$\ln(M_H^2/m_Q^2)$.  Indeed, for a generic scale $\mu$,
$\bar{m}_Q(\mu)$ is given at leading order by:
\begin{eqnarray}
\bar{m}_Q(\mu)_{LO}&=&\bar{m}_Q(m_Q)\left(\frac{\alpha_s(\mu)}{\alpha_s(m_Q)}\right)^
{\frac{2b_0}{\gamma_0}}\\
&=&\bar{m}_Q(m_Q)\left(1-\frac{\alpha_s(\mu)}{4\pi}\ln\left(\frac{\mu^2}{m_Q^2}\right)+
\cdots\right)\,\,\,,\nonumber
\end{eqnarray}
where $b_0$ and $\gamma_0$ are the first coefficients of the $\beta$
and $\gamma$ functions of QCD, while at higher orders it reads:
\begin{equation}
\label{eq:running_mass_lo}
\bar{m}_Q(\mu)=\bar{m}_Q(m_Q)\frac{f\left(\alpha_s(\mu)/\pi\right)}
{f\left(\alpha_s(m_Q)/\pi\right)}\,\,\,,
\end{equation}
where, from renormalization group techniques, the function $f(x)$
is of the form:
\begin{eqnarray}
\label{eq:running_mass_higher_order}
f(x)&=&\left(\frac{25}{6}x\right)^{\frac{12}{25}}\left[1+1.014x+\ldots\right]
\,\,\,\,\,\,\mbox{for}\,\,\,\,\,m_c\!<\!\mu\!<\!m_b\,\,\,,\\
f(x)&=&\left(\frac{23}{6}x\right)^{\frac{12}{23}}\left[1+1.175x+\ldots\right]
\,\,\,\,\,\,\mbox{for}\,\,\,\,\,m_b\!<\!\mu\!<\!m_t\,\,\,,\nonumber\\
f(x)&=&\left(\frac{7}{2}x\right)^{\frac{4}{7}}\left[1+1.398x+\ldots\right]
\,\,\,\,\,\,\mbox{for}\,\,\,\,\,\mu\!>\!m_t\,\,\,.\nonumber
\end{eqnarray}
As we can see from Eqs.~(\ref{eq:running_mass_lo}) and
(\ref{eq:running_mass_higher_order}), by using the $\overline{MS}$
running mass, leading and subleading logarithms up to the order of
the calculation are actually resummed at all orders in $\alpha_s$.

The overall mass factor coming from the quark Yukawa coupling square
is actually the only place where we want to employ a running mass. For
quarks like the $b$ quark this could indeed have a large impact, since,
in going from $\mu\simeq M_H$ to $\mu\simeq m_b$, $\bar{m}_n(\mu)$
varies by almost a factor of two, making therefore almost a factor of
four at the rate level.  All other mass corrections, in the matrix
element and phase space entering the calculation of the $H\rightarrow
Q\bar{Q}$ decay rate, can in first approximation be safely neglected.

\subsubsection{Loop induced Higgs-boson decays: 
\label{subsubsec:sm_higgs_loop_decay}
$H\rightarrow \gamma\gamma,\gamma Z,gg$}
\label{subsubsec:sm_higgs_loop_decays}
As seen in Section~\ref{subsec:higgs_sm}, the $H\gamma\gamma$ and
$H\gamma Z$ couplings are induced at one loop via both a fermion loop
and a W-loop. At the lowest order the decay rate for
$H\rightarrow\gamma\gamma$ can be written as:
\begin{equation}
\label{eq:rate_Hgammagamma_lo}
\Gamma(H\rightarrow\gamma\gamma)=
\frac{G_F\alpha^2M_H^3}{128\sqrt{2}\pi^3}
\left|\sum_f N_c^fQ_f^2A_f^H(\tau_f)+A_W^H(\tau_W)\right|^2\,\,\,,
\end{equation}
where $N_c^f=1,3$ (for $f=l,q$ respectively), $Q_f$ is the charge of
the $f$ fermion species, $\tau_f=4m_f^2/M_H^2$, the function $f(\tau)$
is defined as:
\begin{equation}
\label{eq:f_tau}
f(\tau)=\left\{
\begin{array}{lr}
\arcsin^2\frac{1}{\sqrt{\tau}}&\tau\ge 1\\
-\frac{1}{4}\left[\ln\frac{1+\sqrt{1-\tau}}{1-\sqrt{1-\tau}}-i\pi\right]^2&
\tau<1\,\,\,,
\end{array}\right.
\end{equation}
and the form factors $A_f^H$ and $A_W^H$ are given by:
\begin{eqnarray}
\label{eq:rate_Hgammagamma_lo_form_factors}
A_f^H&=&2\tau\left[1+(1-\tau)f(\tau)\right]\,\,\,,\\
A_W^H(\tau)&=&-\left[2+3\tau+3\tau(2-\tau)f(\tau)\right]\,\,\,.\nonumber
\end{eqnarray}
On the other hand, the decay rate for $H\rightarrow\gamma Z$ is given by:
\begin{equation}
\label{eq:rate_HgammaZ_lo}
\Gamma(H\rightarrow \gamma Z)=
\frac{G_F^2M_W^2\alpha M_H^3}{64\pi^4}\left(1-\frac{M_Z^2}{M_H^2}\right)^3
\left|\sum_f A_f^H(\tau_f,\lambda_f)+A_W^H(\tau_W,\lambda_W)\right|^2\,\,\,,
\end{equation}
where $\tau_i\!=\!4M_i^2/M_H^2$ and $\lambda_i\!=\!4M_i^2/M_Z^2$
($i\!=\!f,W$), and the form factors $A_f^H(\tau,\lambda)$ and
$A_W^H(\tau,\lambda)$ are given by:
\begin{eqnarray}
\label{eq:rate_HgammaZ_lo_form_factors}
A_f^H(\tau,\lambda)&=&2N_c^f\frac{Q_f(I_{3f}-2Q_f\sin^2\theta_W)}{\cos\theta_W}
\left[I_1(\tau,\lambda)-I_2(\tau,\lambda)\right]\,\,\,,\\
A_W^H(\tau,\lambda)&=&\cos\theta_W\left\{
\left[\left(1+\frac{2}{\tau}\right)\tan^2\theta_W-\left(5+\frac{2}{\tau}\right)\right]
I_1(\tau,\lambda)\nonumber\right.\\
&&\left.\phantom{\frac{1}{2}}+4\left(3-\tan^2\theta_W\right)I_2(\tau,\lambda)
\right\}\,\,\,,
\end{eqnarray}
where $N_c^f$ and $Q_f$ are defined after
Eq.~(\ref{eq:rate_Hgammagamma_lo}), and $I_3^f$ is the weak isospin of the
$f$ fermion species. Moreover:
\begin{eqnarray}
\label{eq:rate_Hgammagamma_lo_I1I2}
I_1(\tau,\lambda)&=&\frac{\tau\lambda}{2(\tau-\lambda)}+
\frac{\tau^2\lambda^2}{2(\tau-\lambda)^2}[f(\tau)-f(\lambda)]+
\frac{\tau^2\lambda}{(\tau-\lambda)^2}[g(\tau)-g(\lambda)]\,\,\,,\nonumber\\
I_2(\tau,\lambda)&=&-\frac{\tau\lambda}{2(\tau-\lambda)}[f(\tau)-f(\lambda)]\,\,\,,
\end{eqnarray}
and
\begin{equation}
\label{eq:g_tau}
g(\tau)=\left\{
\begin{array}{lr}
\sqrt{\tau-1}\arcsin\frac{1}{\sqrt{\tau}}&\tau\ge 1\\
\frac{\sqrt{1-\tau}}{2}\left[\ln\frac{1+\sqrt{1-\tau}}{1-\sqrt{1-\tau}}-i\pi\right]&
\tau<1
\end{array}\right.
\end{equation}
while $f(\tau)$ is defined in Eq.~(\ref{eq:f_tau}). QCD and EW
corrections to both $\Gamma(H\rightarrow\gamma\gamma)$ and
$\Gamma(H\rightarrow\gamma Z)$ are pretty small and for their explicit
expression we refer the interested reader to the literature
\cite{Spira:1997dg,Djouadi:2005gi}.

\begin{figure}
\centering
\includegraphics[bb=125pt 470pt 490pt 760pt,scale=0.65]{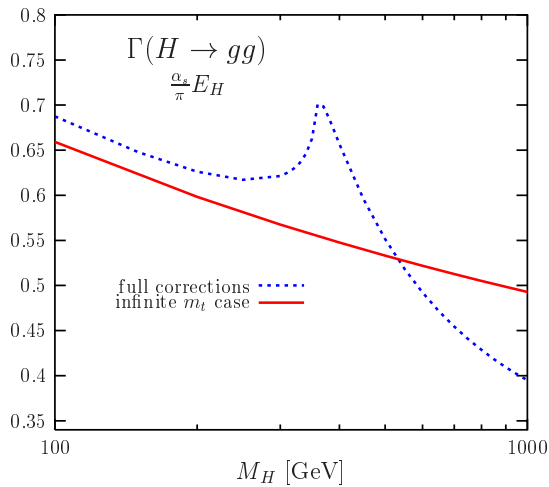}
\vspace{0.3truecm}
\caption[]{The QCD correction factor for the partial width
$\Gamma(H\rightarrow gg)$ as a function of the Higgs-boson mass, in
the full massive case with $m_t\!=\!178$~GeV (dotted line) and in the
heavy-top-quark limit (solid line). The strong coupling constant is
normalized at $\alpha_s(M_Z)\!=\!0.118$. From
Ref.~\cite{Djouadi:2005gi}.\label{fig:hgg_nlo}}
\end{figure}

As far as $H\rightarrow gg$ is concerned, this decay can only be
induced by a fermion loop,
and therefore its rate, at the lowest order, can be written as:
\begin{equation}
\label{eq:rate_Hgg_lo}
\Gamma(H\rightarrow gg)=\frac{G_F\alpha_s^2M_H^3}{36\sqrt{2}\pi^3}
\left|\frac{3}{4}\sum_q A_q^H(\tau_q)\right|\,\,\,,
\end{equation}
where $\tau_q\!=\!4m_q^2/M_H^2$, $f(\tau)$ is defined in
Eq.(\ref{eq:f_tau}) and the form factor $A_q^H(\tau)$ is given in
Eq.~(\ref{eq:rate_HgammaZ_lo_form_factors}).  QCD corrections to
$H\rightarrow gg$ have been calculated up to NNLO in the
$m_t\rightarrow\infty$ limit, as explained in
Section~\ref{subsubsection:sm_higgs_decays_rad_corr_general}. At NLO
the expression of the corrected rate is remarkably simple
\begin{equation}
\Gamma(H\rightarrow gg(g),q\bar{q}g)=\Gamma_{LO}(H\rightarrow gg)
\left[1+E(\tau_Q)\frac{\alpha_s^{(N_L)}}{\pi}\right]\,\,\,,
\end{equation}
where
\begin{equation}
E(\tau_Q)\stackrel{M_H^2\ll 4m_q^2}{\longrightarrow} 
\frac{95}{4}-\frac{7}{6}N_L +
\frac{33-2N_F}{6}\log\left(\frac{\mu^2}{M_H^2}\right)\,\,\,.
\end{equation}
When compared with the fully massive NLO calculation (available in
this case), the two calculations display an impressive $10\%$
agreement, as illustrated in Fig.~\ref{fig:hgg_nlo}, even in regions
where the light Higgs approximation is not justified. This is actually
due to the presence of large constant factors in the first order of
QCD corrections.
We also observe that the first order of QCD corrections has quite a
large impact on the lowest order cross section, amounting to more than
50\% of $\Gamma_{LO}$ on average. This has been indeed the main reason
to prompt for a NNLO QCD calculation of $\Gamma(H\rightarrow gg)$. The
result, obtained in the heavy-top approximation, has shown that NNLO
QCD corrections amount to only 20\% of the NLO cross section,
therefore pointing to a convergence of the $\Gamma(H\rightarrow gg)$
perturbative series. We will refer to this discussion when dealing
with the $gg\rightarrow H$ production mode, since its cross section can be
easily related to $\Gamma(H\rightarrow gg)$.

\subsection{MSSM Higgs-boson branching ratios}
\label{subsec:mssm_higgs_branching_ratios}
The decay patterns of the MSSM Higgs bosons are many and diverse,
depending on the specific choice of supersymmetric parameters. In
particular they depend on the choice of $M_A$ and $\tan\beta$, which
parameterize the MSSM Higgs sector, and they are clearly sensitive to
the choice of other supersymmetric masses (gluino masses, squark
masses, etc.) since this determines the possibility for the MSSM Higgs
bosons to decay into pairs of supersymmetric particles and for the
radiative induced decay channels ($h^0,H^0\rightarrow gg,\gamma\gamma,\gamma
Z$)  to receive supersymmetric loop contributions.

\begin{figure}
\hspace{-0.4truecm}
\begin{tabular}{lr}
\begin{minipage}{0.5\linewidth}
{\includegraphics[scale=0.35]{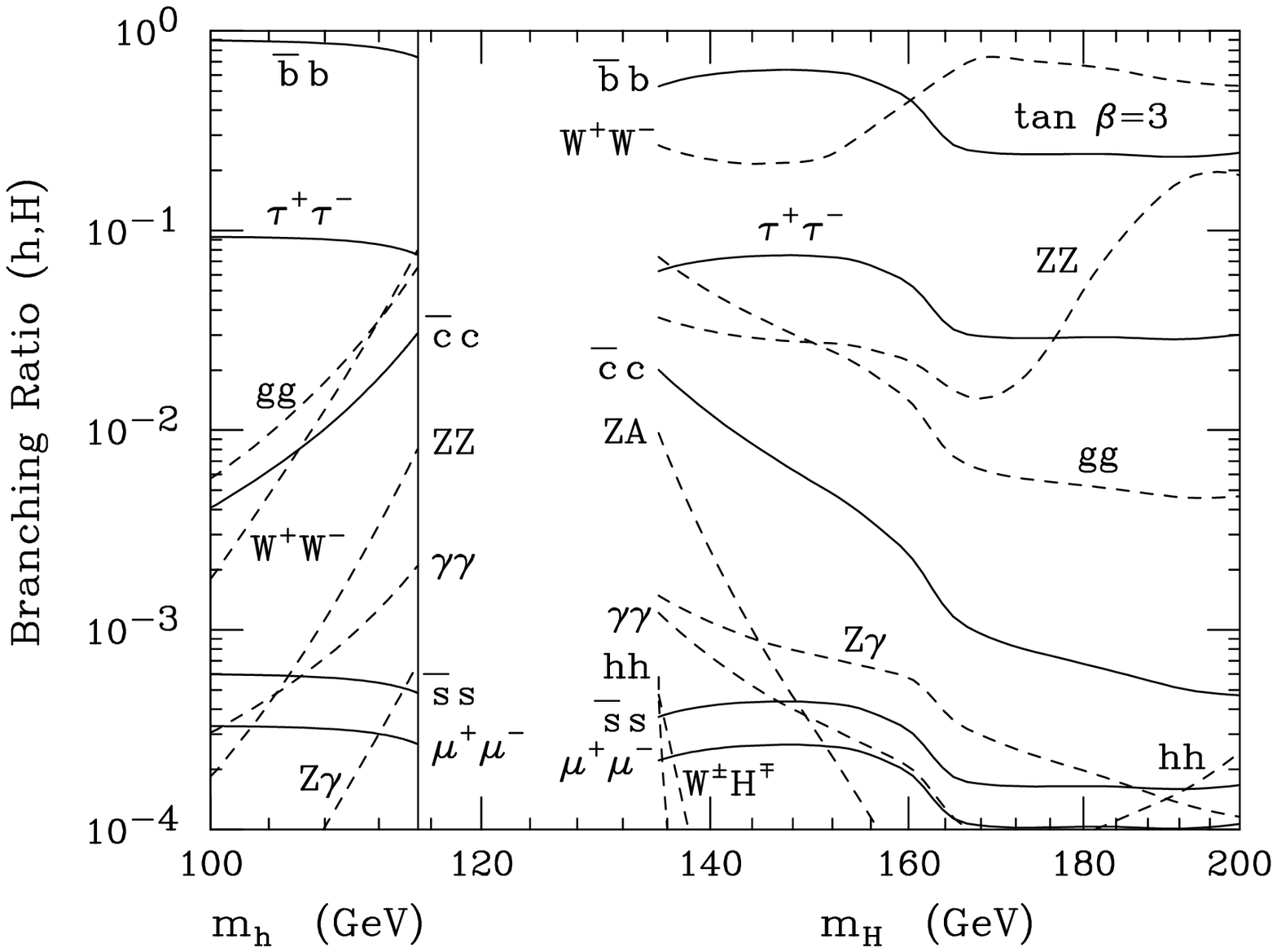}}
\end{minipage} &
\begin{minipage}{0.5\linewidth}
{\includegraphics[scale=0.35]{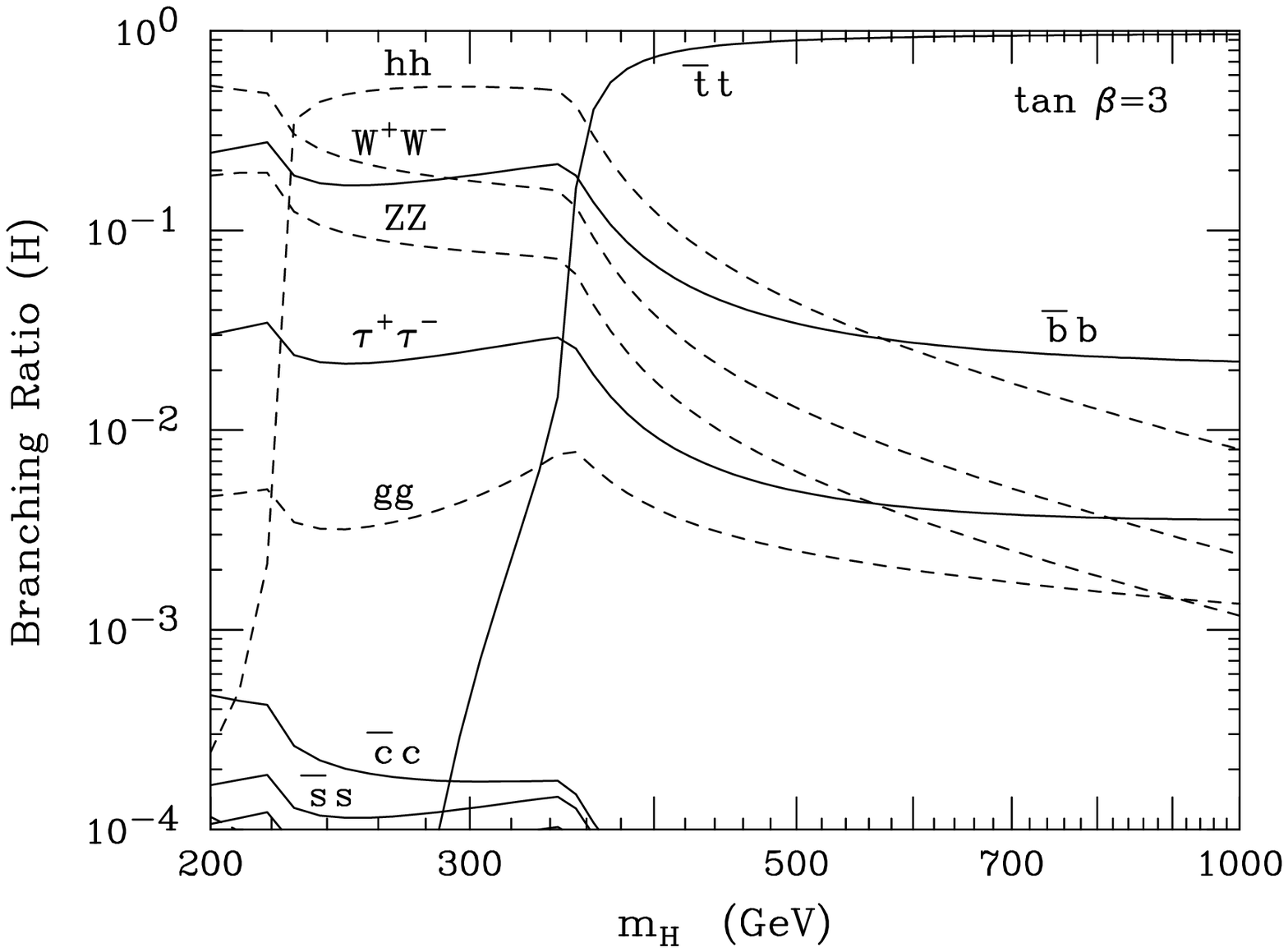}}
\end{minipage} \\
\begin{minipage}{0.5\linewidth}
{\includegraphics[scale=0.35]{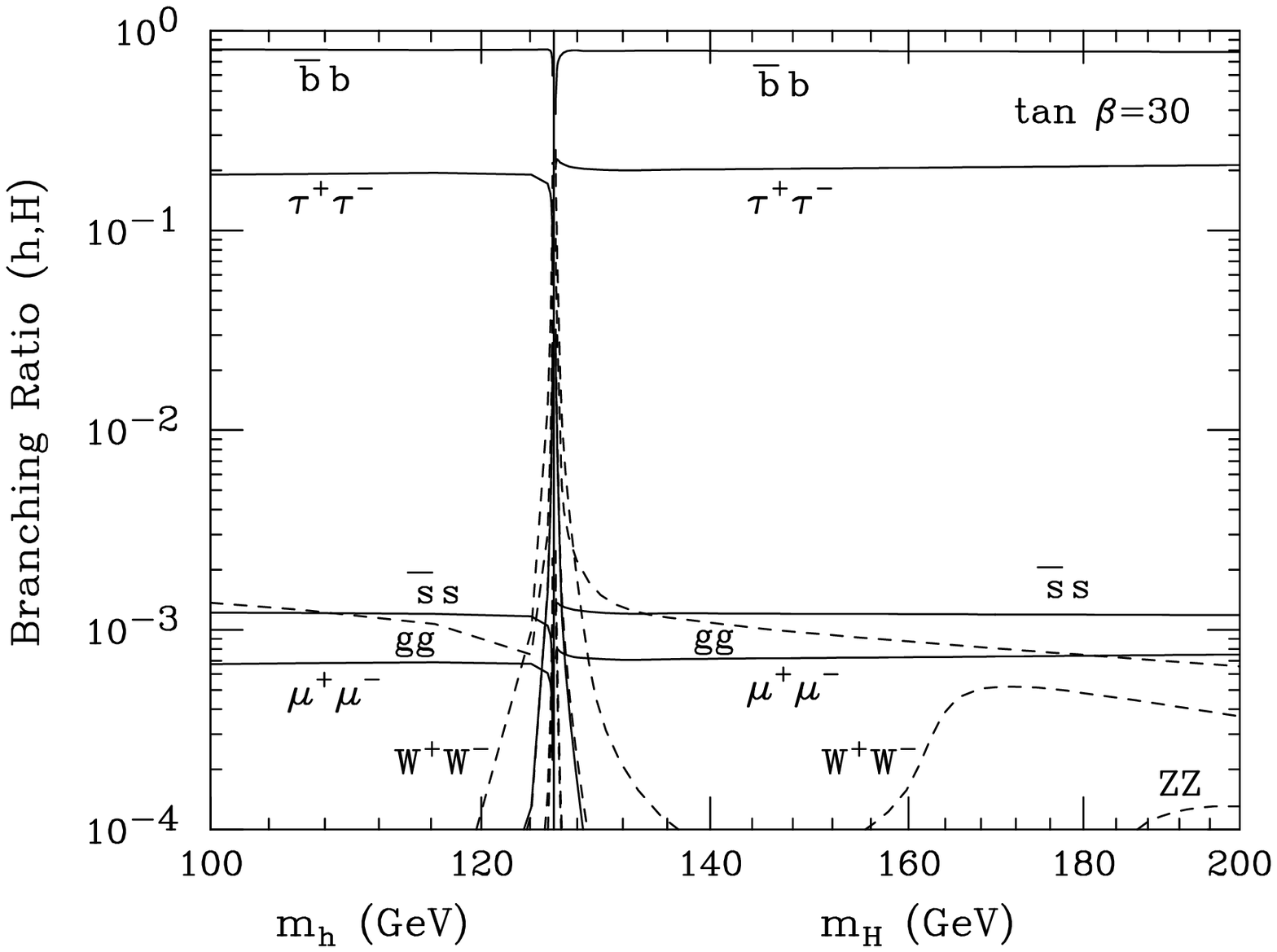}}
\end{minipage} &
\begin{minipage}{0.5\linewidth}
{\includegraphics[scale=0.35]{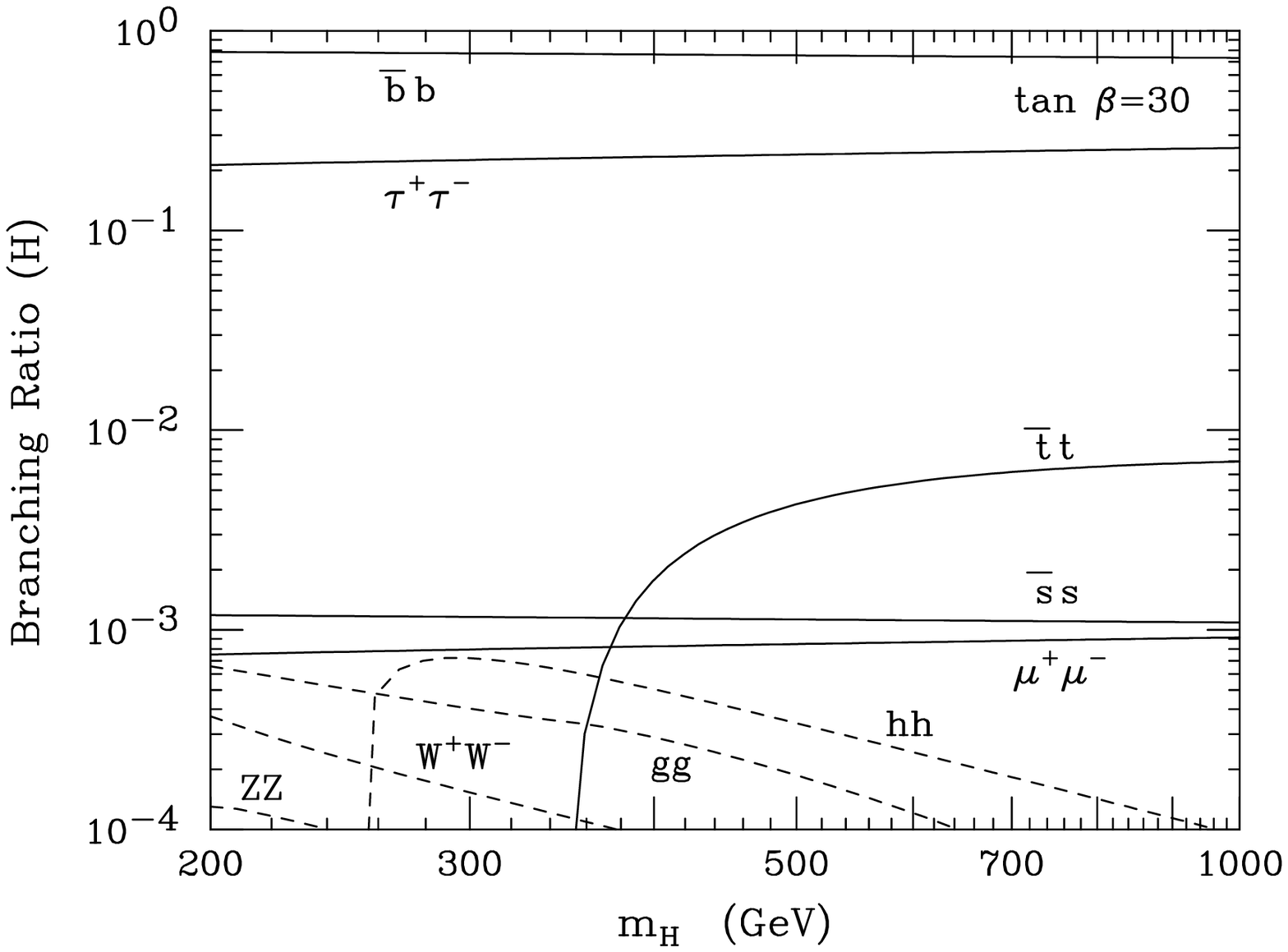}}
\end{minipage} 
\end{tabular}
\caption[]{Branching ratios for the $h^0$ and $H^0$ MSSM Higgs bosons,
for $\tan\beta\!=\!3,30$. The range of $M_H$ corresponds to
$M_A\!=\!90\,\mbox{GeV}-1\,\mbox{TeV}$, in the MSSM scenario discussed
in the text, with maximal top-squark mixing. The vertical line in the
left hand side plots indicates the upper bound on $M_h$, which, for
the given scenario is $M_h^{max}\!=\!115$~GeV ($\tan\beta=3$) or
$M_h^{max}\!=\!125.9$~GeV ($\tan\beta=30$). From
Ref.~\cite{Carena:2002es}.\label{fig:mssm_h_H_br_ratios_tanb3_30} }
\end{figure}
In order to be more specific, let us assume that all supersymmetric
masses are large enough to prevent the decay of the MSSM Higgs bosons
into pairs of supersymmetric particles (a good choice could be
$M_{\tilde g}\!=\!M_Q\!=\!=\!M_U\!=\!M_D\!=\!1$~TeV). Then, we only
need to examine the decays into SM particles and compare with the decay
patterns of a SM Higgs boson to identify any interesting difference.
From the study of the MSSM Higgs-boson couplings in
Sections~\ref{subsubsec:mssm_higgs_couplings_bosons} and
\ref{subsubsec:mssm_higgs_couplings_fermions}, we expect that:
\emph{i)} in the decoupling regime, when $M_A\gg M_Z$, the properties
of the $h^0$ neutral Higgs boson are very much the same as the SM
Higgs boson; while away from the decoupling limit \emph{ii)} the
decay rates of $h^0$ and $H^0$ to electroweak gauge bosons are
suppressed with respect to the SM case, in particular for large Higgs
masses ($H^0$), \emph{iii)} the $A^0\rightarrow VV$ ($V=W^\pm,Z^0$)
decays are absent, \emph{iv)} the decay rates of $h^0$ and $H^0$ to
$\tau^+\tau^-$ and $b\bar{b}$ are enhanced for large $\tan\beta$,
\emph{v)} even for not too large values of $\tan\beta$, due to
\emph{ii)} above, the $h^0,H^0\rightarrow\tau^+\tau^-$ and 
$h^0,H^0\rightarrow b\bar{b}$ decay are large up
to the $t\bar{t}$ threshold, when the decay $H^0\rightarrow t\bar{t}$
becomes dominant, \emph{vi)} for the charged Higgs boson, the decay
$H^+\rightarrow\tau^+\nu_\tau$ dominates over $H^+\rightarrow
t\bar{b}$ below the $t\bar{b}$ threshold, and vice versa above it.

\begin{figure}
\hspace{-0.4truecm}
\begin{tabular}{cc}
\begin{minipage}{0.5\linewidth}
{\includegraphics[scale=0.35]{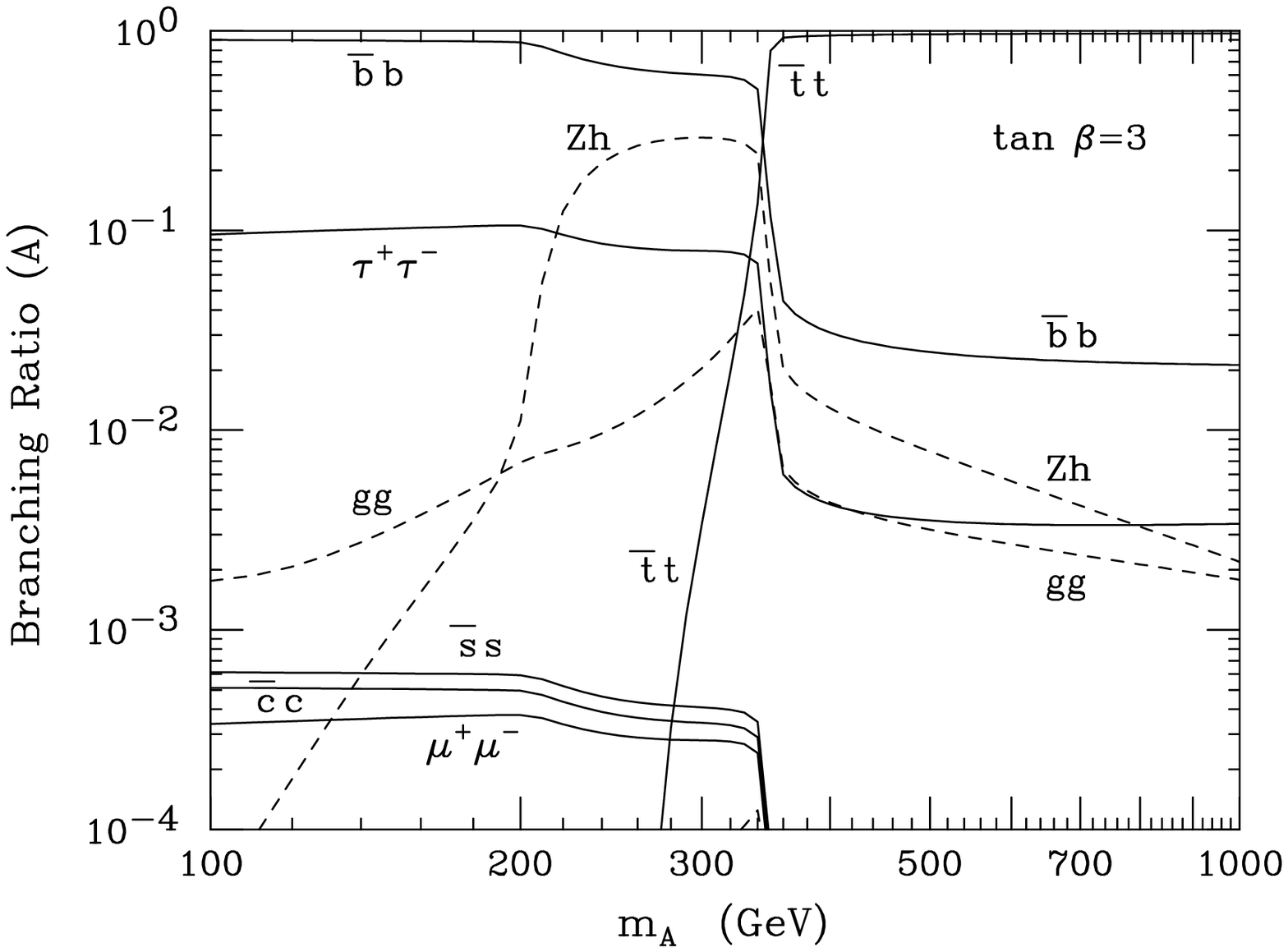}}
\end{minipage} &
\begin{minipage}{0.5\linewidth}
{\includegraphics[scale=0.35]{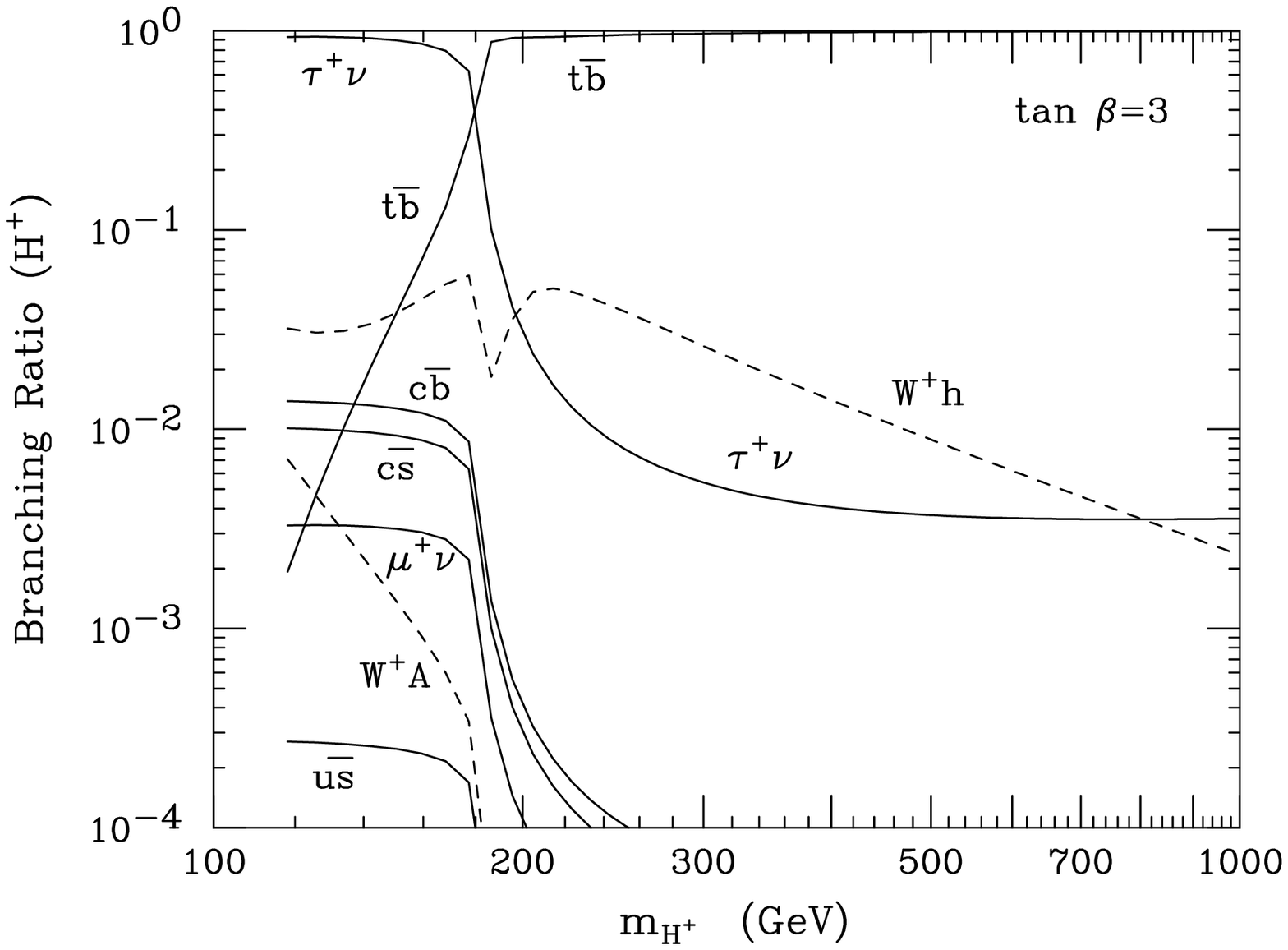}}
\end{minipage} \\
\begin{minipage}{0.5\linewidth}
{\includegraphics[scale=0.35]{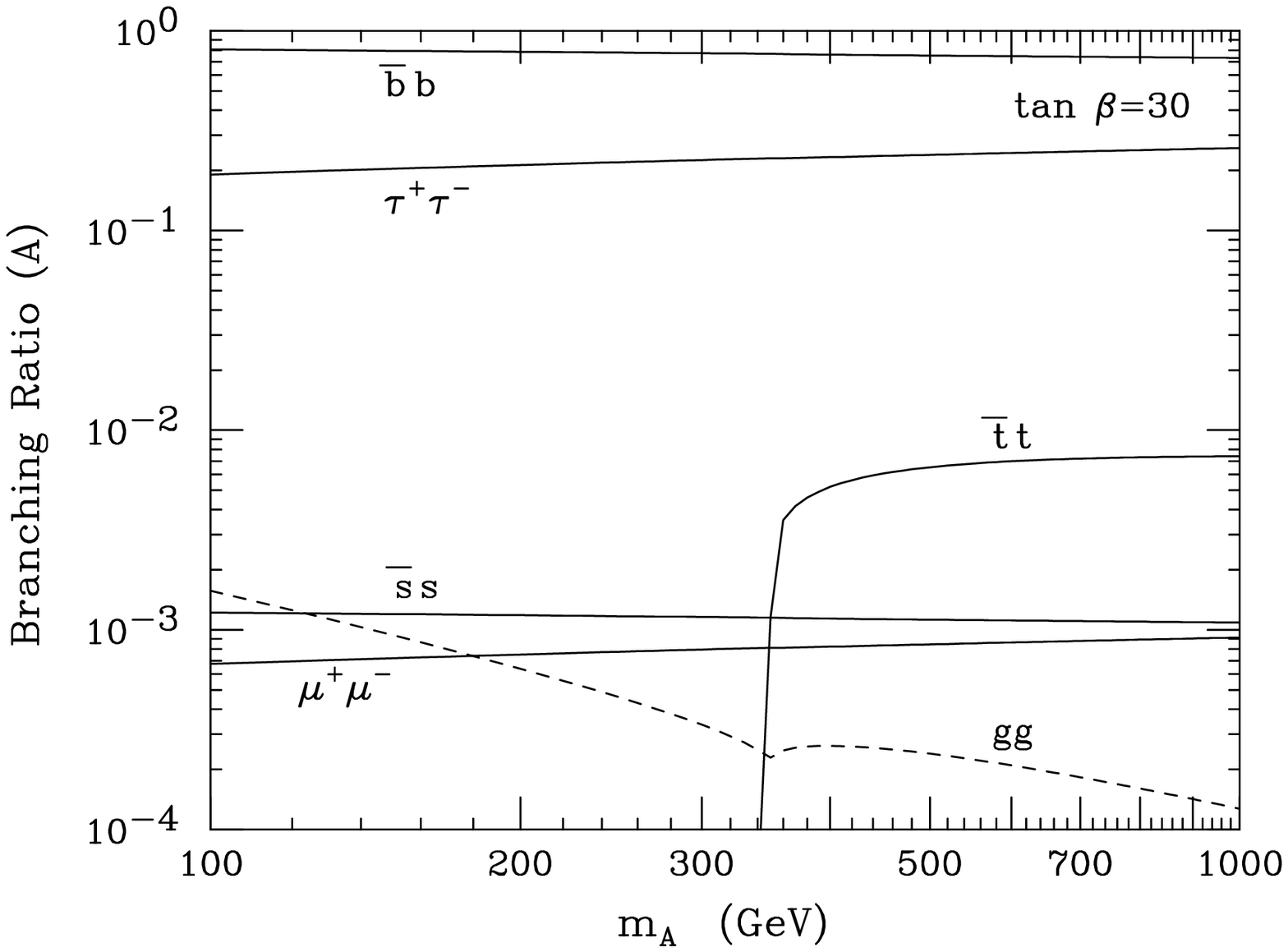}}
\end{minipage} &
\begin{minipage}{0.5\linewidth}
{\includegraphics[scale=0.35]{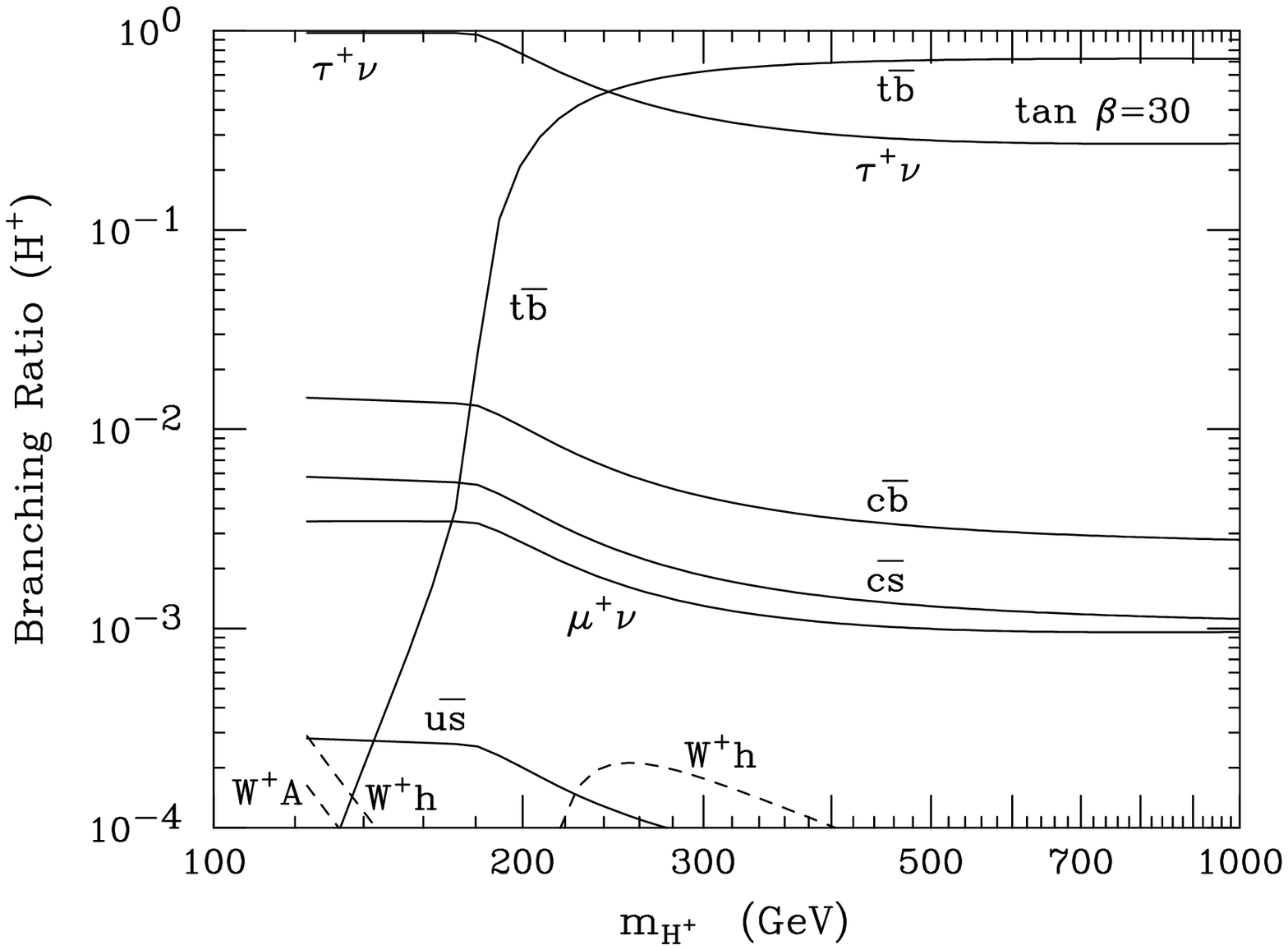}}
\end{minipage} 
\end{tabular}
\caption[]{Branching ratios for the $A^0$ and $H^+$ MSSM Higgs bosons,
for $\tan\beta\!=\!3,30$. The range of $M_{H^\pm}$ corresponds to
$M_A\!=\!90\,\mbox{GeV}-1\,\mbox{TeV}$, in the MSSM scenario discussed
in the text, with maximal top-squark mixing. From
Ref.~\cite{Carena:2002es}.\label{fig:mssm_A_H+_br_ratios_tanb3_30}}
\end{figure}

As far as QCD and EW radiative corrections go, what we have seen in
Sections~\ref{subsubsec:sm_higgs_to_gaugebosons}-\ref{subsubsec:sm_higgs_loop_decays}
for the SM case applies to the corresponding MSSM decays
too. Moreover, the truly MSSM corrections discussed in
Sections~\ref{subsubsec:mssm_higgs_couplings_bosons} and
\ref{subsubsec:mssm_higgs_couplings_fermions} need
to be taken into account and are included in
Figs.\ref{fig:mssm_h_H_br_ratios_tanb3_30} and \ref{fig:mssm_A_H+_br_ratios_tanb3_30}.

\subsection{Direct bounds on both SM and MSSM Higgs bosons from LEP}
\label{subsec:sm_mssm_higgs_direct_bounds}
\begin{figure}
\begin{tabular}{lcr}
\begin{minipage}{.3\linewidth}
\includegraphics[scale=0.5]{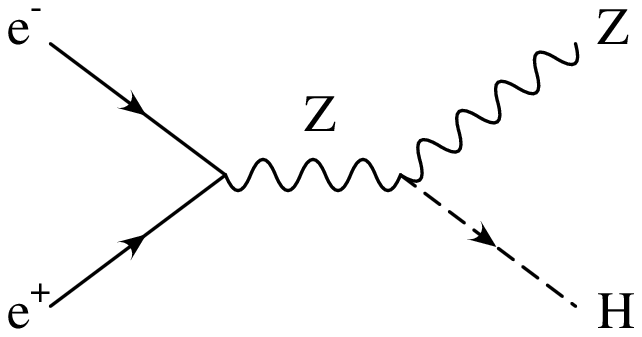}
\end{minipage} &
\begin{minipage}{.3\linewidth}
\includegraphics[scale=0.5]{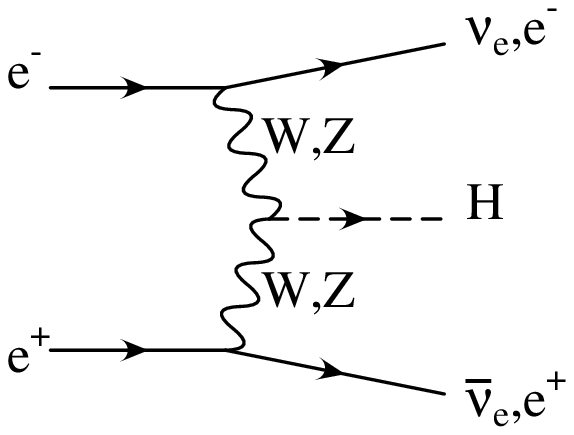}
\end{minipage} &
\begin{minipage}{.3\linewidth}
\includegraphics[scale=0.5]{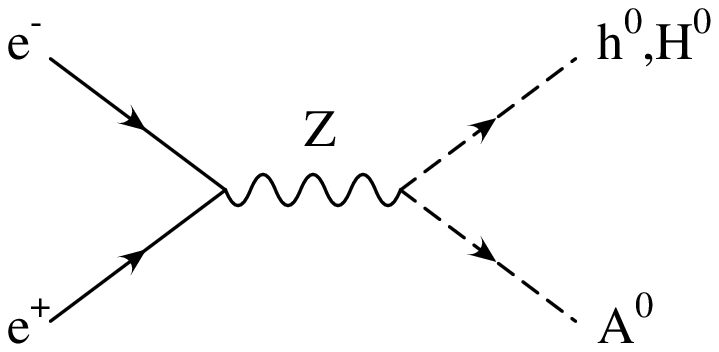}
\end{minipage}
\end{tabular}
\caption[]{SM and MSSM neutral Higgs-boson production channels at LEP2. 
\label{fig:lep2_search_processes}}
\end{figure}

LEP2 has searched for a SM Higgs at center of mass energies between
189 and 209~GeV.  In this regime, a SM Higgs boson is produced mainly
through Higgs-boson strahlung from $Z$ gauge bosons,
$e^+e^-\rightarrow Z^*\rightarrow HZ$, and to a lesser extent through
$WW$ and $ZZ$ gauge boson fusion, $e^+e^-\rightarrow WW,ZZ\rightarrow
H\nu_e\bar{\nu}_e, He^+e^-$ (see
Fig.~\ref{fig:lep2_search_processes}). Once produced, it decays mainly
into $b\bar{b}$ pairs, and more rarely into $\tau^+\tau^-$ pairs. The
four LEP2 experiments have been looking for: \emph{i)} a four jet
final state ($H\rightarrow b\bar{b}$, $Z\rightarrow q\bar{q}$), 
\emph{ii)} a missing
energy final state ($H\rightarrow b\bar{b}$,
$Z\rightarrow\nu\bar{\nu}$), 
\emph{iii)} a
leptonic final state ($H\rightarrow b\bar{b}$, $Z\rightarrow l^+l^-$) and
\emph{iv)} a specific  $\tau$-lepton final state ($H\rightarrow
b\bar{b}$, $Z\rightarrow\tau^+\tau^-$ plus $
H\rightarrow\tau^+\tau^-$, $Z\rightarrow q\bar{q}$).
The absence of any statistical significant signal has set a 95\% CL
lower bound on the SM Higgs boson at
\[
M_{H_{SM}}>114.4\,\,\mbox{GeV}\,\,\,. 
\]

\begin{figure}
\hspace{-1.truecm}
\begin{tabular}{lr}
\begin{minipage}{0.5\linewidth}
{\includegraphics[scale=0.35]{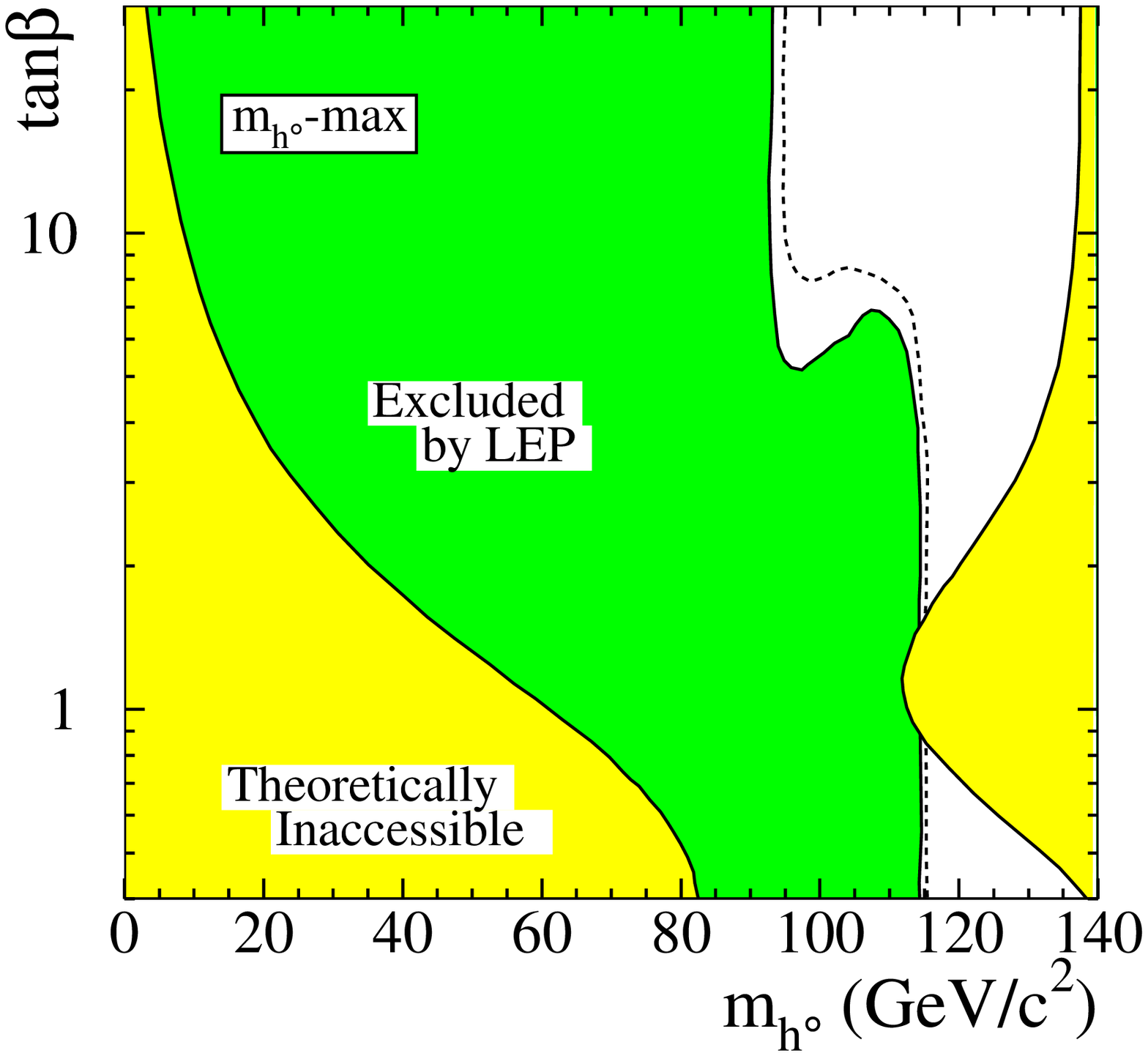}}
\end{minipage} &
\begin{minipage}{0.5\linewidth}
{\includegraphics[scale=0.35]{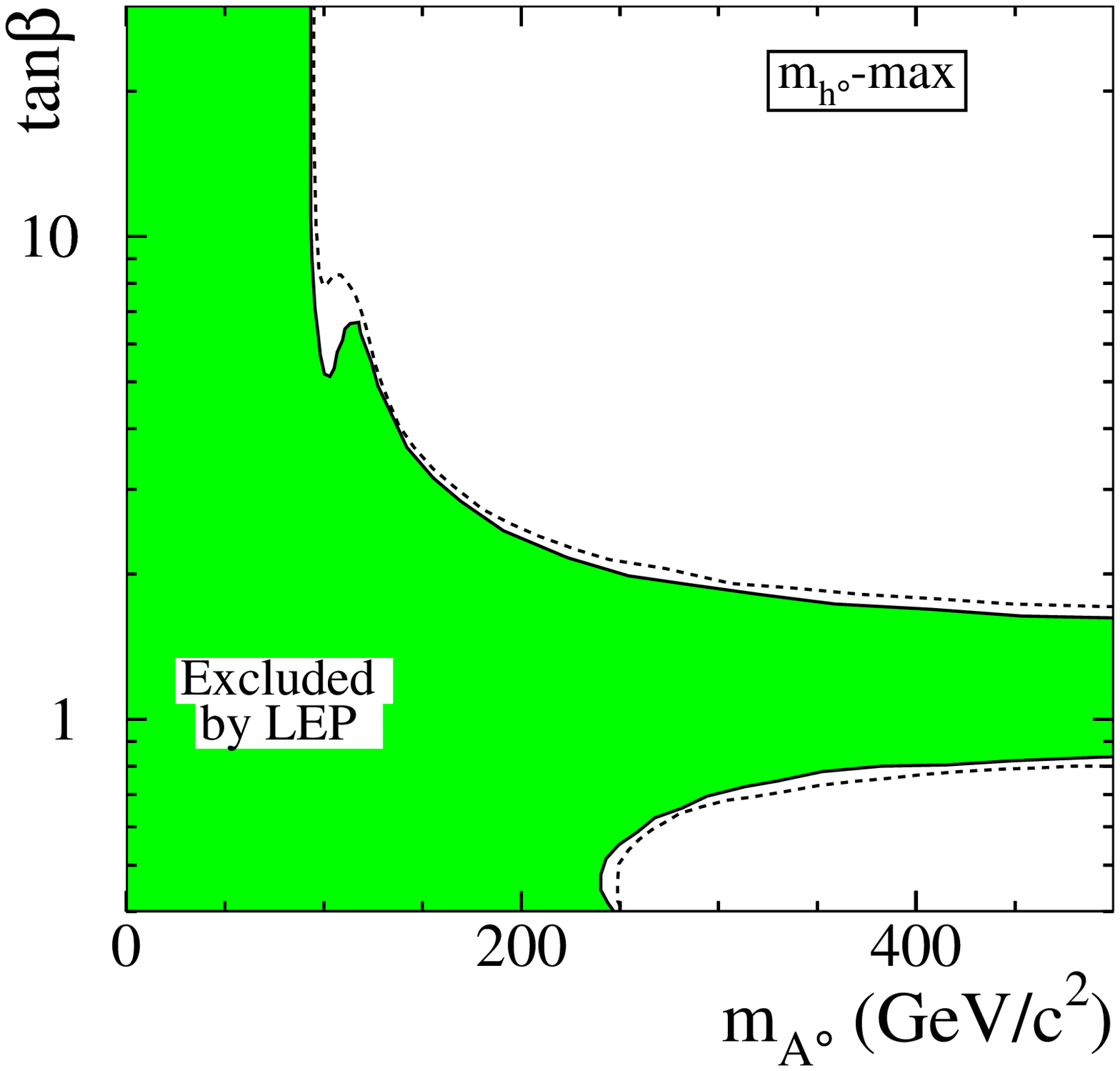}}
\end{minipage}\\
\begin{minipage}{0.5\linewidth}
{\includegraphics[scale=0.35]{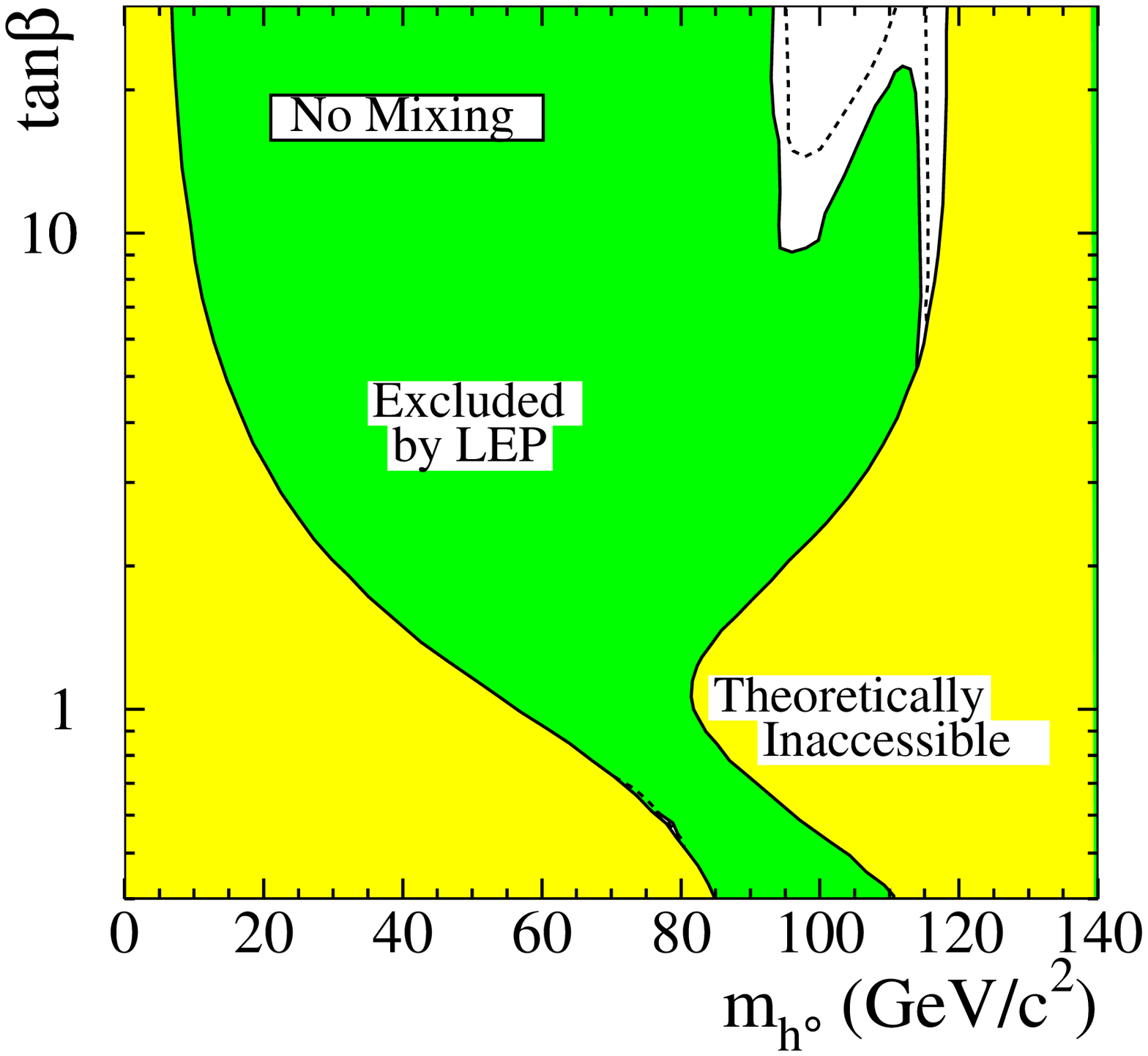}}
\end{minipage} &
\begin{minipage}{0.5\linewidth}
{\includegraphics[scale=0.35]{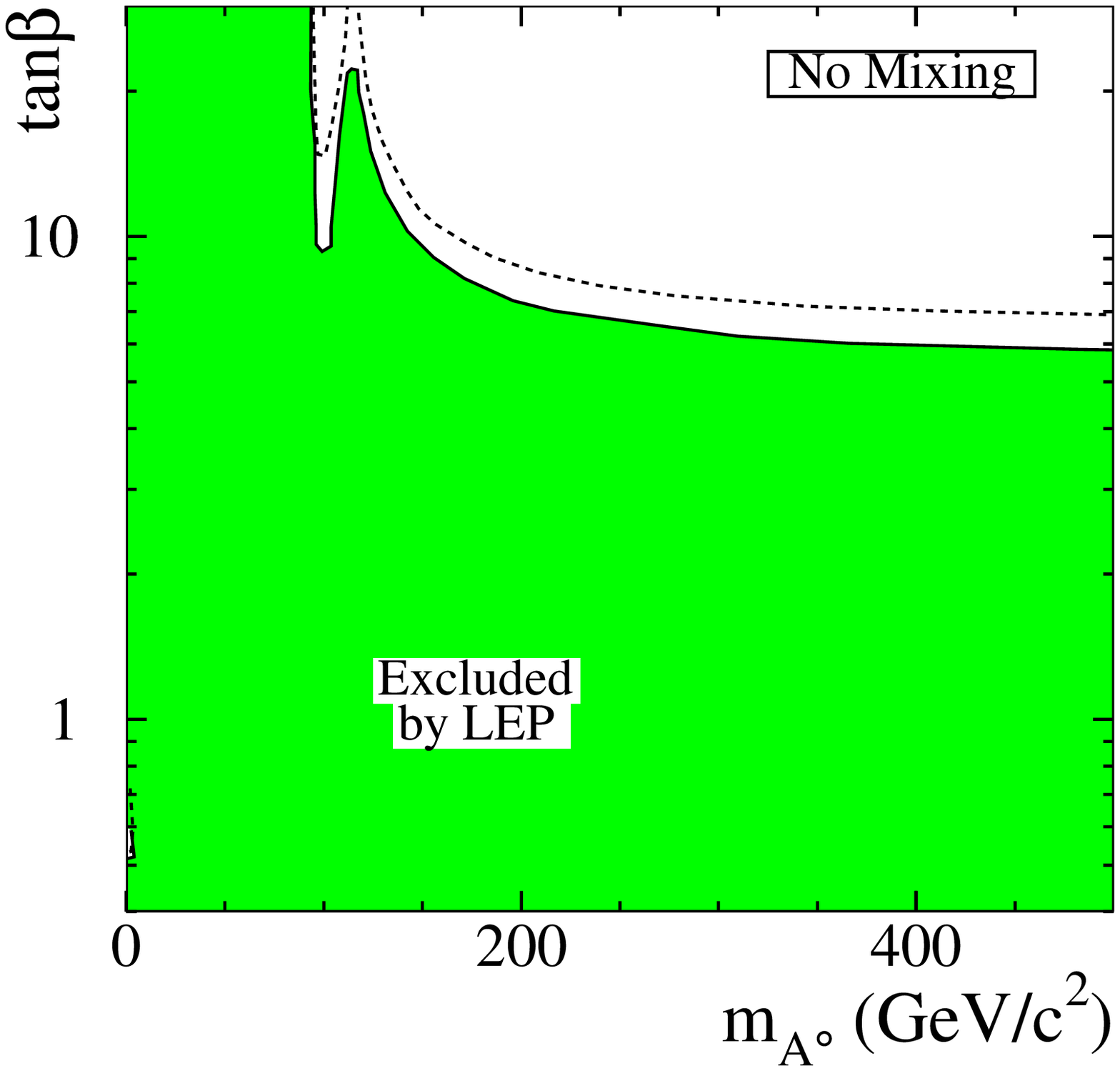}}
\end{minipage}
\end{tabular}
\caption[]{95\% CL exclusion limits for MSSM Higgs parameters from LEP2:
$(M_h,\tan\beta)$ (left) and $(M_A,\tan\beta)$ (right). Both the
maximal and no-mixing scenarios are illustrated, for $M_S\!=\!1$~TeV
and $m_t\!=\!179.3$~GeV. The dashed lines indicate the boundaries that
are excluded on the basis of Monte Carlo simulations in the absence of
a signal. From
Ref.~\cite{Djouadi:2005gj}.\label{fig:mh_mA_tanb_exclusion_from_searches}}
\end{figure}
LEP2 has also looked for the light scalar ($h^0$) and pseudoscalar
($A^0$) MSSM neutral Higgs bosons. In the decoupling regime, when
$A^0$ is very heavy and $h^0$ behaves like a SM Higgs bosons, only
$h^0$ can be observed and the same bounds established for the SM Higgs
boson apply. The bound can however be lowered when $m_A$ is
lighter. In that case, $h^0$ and $A^0$ can also be pair produced
through $e^+e^-\rightarrow Z\rightarrow h^0A^0$ (see
Fig.~\ref{fig:lep2_search_processes}). Combining the different
production channels one can derive plots like those shown in
Fig.~\ref{fig:mh_mA_tanb_exclusion_from_searches}, where the excluded
$(M_h,\tan\beta)$ and $(M_A,\tan\beta)$ regions of the MSSM parameter
space are shown.  The LEP2 collaborations~\cite{lephwg:2004mssm} have
been able to set the following bounds at 95\% CL:
\[
M_{h,A}>93.0\,\,\mbox{GeV}\,\,\,,
\]
obtained in the limit when $\cos(\beta-\alpha)\simeq 1$
(anti-decoupling regime) and for large $\tan\beta$. The plots in
Fig.~\ref{fig:mh_mA_tanb_exclusion_from_searches} have been obtained
in the maximal mixing scenario (explained in
Section~\ref{subsubsec:higgs_mssm}). For no-mixing, the corresponding
plots would exclude a much larger region of the MSSM parameter space.

Finally, the LEP collaborations have looked for the production of the
MSSM charged Higgs boson in the associated production channel:
$e^+e^-\rightarrow\gamma,Z^*\rightarrow H^+H^-$~\cite{lhwg:2001xy}.
An absolute lower bound of
\[
M_{H^\pm}>79.3\,\,\mbox{GeV}\,\,\,
\]
has been set by the ALEPH collaboration, and slightly lower values
have been obtained by the other LEP collaborations.

Both the Tevatron and the LHC have extended these bounds as we will
discuss in Sections~\ref{sec:tev_searches} and \ref{sec:lhc_searches}.

\subsection{SM Higgs production at hadron colliders}
\label{subsec:sm_higgs_hadron_colliders}

\begin{figure}
\begin{tabular}{lcr}
\begin{minipage}{0.3\linewidth}
\includegraphics[scale=0.5]{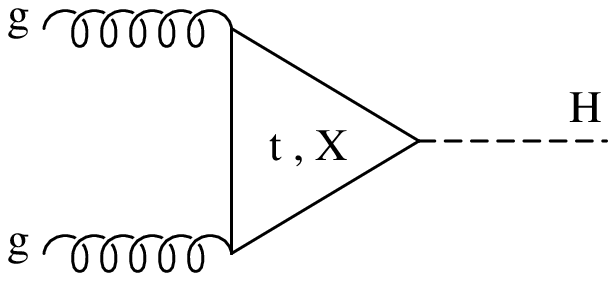}
\end{minipage}&
\begin{minipage}{0.3\linewidth}
\includegraphics[scale=0.5]{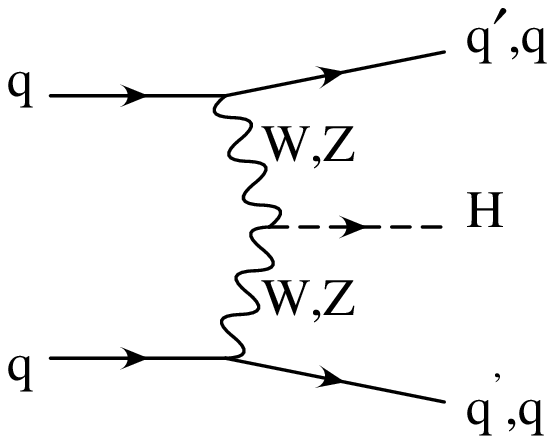}
\end{minipage}&
\begin{minipage}{0.3\linewidth}
\includegraphics[scale=0.5]{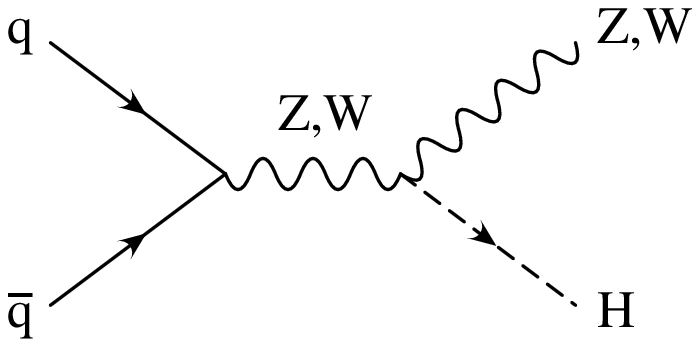}
\end{minipage}
\end{tabular}
\caption[]{Leading Higgs production processes at hadron colliders:
$gg\rightarrow H$, $qq^\prime\rightarrow qq^\prime H$, and $q\bar{q}\rightarrow
WH,ZH$.\label{fig:sm_higgs_production:ggh_qqHZ_qqHW_qqHqq}}
\end{figure}

\begin{figure}
\centering
\includegraphics[scale=0.5]{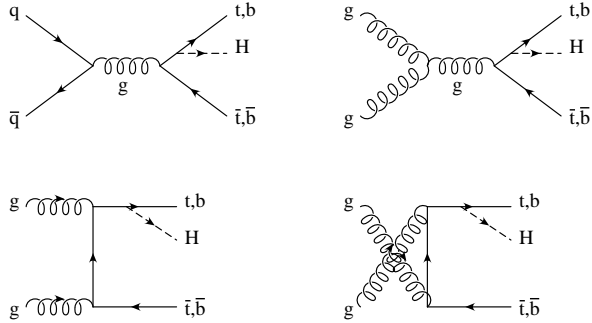}
\caption[]{Higgs production with heavy quarks: sample of Feynman
diagrams illustrating the two corresponding parton level processes
$q\bar{q},gg\rightarrow t\bar{t}H,b\bar{b}H$. Analogous diagrams with
the Higgs-boson leg attached to the remaining top(bottom)-quark
legs are understood.\label{fig:sm_higgs_production:QQh}}
\end{figure}

The parton level processes through which a SM Higgs boson can be
produced at hadron colliders are illustrated in
Figs.~\ref{fig:sm_higgs_production:ggh_qqHZ_qqHW_qqHqq} and
\ref{fig:sm_higgs_production:QQh}.

\begin{figure}
\centering
\includegraphics[scale=0.45,angle=-90]{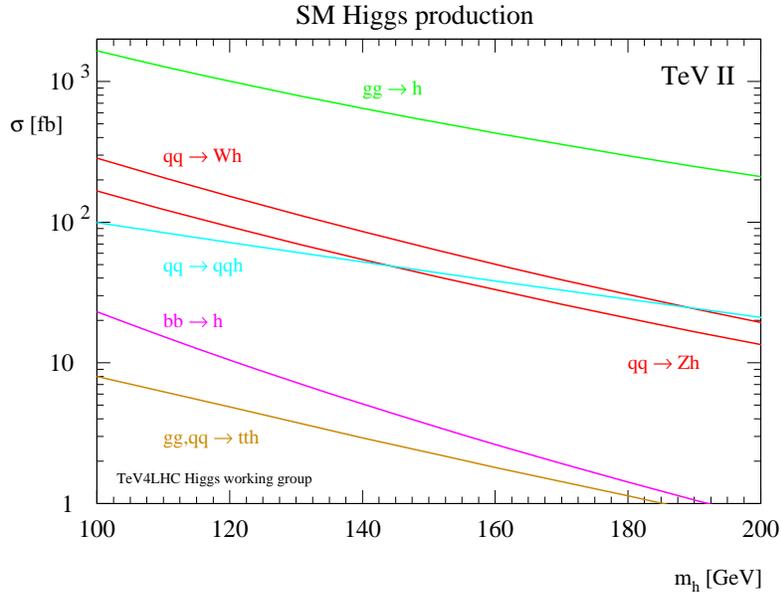}
\caption[]{Cross sections for SM Higgs-boson production
processes at the Tevatron, Run II ($\sqrt{s}\!=\!1.96$~TeV). From
Ref.~\cite{Tev4lhc:hwg}.\label{fig:sm_higgs_tevatron}}
\end{figure}

\begin{figure}
\centering
\includegraphics[scale=0.45,angle=-90]{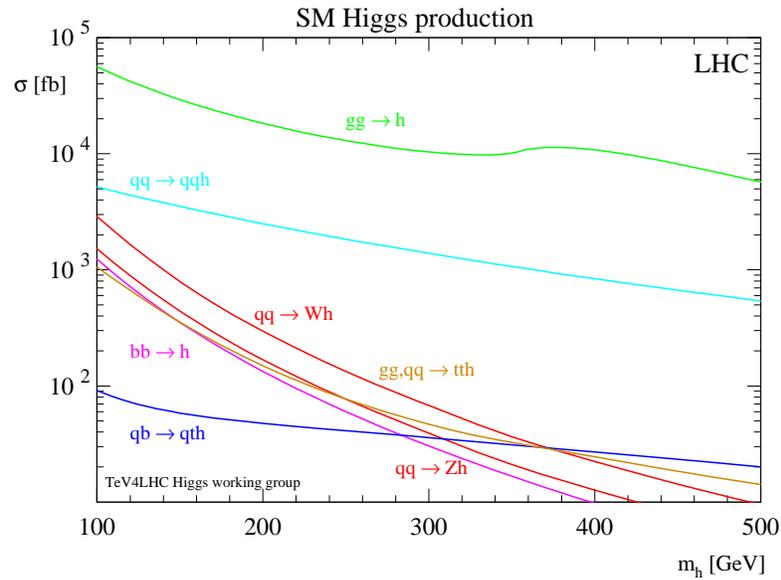}
\caption[]{Cross sections for SM Higgs-boson production
processes at the LHC ($\sqrt{s}\!=\!14$~TeV). From
Ref.~\cite{Tev4lhc:hwg}.\label{fig:sm_higgs_lhc}}
\end{figure}

Figures~\ref{fig:sm_higgs_tevatron} and \ref{fig:sm_higgs_lhc}
summarize the cross sections for all these production modes as
functions of the SM Higgs-boson mass, at the Tevatron with
$\sqrt{s}\!=\!1.96$~TeV and at the LHC with $\sqrt{s}\!=\!14$~TeV.
These figures have been produced during the TeV4LHC
workshop~\cite{Tev4lhc:hwg}, and contain most known orders of QCD
corrections as well as the fairly up to date input parameters. They
serve the purpose of illustrating  the different relevance of
different production processes at both the Tevatron and the LHC and
allows us to discuss some general phenomenological aspects of hadronic
Higgs production.  We postpone further details about QCD corrections
till Section~\ref{sec:theory}, where we will discuss the uncertainties
involved in the prediction of Higgs production cross sections and will
give more accurate plots including both EW and QCD updated effects.

The leading production mode is gluon-gluon fusion, $gg\rightarrow H$
(see first diagram in
Fig.~\ref{fig:sm_higgs_production:ggh_qqHZ_qqHW_qqHqq}). In spite of
being a loop-induced process, it is greatly enhanced by the top-quark
loop. For light- and intermediate-mass Higgs bosons, however, the very
large cross section of this process has to compete against a very
large hadronic background, since the Higgs boson mainly decays to
$b\bar{b}$ pairs, and there is no other non-hadronic probe that can
help distinguishing this mode from the overall hadronic activity in
the detector. To beat the background, one has to employ subleading
Higgs decay modes, like $H\rightarrow\gamma\gamma$, and this
\emph{dilutes} the large cross section to some extent. For larger
Higgs masses, above the $ZZ$ threshold, on the other hand, gluon-gluon
fusion together with $H\rightarrow ZZ$ produces a very distinctive
signal, and make this mode a ``\emph{gold-plated mode}'' for
detection. For this reason, $gg\rightarrow H$ plays a fundamental role
at the LHC over the entire Higgs-boson mass range, but is of very
limited use at the Tevatron, where it can only be considered for Higgs
boson masses very close to the upper reach of the machine ($M_H\simeq
200$~GeV).

Weak boson fusion ($qq^\prime\rightarrow qq^\prime H$, see second diagram in
Fig.~\ref{fig:sm_higgs_production:ggh_qqHZ_qqHW_qqHqq}) and the
associated production with weak gauge bosons ($q\bar{q}\rightarrow
WH,ZH$, see third diagram in
Fig.~\ref{fig:sm_higgs_production:ggh_qqHZ_qqHW_qqHqq}) have also
fairly large cross sections, of different relative size at the
Tevatron and at the LHC. $q\bar{q}\rightarrow WH,ZH$ is particularly
important at the Tevatron, where only a relatively light Higgs boson
($M_H<200$~GeV) is accessible. In this mass region, $gg\rightarrow H,
H\rightarrow \gamma\gamma$ is too small at the Tevatron, while
$qq^\prime\rightarrow qq^\prime H$ is suppressed (because the initial state is
$p\bar{p}$). On the other hand, $qq^\prime\rightarrow qq^\prime H$ becomes
instrumental at the LHC ($pp$ initial state) for low- and
intermediate-mass SM Higgs bosons, where its characteristic final
state configuration, with two very forward jets, has been shown to
greatly help in disentangling this signal from the hadronic
background, using different Higgs decay channels.

Finally, the production of a SM Higgs boson with heavy quarks, in the
two channels $q\bar{q},gg\rightarrow Q\bar{Q}H$ (with $Q\!=\!t,b$, see
Fig.~\ref{fig:sm_higgs_production:QQh}), is sub-leading at both the
Tevatron and the LHC, but has a great physics potential.  The
associated production with $t\bar{t}$ pairs is too small to be
relevant for the Tevatron, but will play an important role at the LHC,
where enough statistics will be available to fully exploit the
signature of a $t\bar{t}H, H\rightarrow b\bar{b}$ final state. Indeed,
this channel has not been used for discovery but will certainly become
important now that the properties of the discovered spin-0 particle
need to be thoroughly investigated, since it offers the unique
possibility of \textit{directly} measuring one of its most important
couplings, namely the coupling to top quarks.  On the other hand, the
production of a SM Higgs boson with $b\bar{b}$ pairs is tiny, since
the SM bottom-quark Yukawa coupling is suppressed by the bottom-quark
mass. Therefore, the $b\bar{b}H,\,H\rightarrow b\bar{b}$ channel is
the ideal candidate to provide evidence of new physics, in particular
of extension of the SM, like supersymmetric models, where the
bottom-quark Yukawa coupling to one or more Higgs bosons is enhanced
(e.g., by large $\tan\beta$ in the MSSM). $b\bar{b}H$ production is
kinematically well within the reach of the Tevatron, RUN II. First
studies from both CDF~\cite{Affolder:2000rg} and
D$\emptyset$~\cite{Abazov:2005yr} have translated the absence of a
$b\bar{b}h^0,H^0,A^0$ signal into an upper bound on the $\tan\beta$
parameter of the MSSM. A difficult channel to measure at the LHC,
because of the large hadronic background, it could however offer a
striking signal of new physics if observed.

\section{Higgs searches at the Tevatron}
\label{sec:tev_searches}

In a recent note \cite{cdfd0:2012cn} the CDF and D0 collaboration
presented combined results of direct searches for the SM Higgs boson
in $p\bar{p}$ collisions at 1.96~TeV. They combined the most recent
results of all the Tevatron Higgs-boson searches in the mass range
$m_H=100-200$~GeV. These analyses sought signals of a SM Higgs boson
produced through associated production with an EW vector boson
($q\bar{q}\rightarrow HW/Z$), through gluon-gluon fusion
($gg\rightarrow H$), and through vector boson fusion ($qq^\prime\rightarrow
Hqq^\prime$), corresponding to integrated luminosities ranging from 5.4 to 10
fb$^{-1}$.  They studied the $H\rightarrow b\bar{b}$, $H\rightarrow
W^+W^-$, $H\rightarrow ZZ$, $H\rightarrow\tau^+\tau^-$, and
$H\rightarrow\gamma\gamma$ decay signatures. The greatest sensitivity
was reached using $H\rightarrow W^+W^-$ (with the $W$s decaying
leptonically) in the $m_H>125$~GeV region and looking for
$q\bar{q}\rightarrow HW/Z$ with $H\rightarrow b\bar{b}$ (with the $W$
or $Z$ decaying leptonically) in the $m_H<125$~GeV mass region.

To quantify the expected sensitivity across the whole mass range, CDF and
D0 have studied the distribution of the Log-Likelihood Ratios (LLR)
for different hypothesis (background only, signal+background)
Results have been presented in terms of $\mathrm{LLR}_b$ and $\mathrm{LLR}_{s+b}$
defined as,
\begin{equation}
\mathrm{LLR}=-2\ln\frac{p(\mathrm{data}|H_1)}{p(\mathrm{data}|H_0)}\,\,\,,
\end{equation}
where $H_1$ denotes the test hypothesis, which admits the presence of
SM backgrounds and a Higgs-boson signal, while $H_0$ is the null
hypothesis, for only SM background, and \textit{data} is either an
ensemble of pseudo-experiment data constructed from the expected
signal and backgrounds, or the actual observed data.  

\begin{figure}
\centering
\includegraphics[scale=0.5]{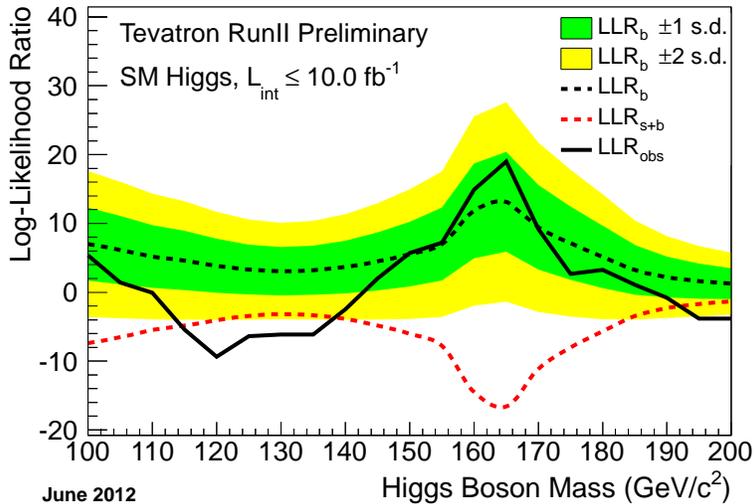}
\caption[]{ Distribution of the LLR as a function of the Higgs-boson
  mass for the combined CDF and D0 analyses (see text for
  interpretation). From
  Ref.~\cite{cdfd0:2012cn}.\label{fig:sm_higgs_tev_comb_LLR}}
\end{figure}
\begin{figure}
\centering
\includegraphics[scale=0.5]{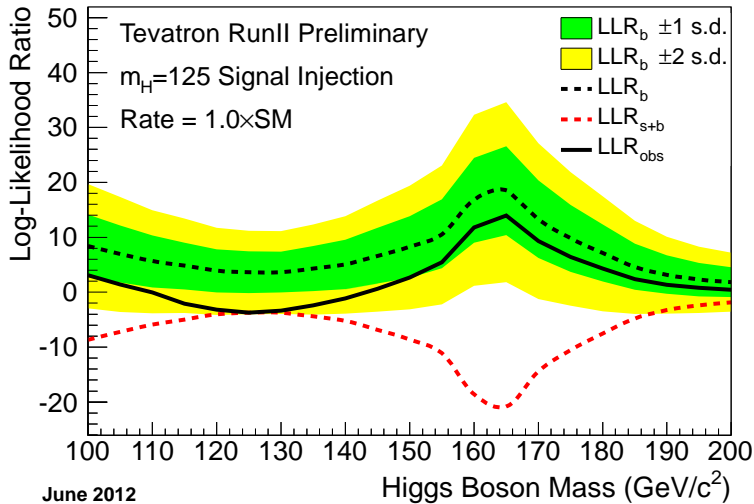}
\caption[]{ Same as Fig.~\ref{fig:sm_higgs_tev_comb_LLR}. The solid
  black line corresponds here to the case of an artificially injected
  SM Higgs-boson signal at $m_H=125$~GeV.  From
  Ref.~\cite{cdfd0:2012cn}.\label{fig:sm_higgs_tev_comb_LLR_125}}
\end{figure}

As an example,
in Fig.~\ref{fig:sm_higgs_tev_comb_LLR} we see the LLR distributions for the combined CDF+D0
analyses as functions of the Higgs-boson mass. The solid black line
corresponds to the observed data ($\mathrm{LLR}_{obs}$). The dashed
black and red lines represent the median for the background-only
hypothesis ($\mathrm{LLR}_b$) and the signal-plus-background hypothesis
($\mathrm{LLR}_{s+b}$). The shaded bands represent the one and two
standard-deviation departures from the median for $\mathrm{LLR}_b$,
assuming that no signal is present and only statistical fluctuations
and systematic effects are present.  We note that the separation
between the medians of the $\mathrm{LLR}_b$ and $\mathrm{LLR}_{s+b}$
distributions provides a measure of the discriminating power of the
search.  Moreover, the value of $\mathrm{LLR}_{obs}$ relative to
$\mathrm{LLR}_{s+b}$ or $\mathrm{LLR}_b$ indicates if the data
distribution resembles more the case in which a signal is present or
not.  With this in mind, Fig.~\ref{fig:sm_higgs_tev_comb_LLR} shows that the data are
consistent with a background-only hypothesis for $m_H>145$~GeV, except
above 190 GeV, where the signal-plus-background and background-only
hypotheses cannot be separated very well.  On the other hand, for
$m_H$ from 110 to 140~GeV we see an excess in the data consistent with
the expectation for a SM Higgs boson in this mass range (red dashed line).
We notice that in this region the ability of separating
$\mathrm{LLR}_{s+b}$ from $\mathrm{LLR}_b$ is at the two-$\sigma$
level. It is interesting to compare these results to what one would obtain by
artificially \textit{injecting} a signal for a SM Higgs with $m_H=125$~GeV.
This is shown in Fig.~\ref{fig:sm_higgs_tev_comb_LLR_125} where the solid black line now represents
the artificial Higgs signal.

The probability of observing a signal-plus-background-like outcome
without the presence of a signal. i.e. the probability that an upward
fluctuation of the background provides a signal-plus-background-like
response as observed data, is defined as,
\begin{equation}
1-\mathrm{CL}_b=p(LLR\le LLR_{obs}|H_0)\,\,\,,
\end{equation}
while the probability of a downward fluctuation of the sum of signal
and background in the data is defined as,
\begin{equation}
\mathrm{CL}_{s+b}=p(LLR\ge LLR_{obs}|H_1)\,\,\,,
\end{equation}
where $\mathrm{LLR}_{obs}$ is the value of the test statistic computed
for the data.
A small value of $\mathrm{CL}_{s+b}$ denotes inconsistency with $H_1$.

\begin{figure}
\centering
\includegraphics[scale=0.5]{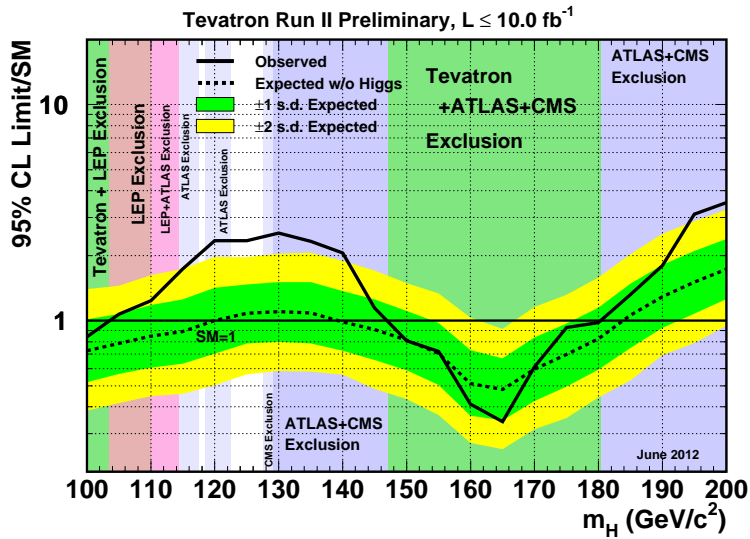}
\caption[]{Observed and expected (background-only hypothesis) 95\%
  C.L. upper limits on the ratios to the SM cross section, as function
  of the Higgs-boson mass, for the combined CDF and D0 analyses. From
  Ref.~\cite{cdfd0:2012cn}.\label{fig:sm_higgs_tev_exclusion}}
\end{figure}

To facilitate comparison with the Standard Model, CDF and D0 have
presented their resulting limit divided by the SM Higgs-boson
production cross section, $\sigma_{SM}$, as a function of the
Higgs-boson mass. This is illustrated in
Fig.~\ref{fig:sm_higgs_tev_exclusion}. This figure is rich of
information and we will discuss it in detail. First of all,
Fig.~\ref{fig:sm_higgs_tev_exclusion} includes all existing limits on
the SM Higgs-boson mass, including previous limits from LEP and limits
provided by the LHC till they recently announced discovery of a spin-0
particle with mass around 126~GeV.
The line patterns and colors have the same meaning as in
Figs.~\ref{fig:sm_higgs_tev_comb_LLR} and
\ref{fig:sm_higgs_tev_comb_LLR_125}, i.e. the black solid line
indicates the observed limit, the black dashed line indicates the
expected limit in the background-only hypothesis, while the green and
yellow bands represent the variation of the expected limit by one and
two standard deviations respectively.

Since the figure shows the 95\% CL of the ratio
$R=\sigma/\sigma_{SM}$, a value of the limit observed ratio that is
less or equal to one excludes that mass at the 95\% C.L. The Tevatron
combined analyses therefore exclude the regions $100<m_H<103$~GeV and
$147<m_H<180$~GeV, as shown in Fig.~\ref{fig:sm_higgs_tev_exclusion}
by the green vertical bands.

On the other hand, if the solid black line is above 1.0 and also
somewhat above the dotted black line (an excess), then there might be
a hint that the Higgs exists with a mass at that value. If the solid
black line is at the upper edge of the yellow band, then there may be
95\% certainty that this is above the expectations. It could be a hint
for a Higgs boson of that mass, or it could be a sign of background
processes or of systematic errors that are not well
understood. Indeed, in Fig.~\ref{fig:sm_higgs_tev_exclusion} we see
that the limit curve goes much above the upper edge of the yellow band
in the region between 115 and 140 GeV, and this could point to the
fact that a Higgs boson may indeed be contributing to the data in that
mass range. Still, in the same region the calculated (expected)
background has not reached yet enough sensitivity since the black
dashed line (as well as the one and two standard deviation bands) goes
above the SM=1 threshold. Therefore the indication of a Higgs-like
fluctuation is in this case statistically weak. In the words of the
experiments, this excess only causes the observed limits not to be as
stringent as expected.

\section{Higgs searches at the LHC}
\label{sec:lhc_searches}

Since it started running in 2010, the LHC has been accumulating an
unprecedented amount of data and has past all expectations in
providing exclusion limits for the SM Higgs boson.  In 2011, the LHC
delivered to ATLAS and CMS up to 5.1~fb$^{-1}$ of integrated luminosity of
$pp$ collisions at 7~TeV center-of-mass energy fulfilling all the data
quality requirements to search for the SM Higgs boson. In 2012 the
center-of-mass energy was increased to 8~TeV, and the accelerator
delivered up to extra 5.9~$\mathrm{fb}^{-1}$ of quality data by July 2012,
when the amazing discovery of a spin-0 particle with mass around
125-127~GeV and SM-like Higgs-boson properties has been delivered to
the world~\cite{:2012gk,:2012gu}.  At the same time the 95\%
C.L. exclusion limits for a SM Higgs boson have been updated to
$110<m_H<122.5$~GeV plus $127<m_H<600$~GeV by CMS~\cite{:2012gu} and
to $110<m_H<122.6$~GeV plus $129.7<m_H<558$~GeV by
ATLAS~\cite{:2012gk}.  Results and details of the search have been
published by the two collaborations in Refs.~\cite{:2012gk,:2012gu}
which update recent analyses appeared earlier in
2012~\cite{ATLAS:2012ae,Chatrchyan:2012tx}.

The LHC experiments have looked for a SM Higgs boson in the wide mass
range between the experimental LEP bound (114~GeV) and about 600 GeV.
The main production mode in this range is gluon fusion ($gg\rightarrow
H$), followed by vector-boson fusion ($qq^\prime\rightarrow
Vqq^\prime$) and the associated productions with weak gauge bosons
($q\bar{q}^{(\prime)}\rightarrow ZH/WH$) and top quarks
($q\bar{q},gg\rightarrow t\bar{t}H$). In all combined analyses so far 
the following channels have been considered:
$H\rightarrow\gamma\gamma$, $H\rightarrow ZZ^*$,
$H\rightarrow WW^{(*)}$,
$H\rightarrow b\bar{b}$, and $H\rightarrow\tau^+\tau^-$

\begin{figure}
\begin{tabular}{lr}
\hspace{-0.8truecm}
\begin{minipage}{0.65\linewidth}
{\includegraphics[scale=0.37]{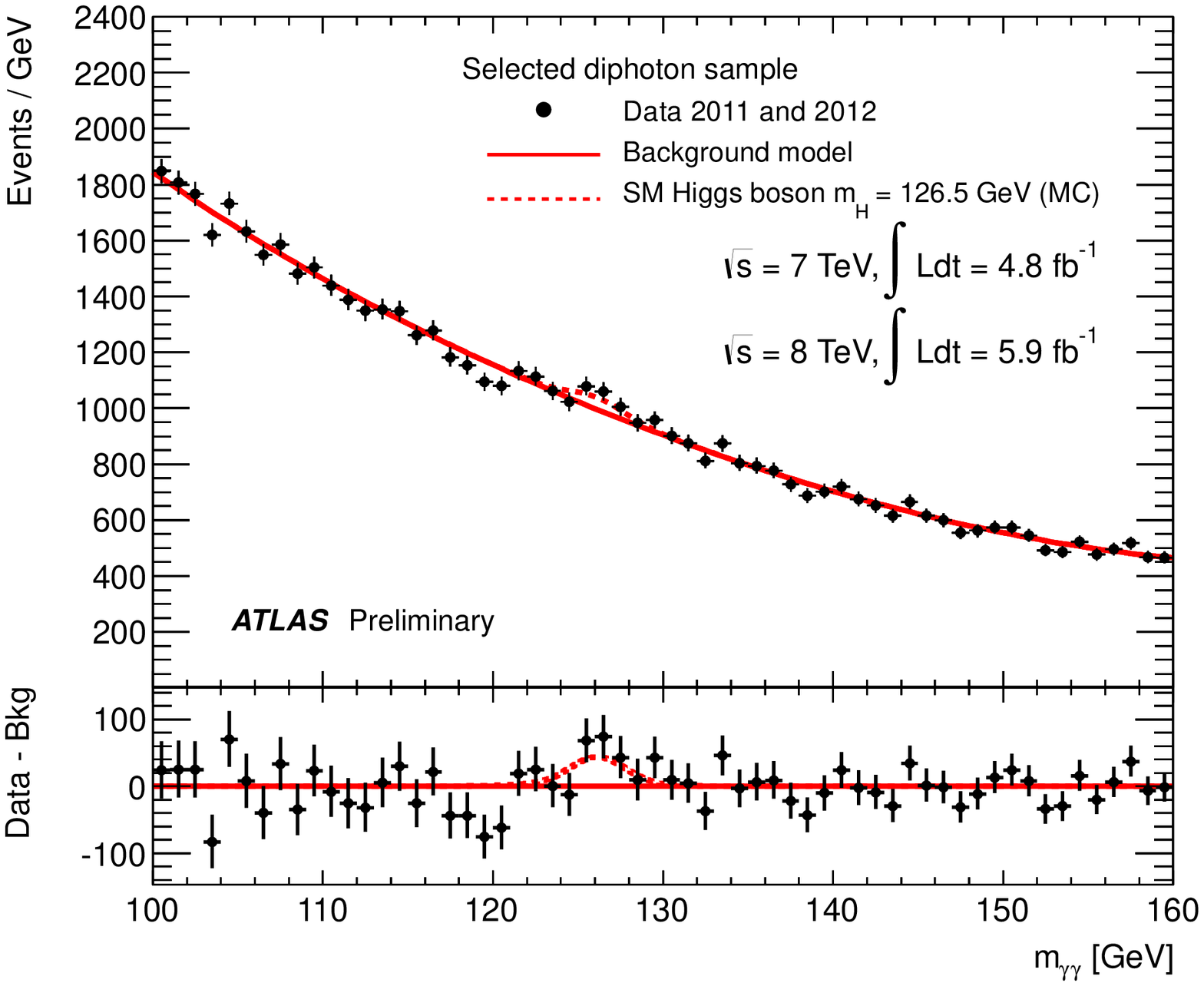}}
\end{minipage} &
\hspace{-1.8truecm}
\begin{minipage}{0.4\linewidth}
{\includegraphics[scale=0.32]{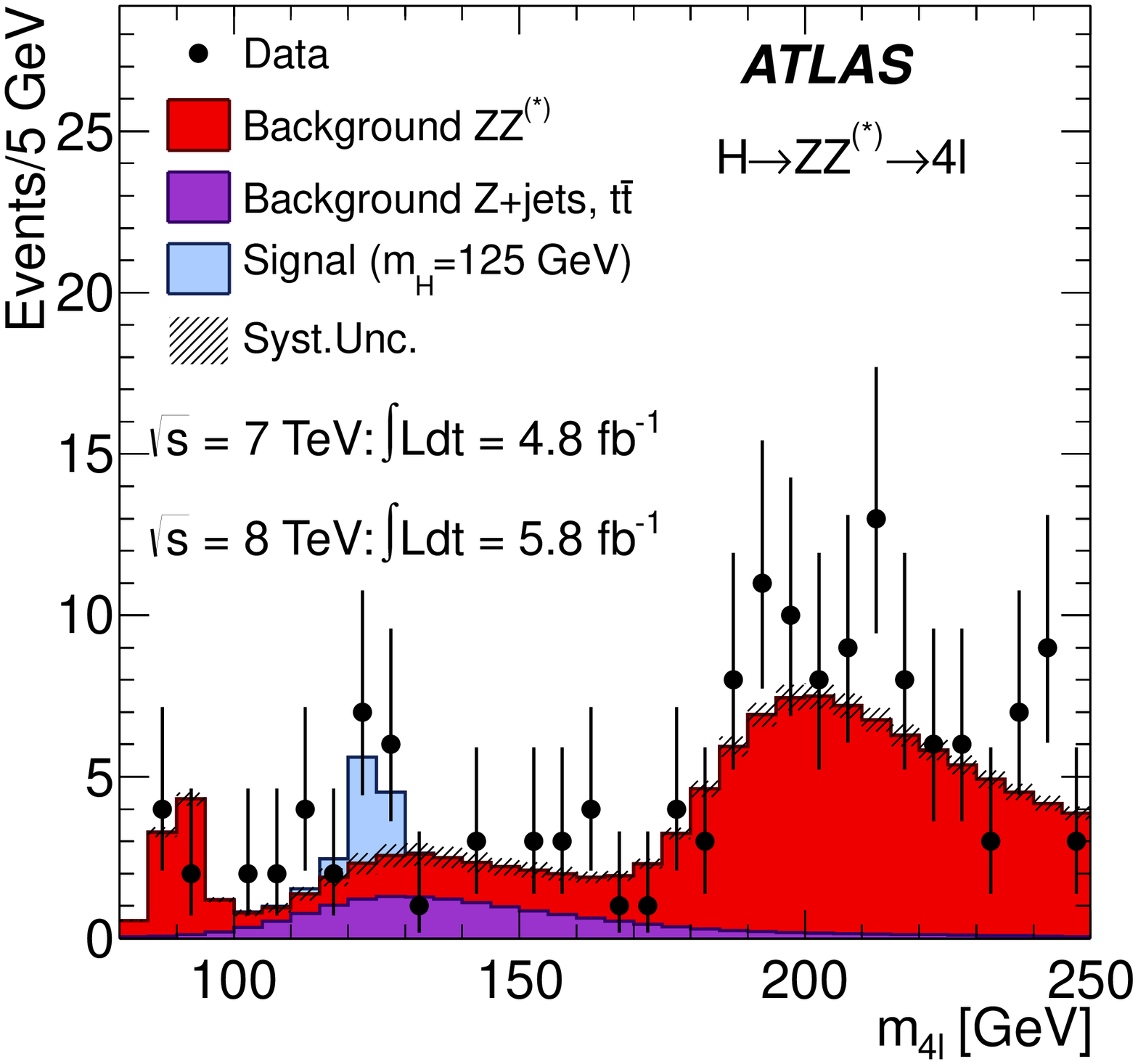}}
\end{minipage}
\end{tabular}
\caption[]{Distributions of the reconstructed invariant for the
  selected candidate events and for the total background and signal
  expected in the $H\rightarrow\gamma\gamma$ (left) and the
  $H\rightarrow ZZ^{(*)}\rightarrow 4l$ (right) channels, in the low
  mass region.  From
  Ref.~\cite{ATLAS:12gaga} and Ref.~\cite{:2012gk}.\label{fig:sm_higgs_discovery_atlas_gaga_ZZ}}
\end{figure}

The crucial channels in the discovery have been the
$H\rightarrow\gamma\gamma$ channel in the low mass region and the
$H\rightarrow ZZ^{(*)}\rightarrow 4l$ over the whole mass range. Both
channels provide a high-resolution invariant mass for fully
reconstructed candidates in the respective mass
regions. The plots in  Fig.~\ref{fig:sm_higgs_discovery_atlas_gaga_ZZ}
are from the ATLAS presentation
at CERN on July 4$^{th}$, 2012: they show the excess of data points around
125-127~GeV in both channels.

\begin{figure}
\begin{tabular}{lr}
\hspace{-1.truecm}
\begin{minipage}{0.5\linewidth}
{\includegraphics[scale=0.38]{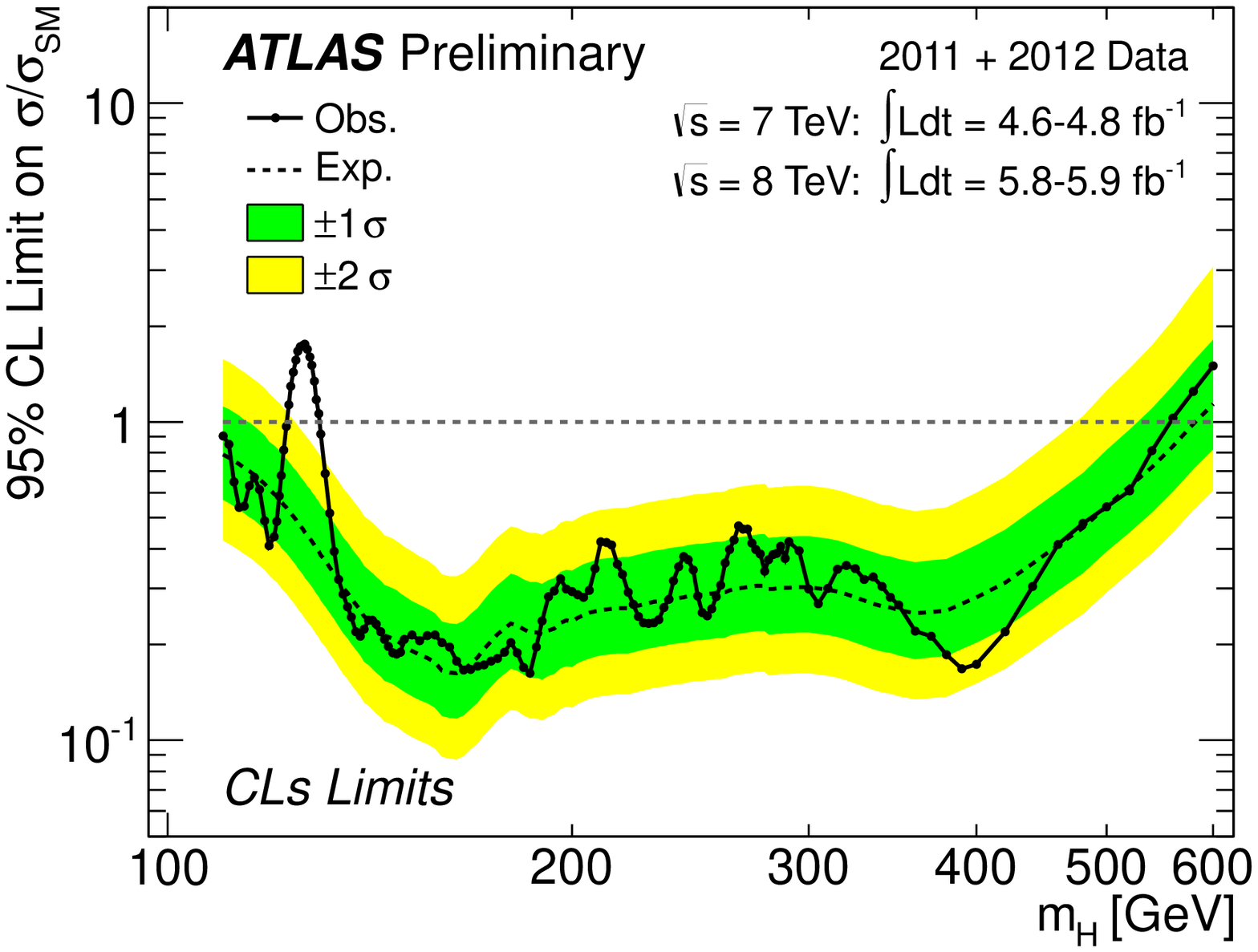}}
\end{minipage} &
\begin{minipage}{0.5\linewidth}
{\hspace{-0.5truecm}\includegraphics[scale=0.38]{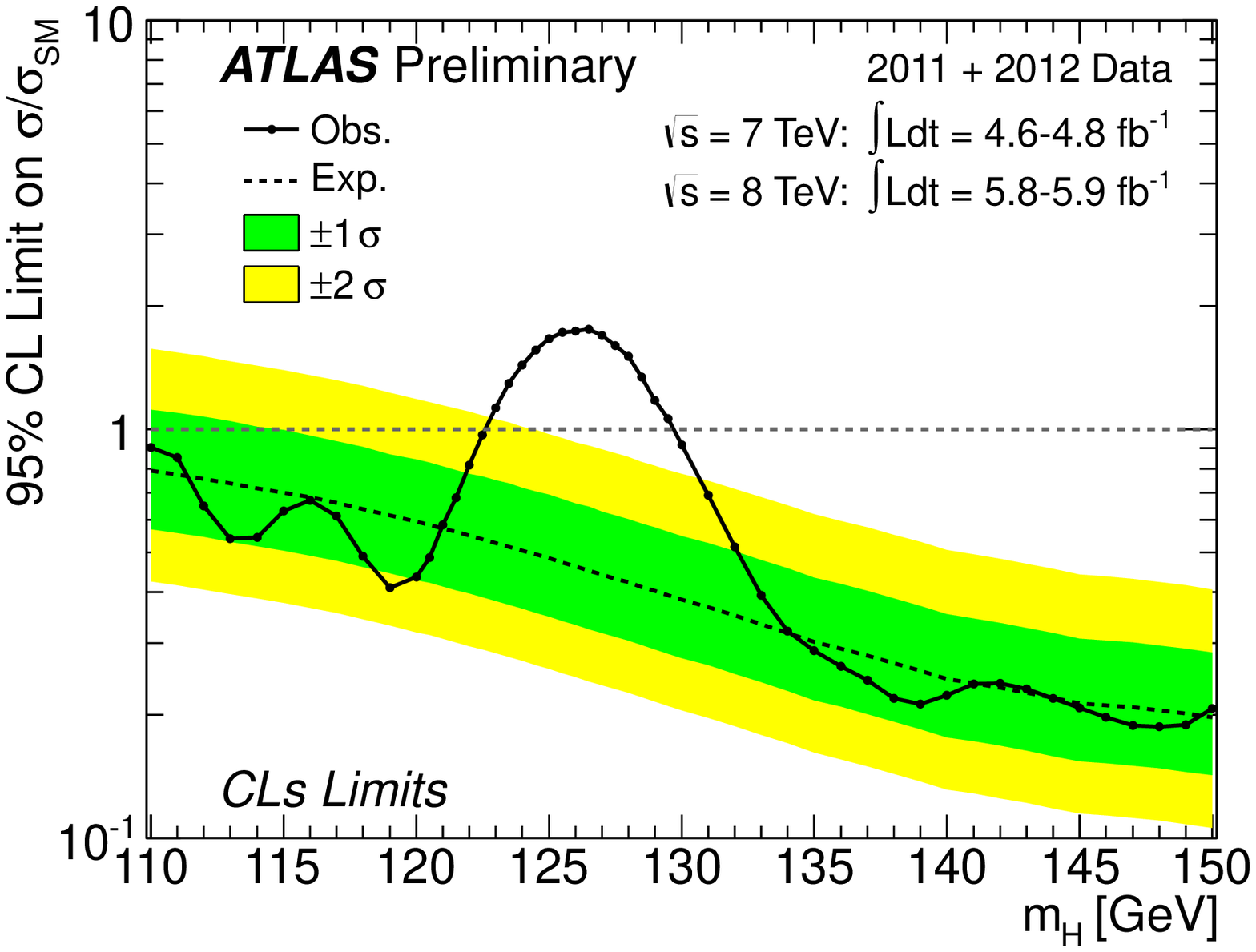}}
\end{minipage}\\
\hspace{-0.5truecm}
\begin{minipage}{0.5\linewidth}
{\includegraphics[scale=0.3]{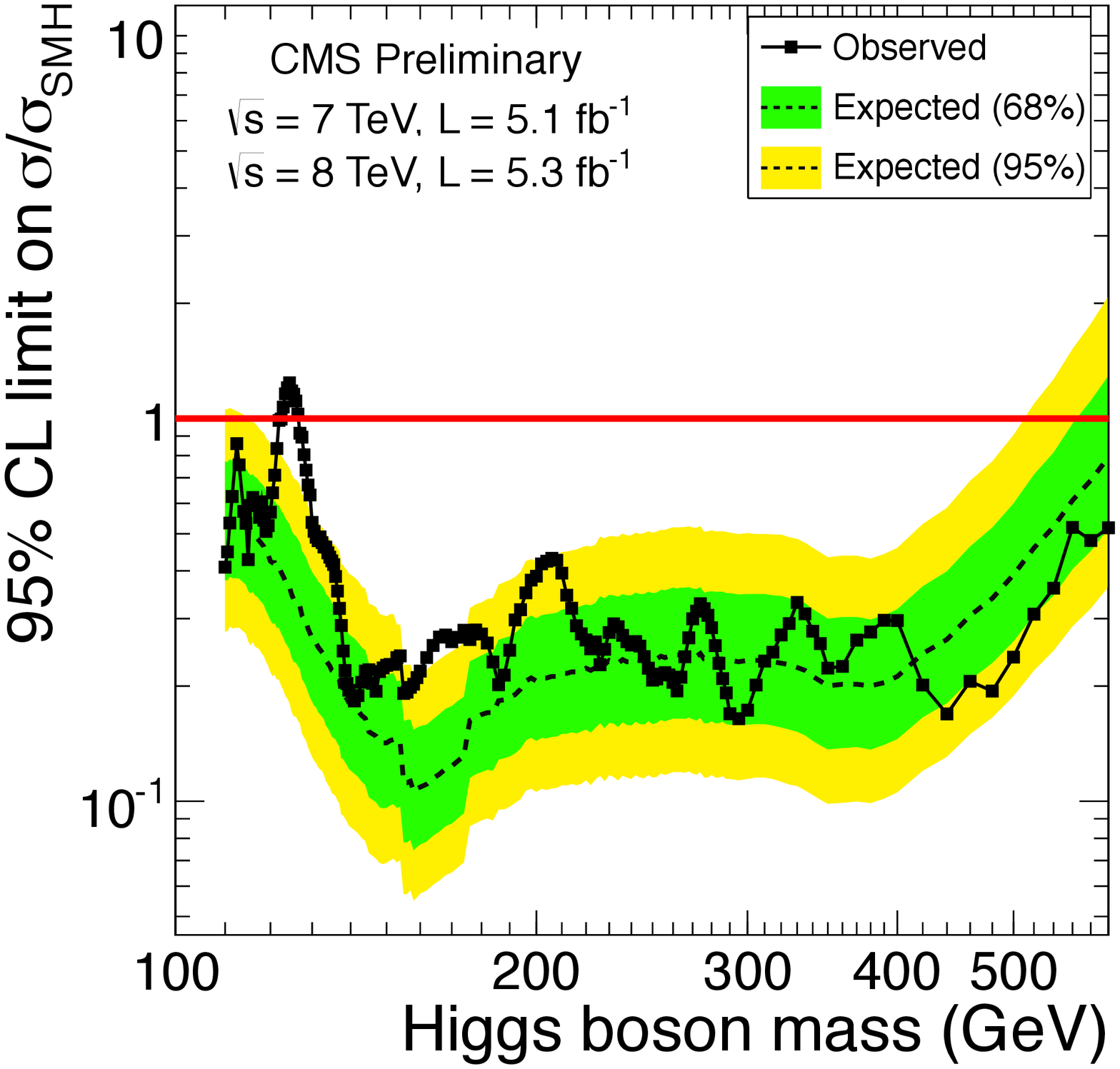}}
\end{minipage} &
\hspace{-0.5truecm}
\begin{minipage}{0.5\linewidth}
{\includegraphics[scale=0.3]{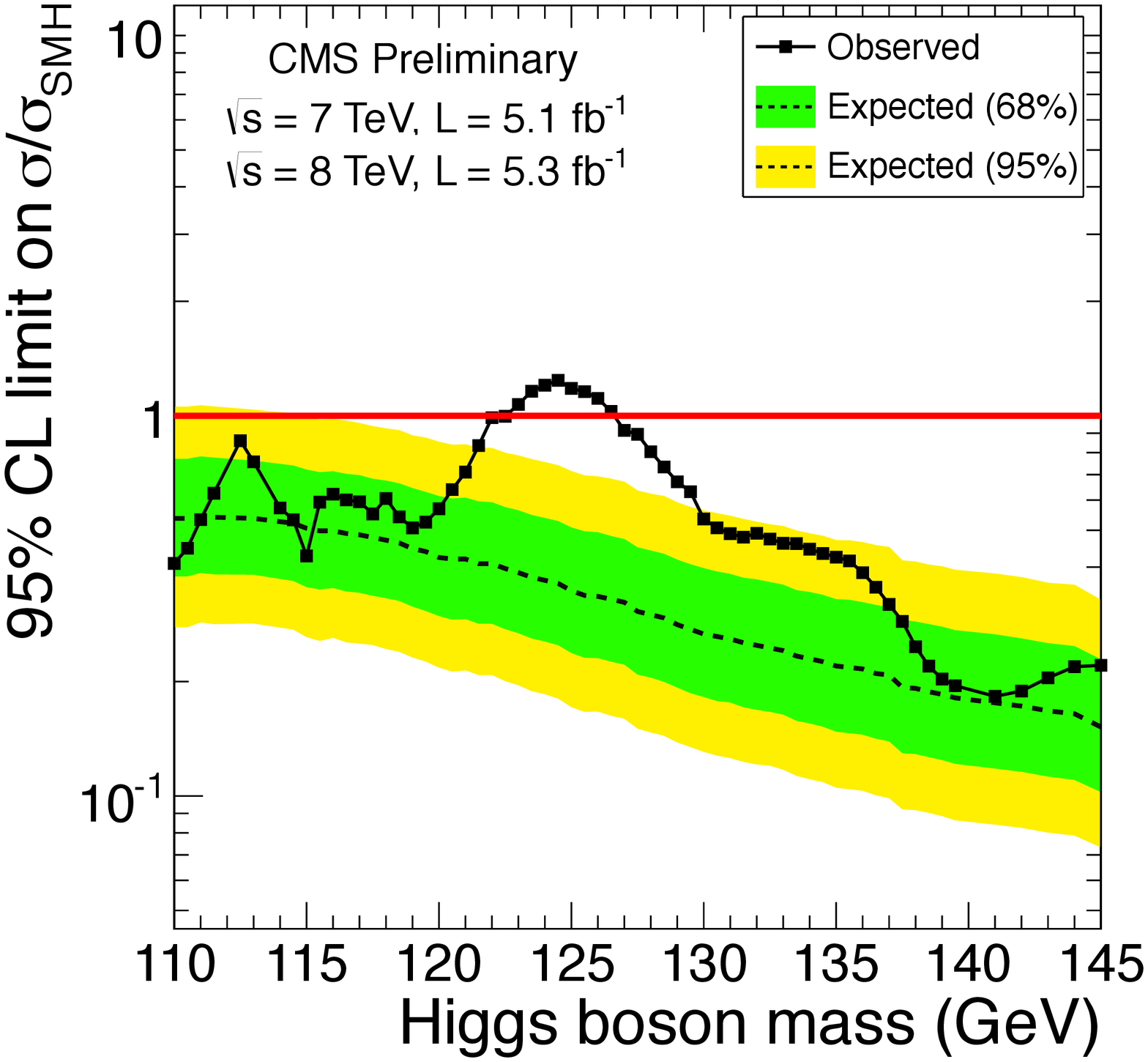}}
\end{minipage}
\end{tabular}
\caption[]{ATLAS and CMS results for the 95\% confidence upper limit on the
  signal strength as a function of $m_H$(full mass range and low mass
  range only, respectively). The black solid curve indicates the
  observed limit and the black dashed curve illustrates the median
  expected limit in the absence of a signal together with the one
  standard deviation (green) and two standard deviation (yellow)
  bands. From
  Refs.~\cite{:2012gk,:2012gu}.\label{fig:sm_higgs_discovery_atlas_excess}}
\end{figure}

The dominant systematic uncertainties are those on the measurement of
the integrated luminosity and on the theoretical predictions of the
signal production cross sections and decay branching ratios, as well
as those related to detector response that impact the reconstruction
analyses in various reconstructing procedures.  More details on both
the uncertainties from the measure of the integrated luminosity and
the detector response can be find in the ATLAS and CMS
papers~\cite{Chatrchyan:2012tx,ATLAS:2012ae} as well as in the
discovery papers~\cite{:2012gk,:2012gu}. The degree of accuracy
reached in the theoretical predictions of both production cross
sections and branching ratios will be discussed in
Sec.~\ref{sec:theory}.  Fig.~\ref{fig:sm_higgs_discovery_atlas_excess}
shows the 95\% CL upper limits on the signal strength. The various
curves and bands have the same meaning as in
Fig.~\ref{fig:sm_higgs_tev_comb_LLR}, as reminded in the figure
caption. From our discussion of Fig.~\ref{fig:sm_higgs_tev_comb_LLR}
we can clearly see that the observed cross section exceeds the
expected background well beyond the two standard-deviation level, in a
region where the expected background is determined with enough
sensitivity to test the SM-Higgs boson hypothesis.

This amazing result brought us by the LHC experiments opens a new
chapter in the study of EWSB. The newly discovered spin-0 particles
will have to be studied in all its properties, in particular its
couplings to gauge bosons and fermions to disentangle possible hints
of new physics through subtle deviations from the SM-Higgs boson
pattern. For instance, it will important to test deviations that may
discriminate the supersymmetric nature of the discovered spin-0 particle.
With this respect, all production and decay modes will have to be
used in order to control the largest possible number of couplings and
determine them from multiple sources. Cleaver strategies and accurate
knowledge of both production cross sections and branching ratios will
be crucial to a successful implementation of the Higgs-boson
physics program in the future of the LHC experiments.

\section{Theoretical predictions for SM Higgs production at hadron
  colliders}
\label{sec:theory}

Given the elusive nature of a Higgs-like signal, a precise theoretical
prediction of both signal and background total cross-sections and
distributions is vital to the success of the LHC program. For this
reason, all Higgs production channels have received a lot of attention
in recent years and they are nowadays calculated including
Next-to-Leading-Order (NLO) and sometimes Next-to-Next-to-Lading Order
(NNLO) QCD corrections, as well as, in some cases, the first order of
electroweak (EW) corrections. More recently, the attention has been
shifting on controlling background processes at the same level of
accuracy. Thanks to the enormous progress in higher-order perturbative
calculations during the last few years, several processes involving
multiple jets, multiple gauge bosons, as well as several massive
fermions, are now predicted at NLO in QCD. For a thorough introduction
to higher-order calculation in quantum field theory and their recent
applications I refer to John Campbell's lectures at this
school~\cite{campbell:tasi11}.  In this context, another crucial
progress of the last decade has been the development of a consistent
interface between NLO parton level calculations and Parton Shower (PS)
Monte Carlo generators, the tools commonly used in experimental
analyses to model the evolution of high energy hadronic collisions. We
nowadays have some main frameworks, namely
MC@NLO~\cite{Frixione:2002ik},
POWHEG~\cite{Nason:2004rx,Frixione:2007vw}, and
SHERPA~\cite{Hoche:2010kg}, within which a parton level calculation
can be consistently matched to the process of radiation emission
implemented in PS Monte Carlo programs like
PYTHIA~\cite{Sjostrand:2006za} and HERWIG~\cite{Corcella:2000bw},
including the first order of QCD corrections. This allows for more
reliable comparisons with data both in terms of kinematic
distributions and overall cross sections. In particular, the impact of
QCD corrections in the presence of kinematic cuts and vetos on
specific final state particles and/or decay products can be studied
more accurately.

\begin{figure}
\begin{center}
\begin{tabular}{cc}
\begin{minipage}{0.5\linewidth}
\hspace{-1.truecm}
{\includegraphics[scale=0.35]{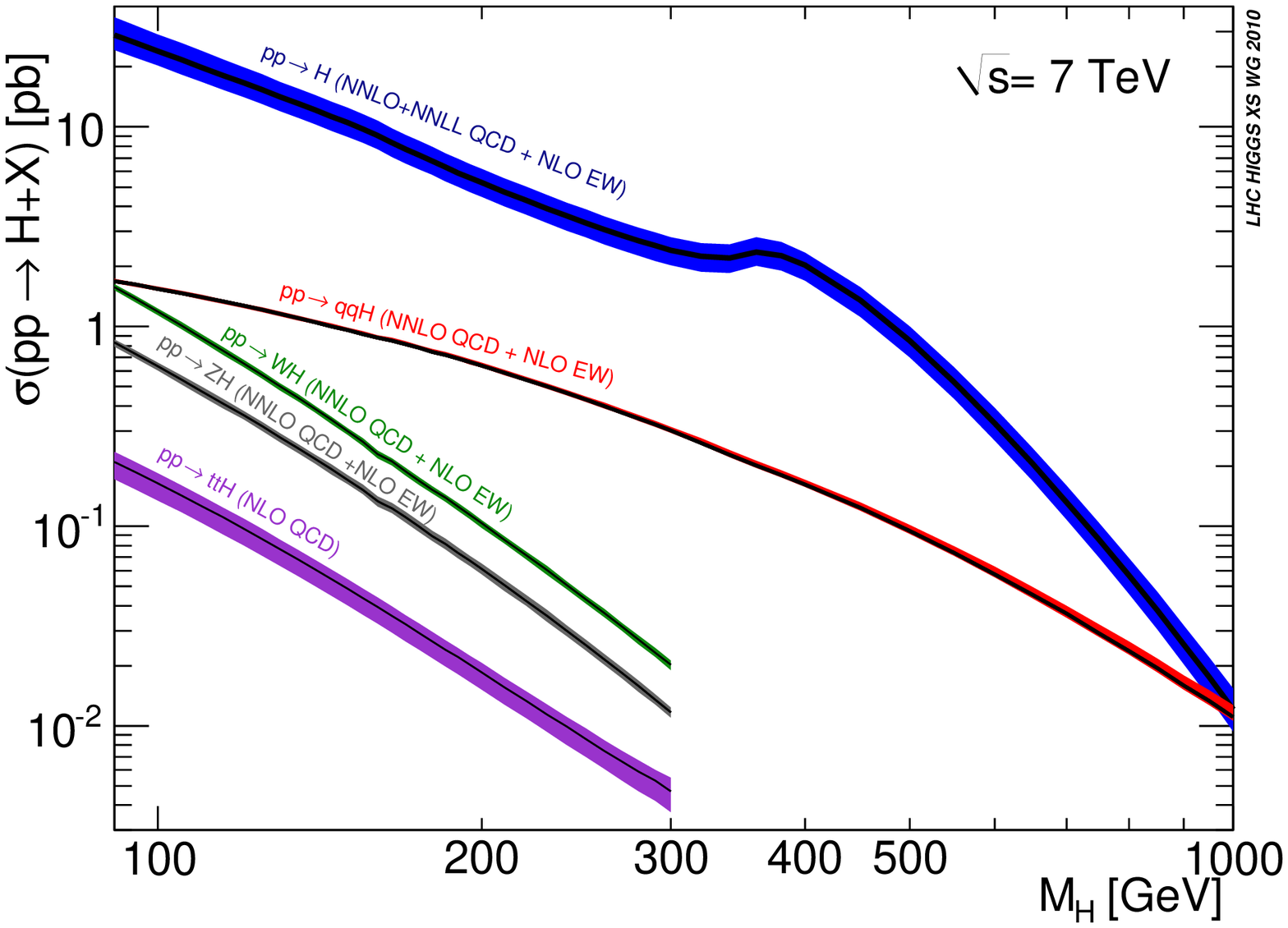}}
\end{minipage}&
\hspace{-1.2truecm}
\begin{minipage}{0.5\linewidth}
{\includegraphics[scale=0.35]{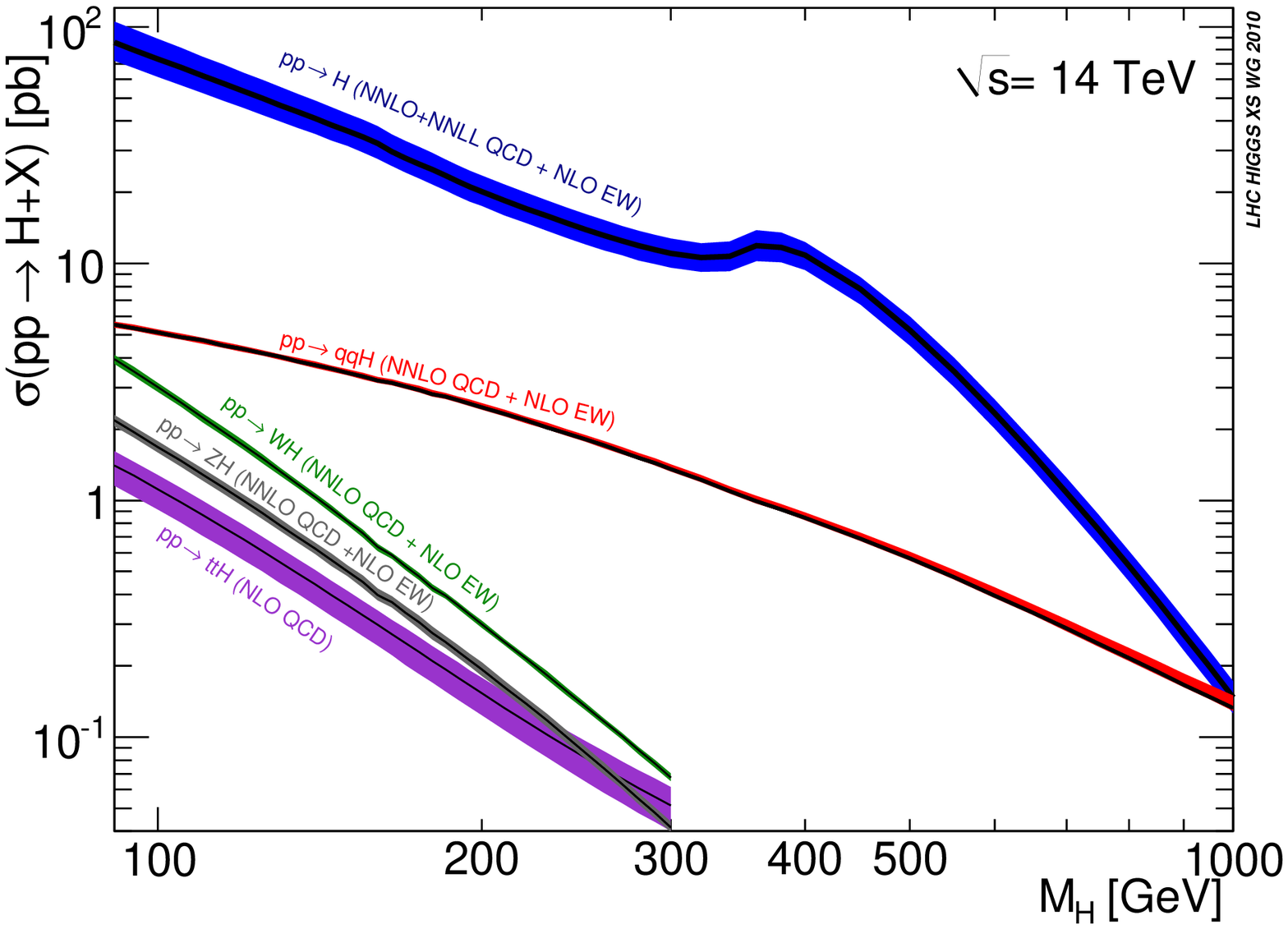}}
\end{minipage}
\end{tabular}
\end{center}
\caption[]{SM-Higgs production cross sections, including parametric and
  systematic theoretical errors, for the LHC at center of mass energies
  of 7~TeV (left) and 14~TeV (right) respectively. From
  Ref.~\cite{Dittmaier:2011ti}.\label{fig:sm_higgs_xsec_bands_yr1}}
\end{figure}

The main references for all SM Higgs-boson production channels are
collected in Table~\ref{tab:nlo_nnlo} for the reader's
convenience. They correspond to the original parton-level NLO/NNLO
calculations, while we refer
to~\cite{Dittmaier:2011ti,Dittmaier:2012vm} for further developments,
including comparison between different calculations as well as results
for the NLO interface with parton-shower Monte Carlo programs.
\begin{table}
\caption{Existing QCD corrections for various SM Higgs production
processes. \label{tab:nlo_nnlo}}
\begin{center}
\begin{tabular}{c|p{10cm}}
process & $\sigma_{NLO,NNLO}$ by\\
\hline
$gg\rightarrow H$ & 
\begin{minipage}{1.\linewidth}{
\scriptsize
\vspace{0.3truecm}
\begin{list}{}
  {\setlength{\topsep}{0.5truecm}\setlength{\parskip}{-0.8truecm}
    \setlength{\itemsep}{0.truecm}
    \setlength{\leftmargin}{0.0truecm}\setlength{\labelwidth}{1.5truecm}}
\item S.Dawson, NPB 359 (1991), 
      A.Djouadi, M.Spira, P.Zerwas, PLB 264 (1991)
\item C.J.Glosser \textit{et al.}, JHEP 0212 (2002); 
      V.Ravindran \textit{et al.}, NPB 634 (2002)
\item D. de Florian \textit{et al.}, PRL 82 (1999)
\item R.Harlander, W.Kilgore, PRL 88 (2002) (NNLO)
\item C.Anastasiou, K.Melnikov, NPB 646 (2002) (NNLO)
\item V.Ravindran \textit{et al.}, NPB 665 (2003) (NNLO)
\item S.Catani \textit{et al.} JHEP 0307 (2003) (NNLL),
\item G.Bozzi \textit{et al.}, PLB 564 (2003), NPB 737 (2006) (NNLL)
\item C.Anastasiou, R.Boughezal, F.Petriello, JHEP (2008) (QCD+EW)
\end{list}}
\end{minipage}\\\\
\hline
$q\bar{q}\rightarrow (W,Z)H$ & 
\begin{minipage}{1.\linewidth}{
\scriptsize
\vspace{0.3truecm}
\begin{list}{}
  {\setlength{\topsep}{0.5truecm}\setlength{\parskip}{-0.8truecm}
    \setlength{\itemsep}{0.truecm}
    \setlength{\leftmargin}{0.0truecm}\setlength{\labelwidth}{1.5truecm}}
\item T.Han, S.Willenbrock, PLB 273 (1991)
\item M.L.Ciccolini, S.Dittmaier, and M.Kr\"amer (2003)  (EW)
\item O.Brien, A.Djouadi, R.Harlander, PLB 579 (2004) (NNLO)
\end{list}}
\end{minipage}\\\\
\hline
$q\bar{q}\rightarrow q\bar{q}H$ & 
\begin{minipage}{1.\linewidth}{
\scriptsize
\vspace{0.3truecm}
\begin{list}{}
  {\setlength{\topsep}{0.5truecm}\setlength{\parskip}{-0.8truecm}
    \setlength{\itemsep}{0.truecm}
    \setlength{\leftmargin}{0.0truecm}\setlength{\labelwidth}{1.5truecm}}
\item T.Han, G.Valencia, S.Willenbrock, PRL 69 (1992)
\item T.Figy, C.Oleari, D.Zeppenfeld, PRD 68 (2003)
\item M.L.Ciccolini, A.Denner,S.Dittmaier (2008) (QCD+EW)
\item P.Bolzoni, F.Maltoni, S.O.Moch, and M.Zaro (2010) (NNLO)
\end{list}}
\end{minipage}\\\\
\hline
$q\bar{q},gg\rightarrow t\bar{t}H$ & 
\begin{minipage}{1.\linewidth}{
\scriptsize
\vspace{0.3truecm}
\begin{list}{}
  {\setlength{\topsep}{0.5truecm}\setlength{\parskip}{-0.8truecm}
    \setlength{\itemsep}{0.truecm}
    \setlength{\leftmargin}{0.0truecm}\setlength{\labelwidth}{1.5truecm}}
\item W.Beenakker \textit{et al.}, PRL 87 (2001), NPB 653 (2003)
\item S.Dawson \textit{et al.}, PRL 87 (2001), PRD 65 (2002), PRD 67,68 (2003)
\end{list}}
\end{minipage}\\\\
\hline
$q\bar{q},gg\rightarrow b\bar{b}H$ & 
\begin{minipage}{1.\linewidth}{
\scriptsize
\vspace{0.3truecm}
\begin{list}{}
  {\setlength{\topsep}{0.5truecm}\setlength{\parskip}{-0.8truecm}
    \setlength{\itemsep}{0.truecm}
    \setlength{\leftmargin}{0.0truecm}\setlength{\labelwidth}{1.5truecm}}
\item S.Dittmaier, M.Kr\"amer, M.Spira, PRD 70 (2004)
\item S.Dawson \textit{et al.}, PRD 69 (2004), PRL 94 (2005)
\end{list}}
\end{minipage}\\\\
\hline
\begin{minipage}{0.2\linewidth}{
\vspace{0.3truecm}
$gb(\bar{b})\rightarrow b(\bar{b})H$}
\end{minipage}
 & 
\begin{minipage}{1.\linewidth}{
\scriptsize
\vspace{0.3truecm}
J.Campbell \textit{et al.}, PRD 67 (2003)}
\end{minipage}\\\\
\hline
$b\bar{b}\rightarrow H$ & 
\begin{minipage}{1.\linewidth}{
\scriptsize
\vspace{0.3truecm}
\begin{list}{}
  {\setlength{\topsep}{0.5truecm}\setlength{\parskip}{-0.8truecm}
    \setlength{\itemsep}{0.truecm}
    \setlength{\leftmargin}{0.0truecm}\setlength{\labelwidth}{1.5truecm}}
\item D.A.Dicus \textit{et al.} PRD 59 (1999); 
      C.Balasz \textit{et al.}, PRD 60 (1999).
\item R.Harlander, W.Kilgore, PRD 68 (2003) (NNLO)
\end{list}}
\end{minipage}\\\\
\hline
\end{tabular}
\end{center}
\end{table}
These calculations and their developments have been the official
reference for Higgs searches at the Tevatron and the LHC. In
particular they have been at the core of an extended program of
providing consistent state-of-the-art theoretical predictions for
Higgs-boson production during the different phases of the LHC (with
center-of-mass energies 7~TeV, 8~TeV, and 14~TeV respectively),
summarized in the work of the LHC Higgs Cross Section Working Group
(LHC-HXSWG)~\cite{Dittmaier:2011ti,Dittmaier:2012vm}. In this context,
common prescriptions to estimate the uncertainties of theoretical
predictions deriving from input parameters, parton distribution
functions, $\alpha_s$, and residual unknown perturbative orders have
been discussed and applied consistently to all the SM (and MSSM)
Higgs-production channels, providing a common background to the LHC
experiments for their analyses. Pictorially this is illustrated in
Fig.~\ref{fig:sm_higgs_xsec_bands_yr1} where the SM Higgs-boson
inclusive cross sections are plotted as functions of its mass $m_H$
over the entire mass range accessible at the LHC, with
center-of-mass-energies 7~TeV (left) and 14~TeV (right) respectively.
The theoretical and parametric uncertainties are included and result
in the color bands around each central curve.  The order of
perturbative corrections included is indicated along each curve.
In its second round of activity ~\cite{Dittmaier:2012vm}, the LHC-HXSWG
focused on investigating the shape of exclusive observables when
higher-order corrections are included and more exclusive cuts,
dictated by experimental analyses, are applied. At the same time, the
interface of most NLO calculation with PS have been implemented and
the most important available background processes have been studied
following the same criteria used for the signal processes.
Considerable more progress in this direction is expected in the months
to come and will be at the core of the future LHC-HXSWG activity.
Given the recent discovery, the focus is expected to shift towards
investigating the properties of the newly discovered particle using
the most sophisticated available tools.

In the following I would like to discuss the relevance of including
different layers of QCD corrections in the calculation of Higgs-boson
production cross sections and  illustrate it with a prototype
example, i.e. the case of gluon-gluon fusion.

\subsection{$gg\rightarrow H$ at NNLO: a prototype example}
\label{subsec:gg_to_H_nnlo}

The gluon-fusion process offers a true learning ground to understand
the complexity of hadronic cross sections. We can learn about the need
of improving the theoretical predictions beyond the LO and even the NLO,
the importance of resumming sets of large corrections at all orders,
the subtleties of interfacing the NNLO calculation with a PS Monte Carlo.
\begin{figure}
\centering
\includegraphics[scale=0.7]{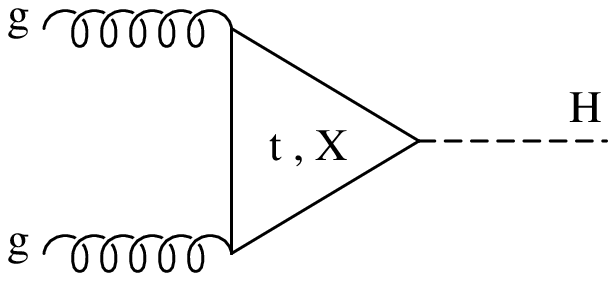}
\caption[]{The $gg\rightarrow H$ production process at lowest order.\label{fig:gg_to_H}}
\end{figure}

Most of the basic ideas that motivate the techniques used in the NNLO
calculation of the cross section for the $gg\rightarrow H$ production
process have been already introduced in
Section~\ref{subsubsec:sm_higgs_loop_decay}, where we discussed the
$H\rightarrow gg$ loop-induced decay.  In particular we know that in
the SM, the main contribution to $gg\rightarrow H$ comes form the
top-quark loop (see Fig.~\ref{fig:gg_to_H}) since:
\begin{equation}
\label{eq:gg_to_H_lo}
\sigma_{LO}=
\frac{G_F\alpha_s(\mu)^2}{288\sqrt{2}\pi}
\left|\sum_q A_q^H(\tau_q)\right|^2\,\,\,,
\end{equation}
where $\tau_q=4m_q^2/M_H^2$ and $A_q^H(\tau_q)\le 1$ with   
$A_q^H(\tau_q)\rightarrow 1$ for $\tau_q\rightarrow\infty$. 
\begin{figure}
\hspace{1.truecm}
\begin{minipage}{0.4\linewidth}
\centering
\includegraphics[scale=0.28]{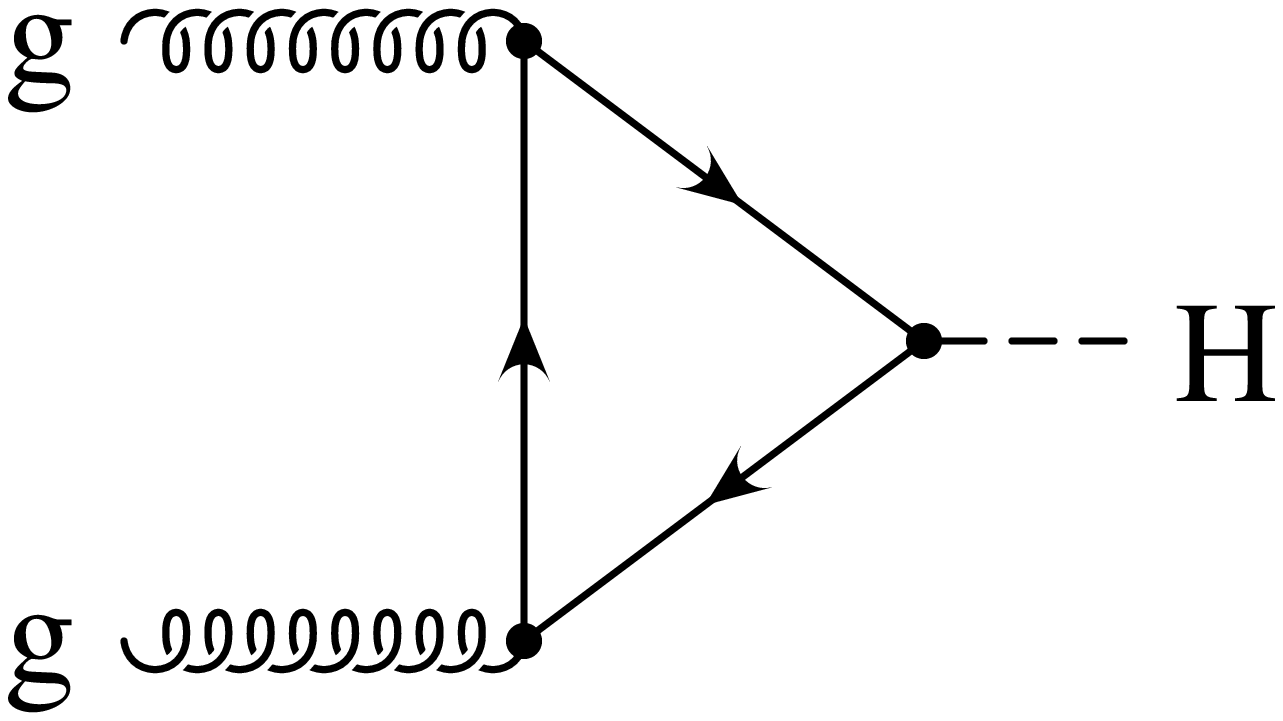}
\end{minipage}
$\longrightarrow$
\begin{minipage}{0.4\linewidth}
\centering
\includegraphics[scale=0.28]{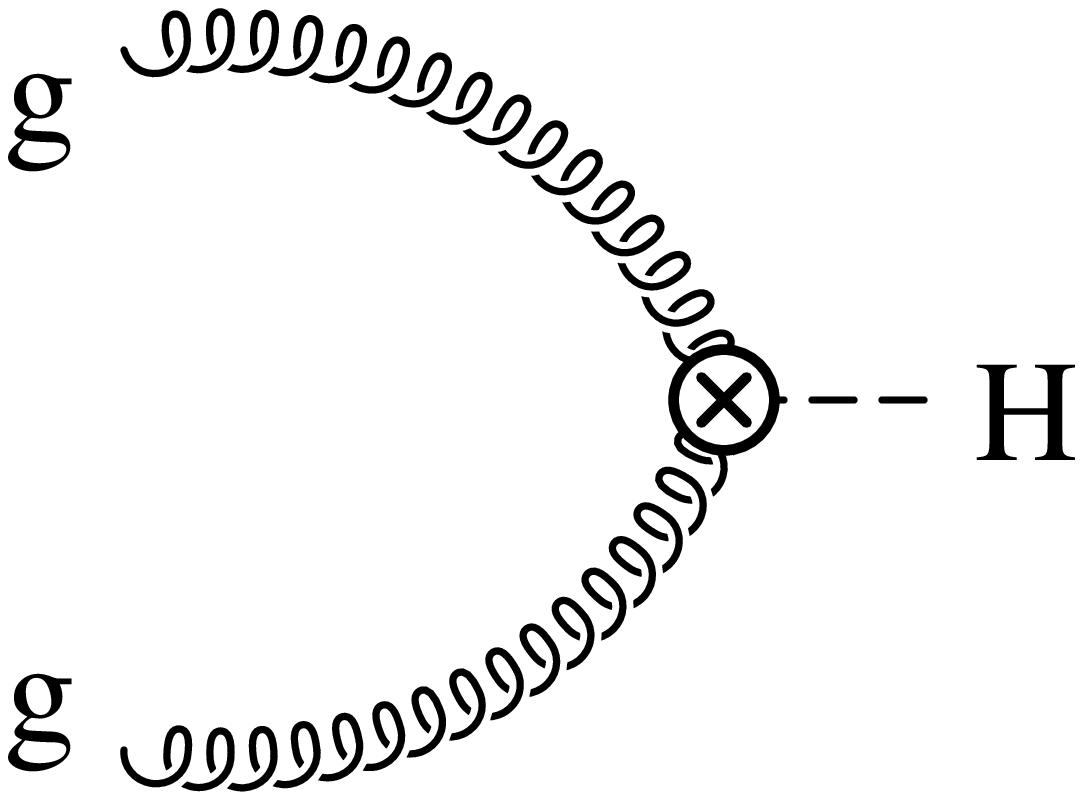}
\end{minipage}
\caption[]{The top-quark loop contribution to $gg\rightarrow H$ gives origin 
to a $ggH$ effective vertex in the $m_t\rightarrow\infty$ limit. 
\label{fig:gg_to_H_effective_vertex}}
\end{figure}

As we saw in Section~\ref{subsubsec:sm_higgs_loop_decay}, one can 
work in the infinite top-quark mass 
limit and reduce the one-loop $Hgg$ vertex to a tree level effective 
vertex, derived from an effective Lagrangian of the form:
\begin{equation}
\label{eq:gg_to_H_effective_lagrangian}
\mathcal{L}_{eff}=
\frac{H}{4v}C(\alpha_s)G^{a\mu\nu}G^a_{\mu\nu}\,\,\,,
\end{equation}
where the coefficient $C(\alpha_s)$, including NLO and NNLO QCD 
corrections, can be written as:
\begin{equation}
\label{eq:gg_to_H_wilson_coefficient}
C(\alpha_s)=\frac{1}{3}\frac{\alpha_s}{\pi}\left[1+c_1\frac{\alpha_s}{\pi}+
c_2\left(\frac{\alpha_s}{\pi}\right)^2+\cdots\right]\,\,\,.
\end{equation}
NLO and NNLO QCD corrections to $gg\rightarrow H$ can then be
calculated as corrections to the effective $Hgg$ vertex, and the complexity
of the calculation is reduced by one order of loops. 

\begin{figure}
\centering
\includegraphics[scale=0.45]{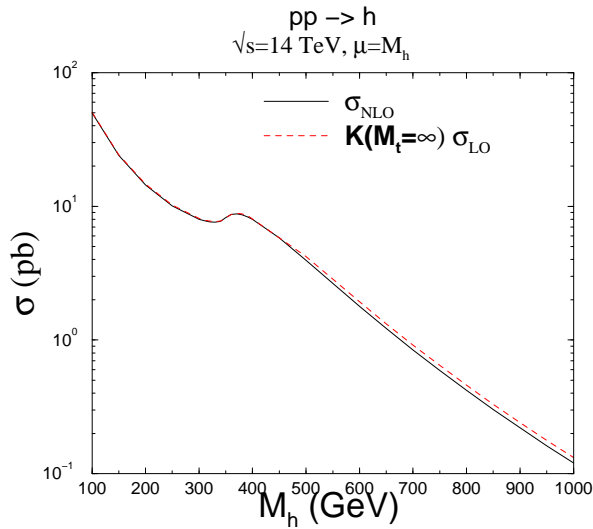}
\caption[]{The NLO cross section for 
  $gg\rightarrow H$ as a function of $M_H$. The two curves represent
  the results of the exact calculation (solid) and of the infinite
  top-quark mass limit calculation (dashed), where the NLO cross
  section has been obtained as the product of the $K$-factor
  ($K\!=\!\sigma_{NLO}/\sigma_{LO}$) calculated in the
  $m_t\!\rightarrow\!\infty$ limit times the LO cross section. From
  Ref.~\cite{Dawson:1998yi}.  \label{fig:gg_to_H_Kfactor}}
\end{figure}

The NLO order of QCD corrections has actually been calculated both
with and without taking the infinite top-quark mass limit. The
comparison between the exact and approximate calculation shows an
impressive agreement at the level of the total cross section, and, in
particular, at the level of the $K$-factor, i.e. the ratio between NLO
and LO total cross sections ($K\!=\!\sigma_{NLO}/\sigma_{LO}$), as
illustrated in Fig.~\ref{fig:gg_to_H_Kfactor}.
It is indeed expected that methods like the infinite top quark mass
limit may not reproduce the correct kinematic distributions of a given
process at higher order in QCD, but are very reliable at the level of
the total cross section, in particular when the cross section receives
large momentum independent contribution at the first order of QCD
corrections. As for the $H\rightarrow gg$ decay process, the NLO
corrections to $gg\rightarrow H$ are very large, changing the LO cross
section by more than 50\%. Since the $gg\rightarrow H$ is the leading
Higgs-boson production mode at hadron colliders, it has been clear for
quite a while that a NNLO calculation was needed in order to
understand the behavior of the perturbatively calculated cross
section, and if possible, in order to stabilize its theoretical
prediction.

\begin{figure}
\hspace{2.truecm}
\begin{minipage}{0.3\linewidth}{
\includegraphics[scale=0.28]{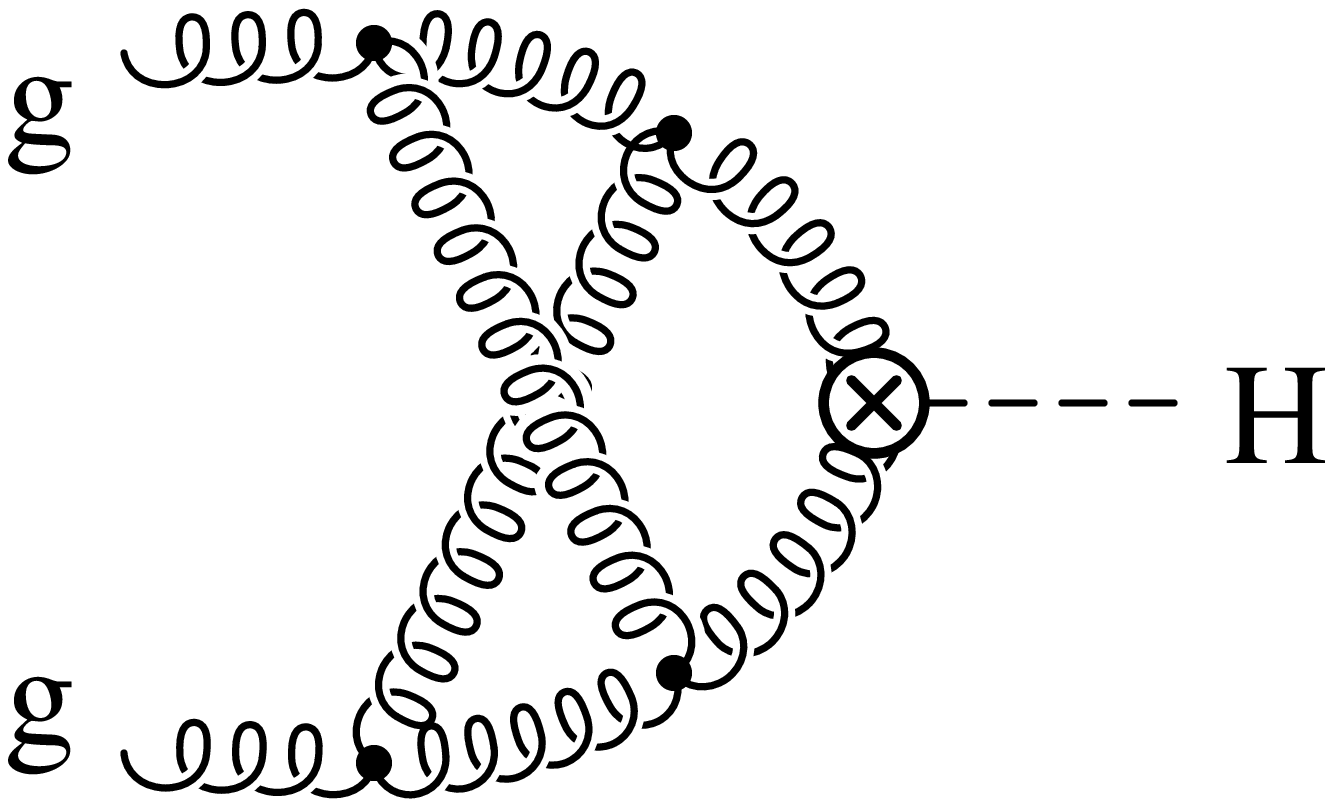}}
\end{minipage}
\hspace{0.5truecm}
\begin{minipage}{0.3\linewidth}{
\includegraphics[scale=0.28]{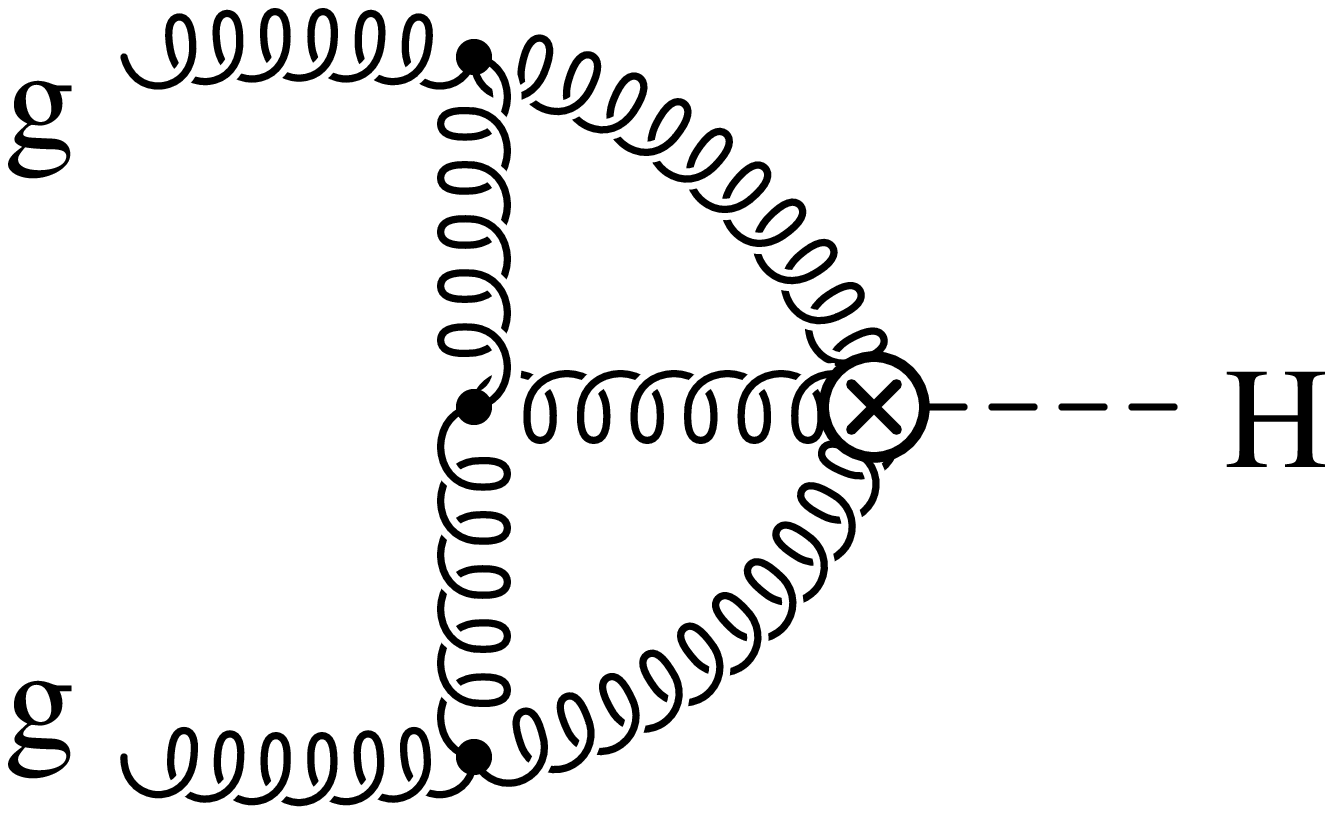}}
\end{minipage}
\caption[]{Two-loop diagrams that enter the NNLO QCD corrections to
  $gg\rightarrow H$. \label{fig:gg_to_H_nnlo}}
\end{figure}

The NNLO corrections to the total cross section have been calculated
using the infinite top-quark mass limit (see
Table~\ref{tab:nlo_nnlo}). The calculation of the NNLO QCD corrections
involves then 2-loop diagrams like the ones shown in
Fig.~\ref{fig:gg_to_H_nnlo}, instead of the original 3-loop diagrams
(A quite formidable task!).
\begin{figure}
\centering
\includegraphics[scale=0.6]{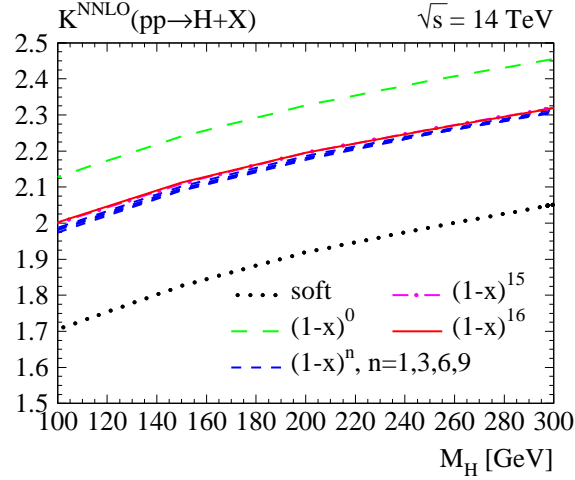}
\caption[]{$K$-factor for $gg\rightarrow H$ at the LHC
  ($\sqrt{s}\!=\!14$~TeV), calculated adding progressively more terms
  in the expansion of Eq.~(\ref{eq:gg_to_H_sigma_hat_ij_nth_term}).
  From Harlander and Kilgore as given in Table~\ref{tab:nlo_nnlo}.
  \label{fig:gg_to_H_soft_expansion_convergence}}
\end{figure}
Moreover, thanks to the $2\rightarrow 1$ kinematic of the
$gg\rightarrow H$ process, the cross section has in one case be calculated in
the so called \emph{soft limit}, i.e. as an expansion in the parameter
$x\!=\!M_H^2/\hat{s}$ about $x\!=\!1$, where $\hat{s}$ is the partonic
center of mass energy (see paper by Harlander and Kilgore in 
Table~\ref{tab:nlo_nnlo}). The $n$-th term in the expansion of 
the partonic cross section $\hat\sigma_{ij}$,
\begin{equation}
\label{eq:gg_to_H_sigma_hat_ij_exp}
\hat{\sigma}_{ij}=\sum_{n\ge 0}\left(\frac{\alpha_s}{\pi}\right)^n
\hat{\sigma}^{(n)}_{ij}\,\,\,,
\end{equation}
can then be written in the soft limit ($x\rightarrow 1$) as follows:
\begin{equation}
\label{eq:gg_to_H_sigma_hat_ij_nth_term}
\hat{\sigma}^{(n)}_{ij}= \underbrace{
  a^{(n)}\delta(1-x)+\sum_{k=0}^{2n-1}b_k^{(n)}
  \left[\frac{\ln^k(1-x)}{1-x}\right]_+}_{\mbox{purely soft terms}} +
\underbrace{
\sum_{l=0}^{\infty}\sum_{k=0}^{2n-1}c_{lk}^{(n)}(1-x)^l\ln^k(1-x)}_
{\mbox{collinear+hard terms}}
\end{equation}
where we have made explicit the origin of different terms in the
expansion. The NNLO cross section is then obtained by calculating the
coefficients $a^{(2)}$, $b^{(2)}_k$ , and $c^{(2)}_{lk}$,
 for $l\ge 0$ and $k\!=\!0,\ldots,3$.
\begin{figure}
\centering
\includegraphics[scale=0.6]{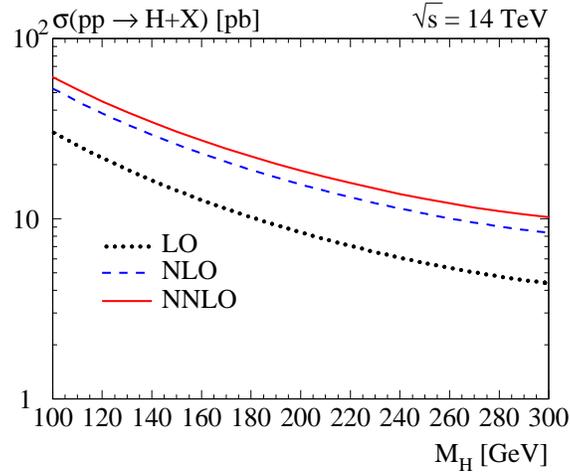}
\caption[]{Cross section for $gg\rightarrow H$ at the LHC 
($\sqrt{s}\!=\!14$~TeV), calculated at LO, NLO and NNLO of
  QCD corrections, as a function of $M_H$, for
  $\mu_F\!=\!\mu_R\!=\!M_H/4$. From Harlander and Kilgore in
  Table~\ref{tab:nlo_nnlo}. \label{fig:gg_to_H_nnlo_nlo_lo}}
\end{figure}
In Fig.~\ref{fig:gg_to_H_soft_expansion_convergence} we see
the convergence behavior of the expansion in
Eq.~(\ref{eq:gg_to_H_sigma_hat_ij_nth_term}). Just adding the first few
terms provides a remarkably stable $K$-factor. The results shown in
Fig.~\ref{fig:gg_to_H_soft_expansion_convergence} have been indeed
confirmed by a full calculation~\cite{Anastasiou:2002yz}, 
where no soft approximation has been used. 

\begin{figure}
\hspace{-0.4truecm}
\begin{tabular}{cc}
\begin{minipage}{0.44\linewidth}{
\includegraphics[bb=100 265 470 565,scale=0.5]{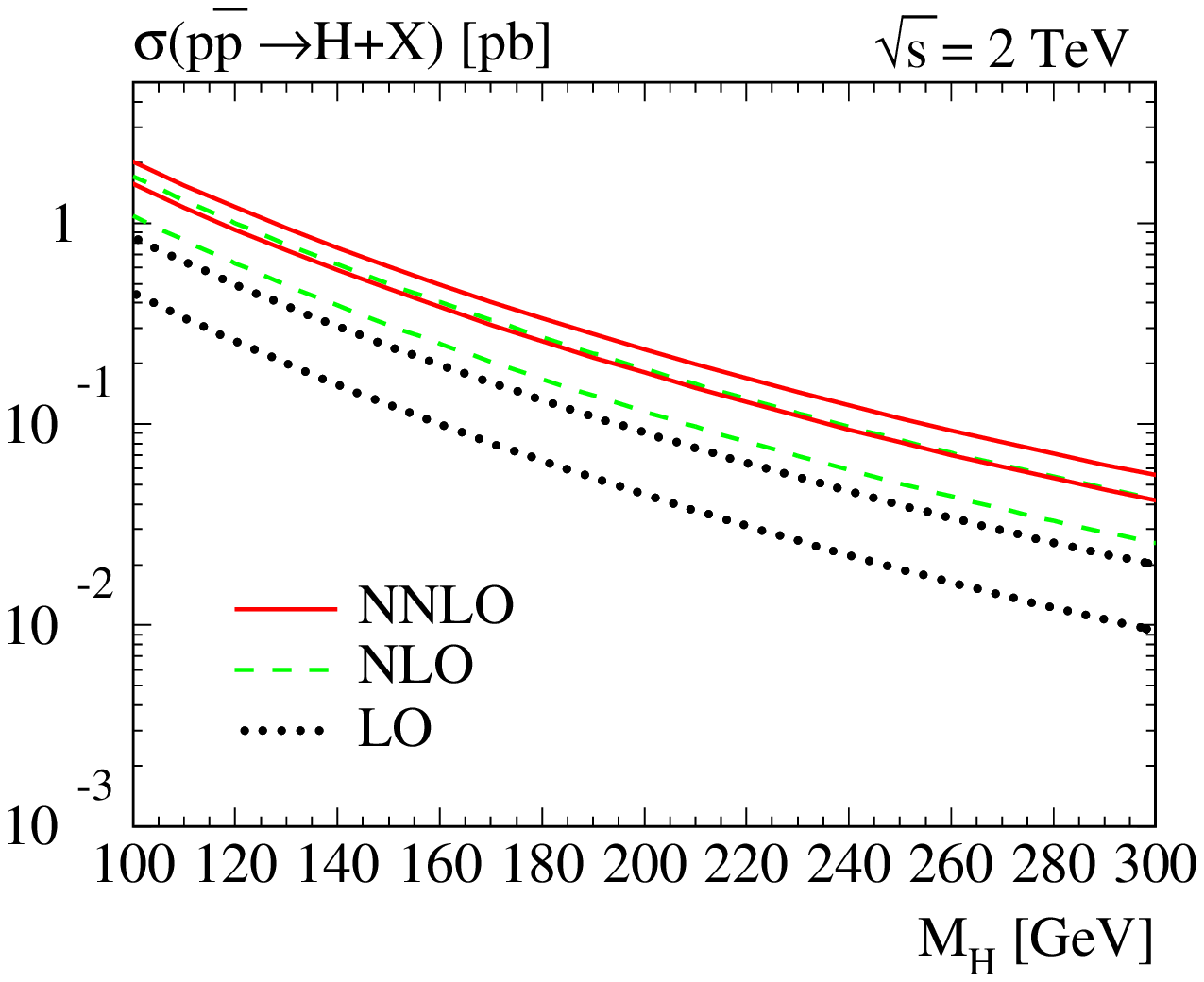}}
\end{minipage}&
\hspace{0.4truecm}
\begin{minipage}{0.44\linewidth}{
\includegraphics[bb=100 265 470 565,scale=0.5]{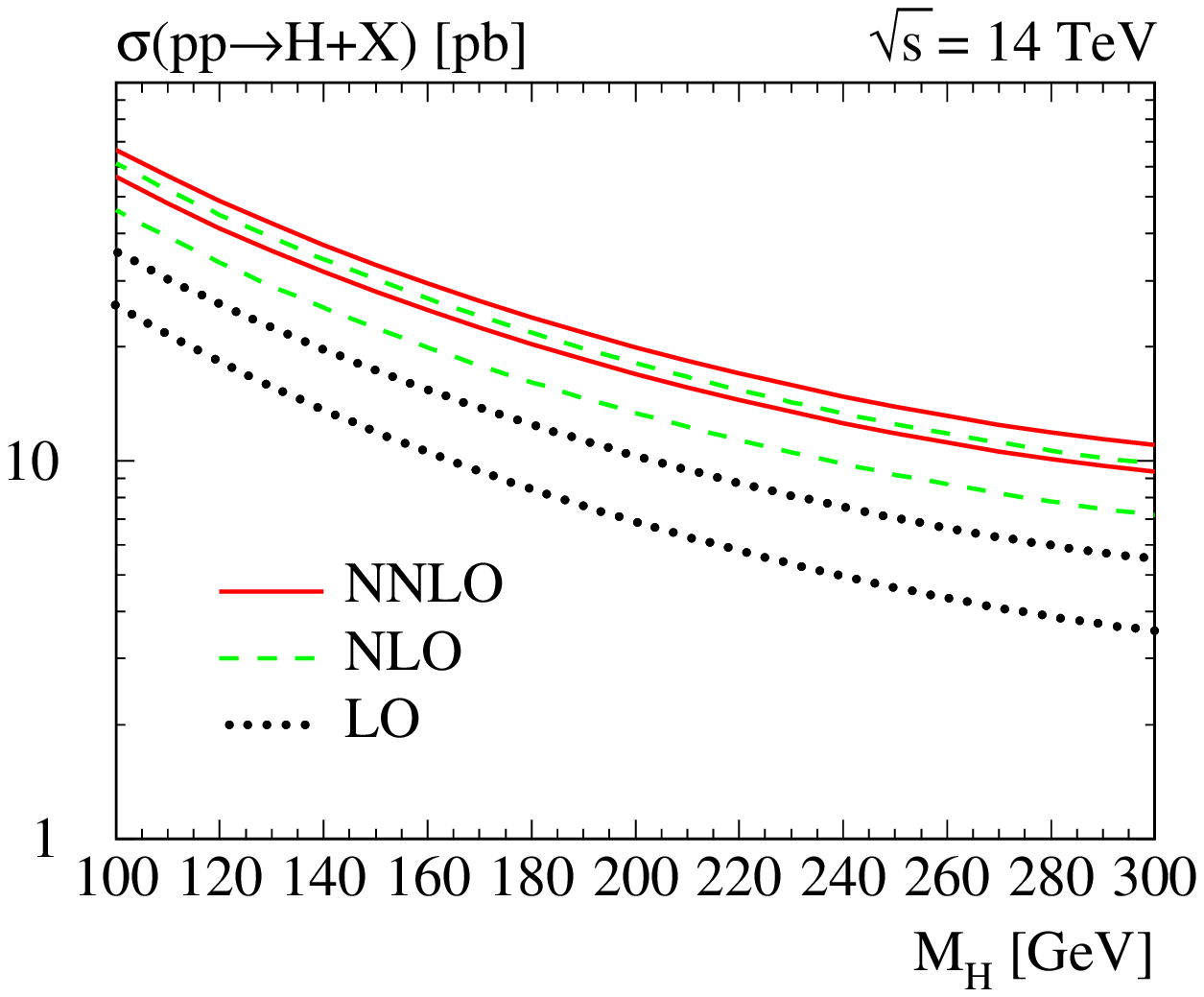}}
\end{minipage}
\end{tabular}
\caption[]{Residual renormalization/factorization scale dependence of
  the LO, NLO, and NNLO cross section for $gg\rightarrow H$, at the
  Tevatron ($\sqrt{s}\!=\!2$~TeV) and at the LHC
  ($\sqrt{s}\!=\!14$~TeV), as a function of $M_H$. The bands are
  obtained by varying $\mu_R\!=\!\mu_F$ by a factor of 2 about the
  central value $\mu_F\!=\!\mu_R\!=\!M_H/4$. 
 From Ref.~\cite{Harlander:2002wh}. \label{fig:gg_to_H_nnlo_nlo_lo_mu_dep}}
\end{figure}

\begin{figure}
\begin{tabular}{lr}
\hspace{-0.5truecm}
\begin{minipage}{0.5\linewidth}{
\includegraphics[scale=0.37]{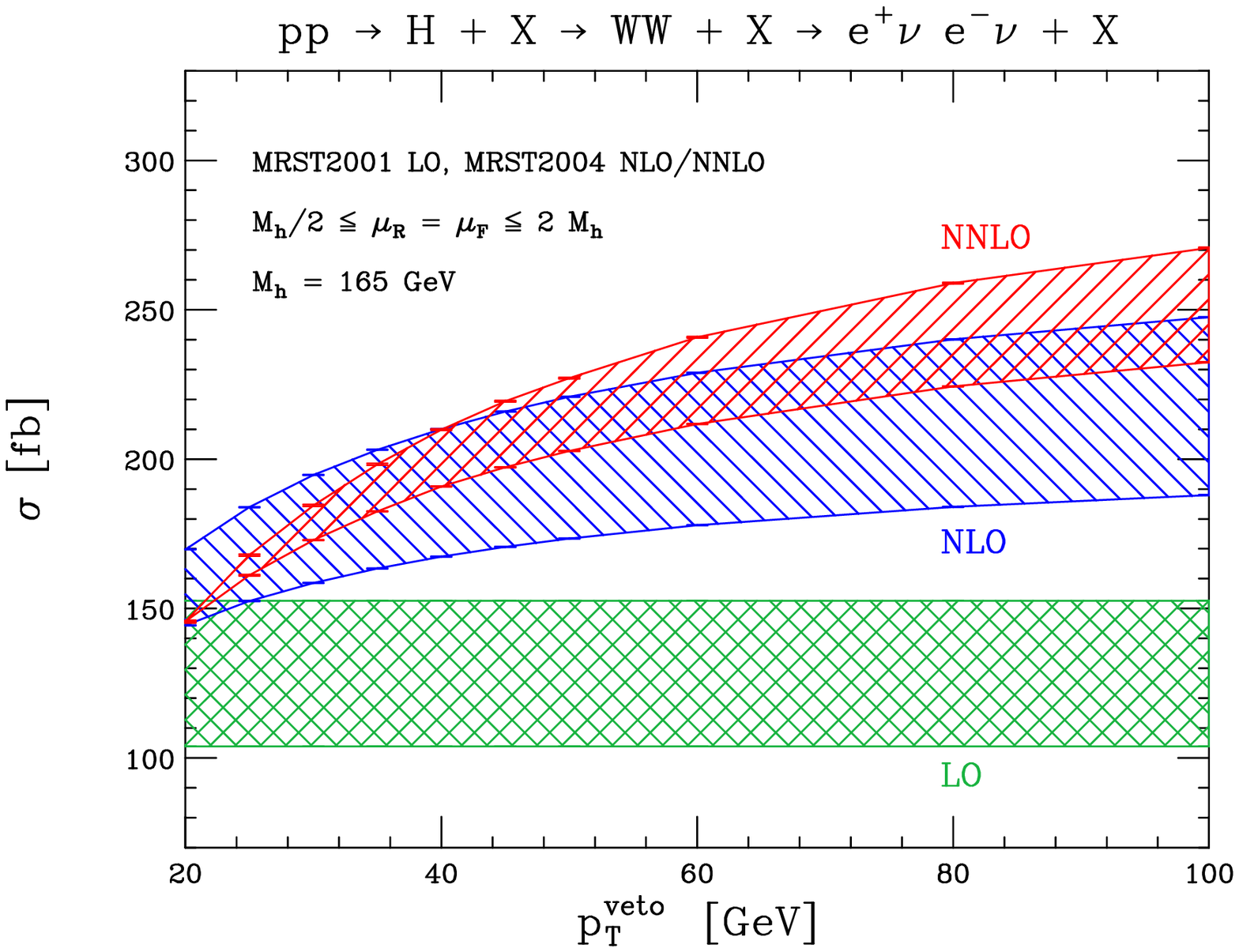}}
\end{minipage} &
\hspace{-0.5truecm}
\begin{minipage}{0.5\linewidth}{
\includegraphics[scale=0.37]{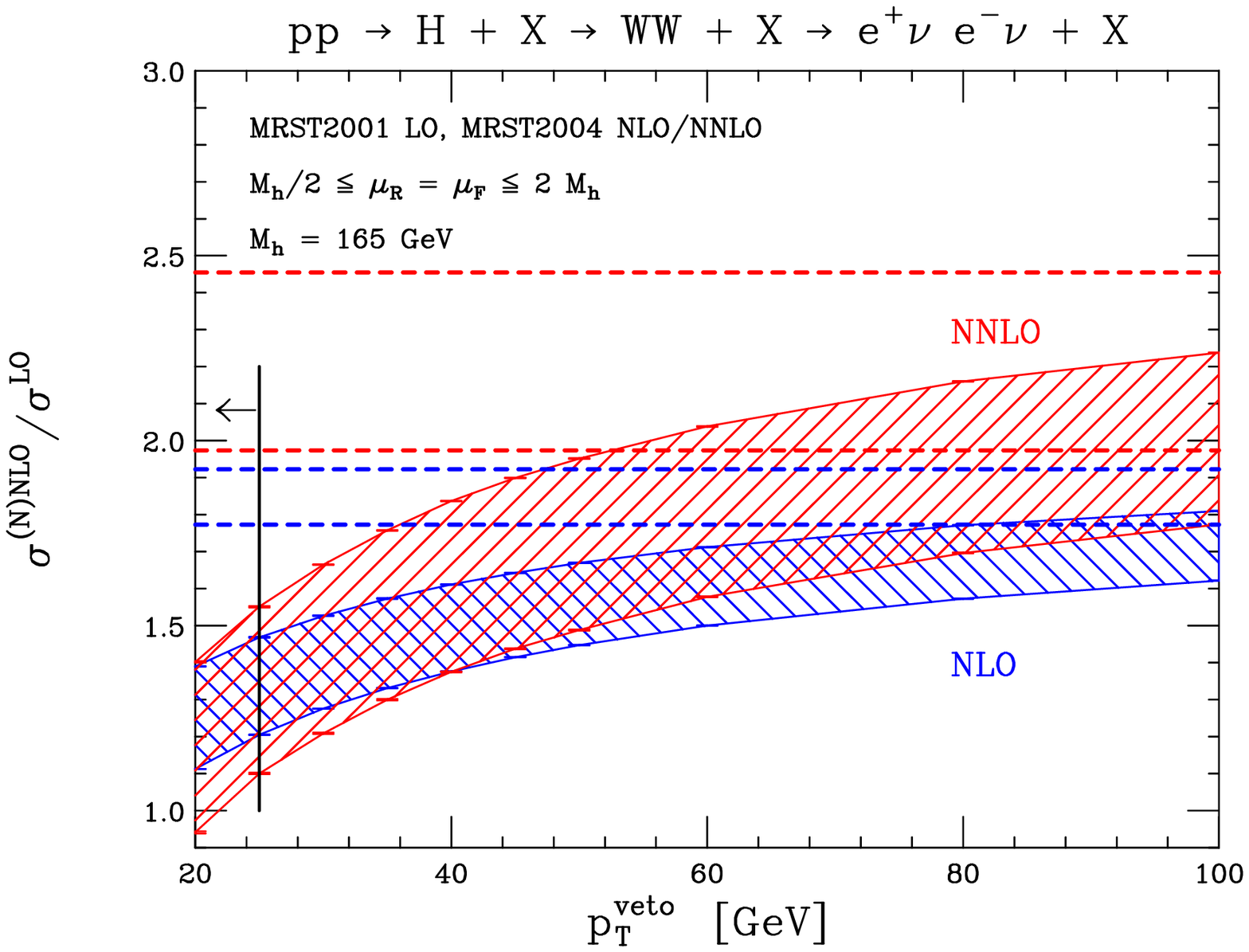}}
\end{minipage} 
\end{tabular}
\caption[]{Dependence of the LO, NLO, and NNLO cross section from
  vetoing events with jets in the central region $|\eta|< 2.5$ and
  $p_T^{jet} > p_T^{\mathrm{veto}}$. The right plot shows the $K$-factor
  as a function of $p_T^{\mathrm{veto}}$. The dashed horizontal lines
    correspond to the NLO and NNLO $K$-factors for the inclusive
    cross-section. From
    Ref.~\cite{Anastasiou:2007mz}.\label{fig:gg_to_H_nnlo_exclusive_cuts}}
\end{figure}

The results of the NNLO calculation~\cite{Harlander:2002wh,Anastasiou:2002yz}
are illustrated in Figs.~\ref{fig:gg_to_H_nnlo_nlo_lo} and
\ref{fig:gg_to_H_nnlo_nlo_lo_mu_dep}.
\begin{figure}
\hspace{0.3truecm}
\begin{tabular}{cc}
\begin{minipage}{0.45\linewidth}{\hspace{-1.truecm}
\includegraphics[scale=0.55]{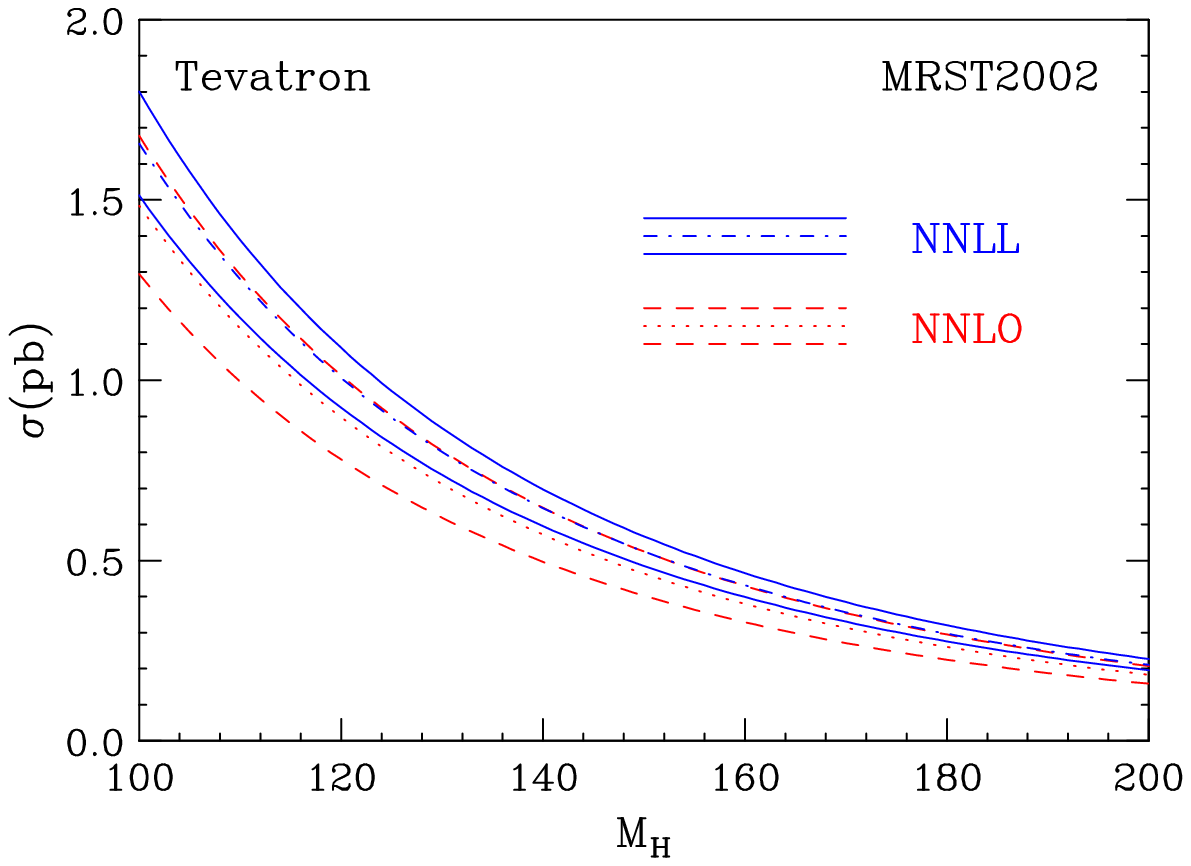}}
\end{minipage}&
\hspace{-0.5truecm}
\begin{minipage}{0.45\linewidth}{
\includegraphics[scale=0.55]{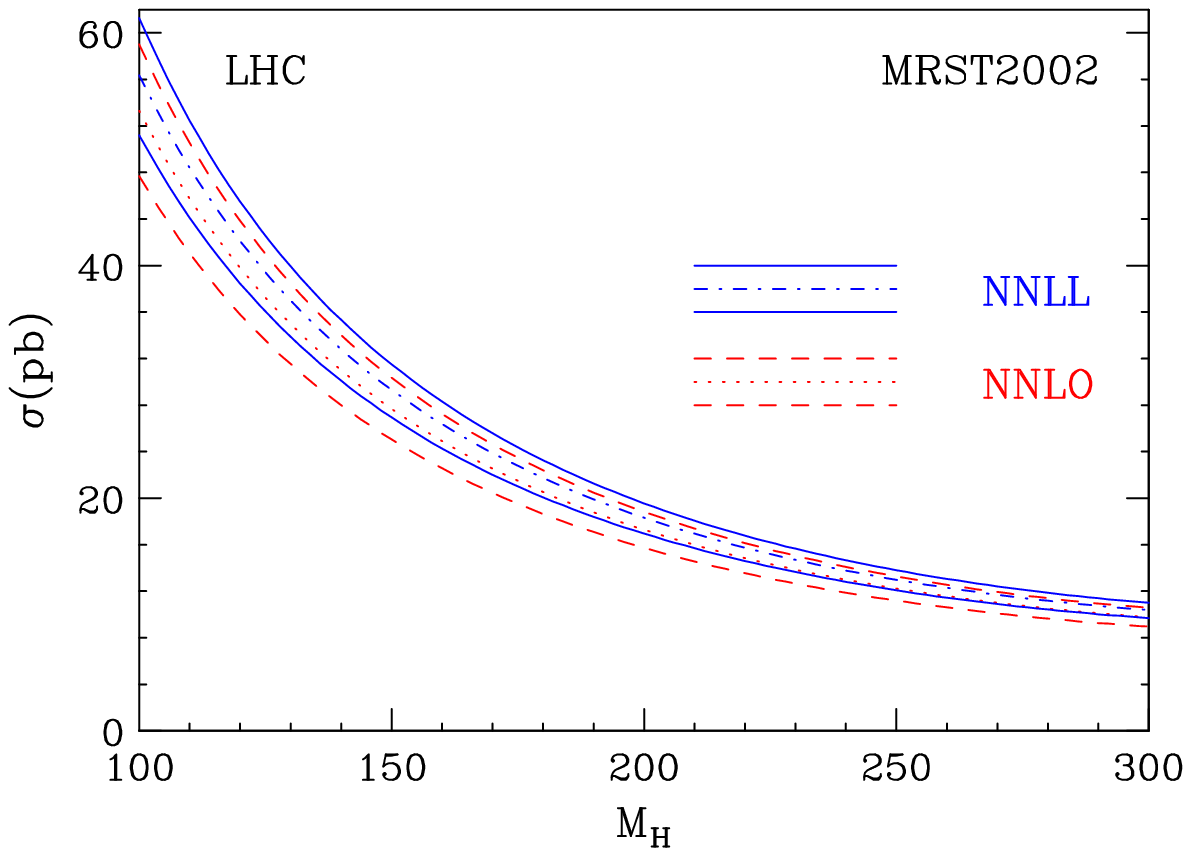}}
\end{minipage}
\end{tabular}
\caption[]{NNLL and NNLO cross sections for Higgs-boson production via
gluon-gluon fusion at both the Tevatron and the LHC. From
Ref.~\cite{Catani:2003zt}\label{fig:gg_to_H_nnll_total_xsc}}
\end{figure}
In Fig.~\ref{fig:gg_to_H_nnlo_nlo_lo} we can observe the convergence
of the perturbative calculation of $\sigma(gg\rightarrow H$), since
the difference between NLO and NNLO is much smaller than the original
difference between LO and NLO. This is further confirmed in
Fig.~\ref{fig:gg_to_H_nnlo_nlo_lo_mu_dep}, where we see that the
uncertainty band of the NNLO cross section overlaps with the
corresponding NLO band. Therefore the NNLO term in the perturbative
expansion only modify the NLO cross section within its NLO theoretical
uncertainty. This is precisely what one would expect from a good
convergence behavior. Moreover, the narrower NNLO bands in
Fig.~\ref{fig:gg_to_H_nnlo_nlo_lo_mu_dep} shows that the NNLO result
is pretty stable with respect to the variation of both renormalization
and factorization scales. This has actually been checked thoroughly in
the original papers, by varying both $\mu_R$ and $\mu_F$ independently
over a range broader than the one used in
Fig.~\ref{fig:gg_to_H_nnlo_nlo_lo_mu_dep}. The NNLO calculation has
been more recently implemented into the HNNLO
code~\cite{Catani:2008me} and subsequently extended by including the
$H\rightarrow\gamma\gamma$, $H\rightarrow WW/ZZ\rightarrow 4l$, with
the possibility to apply arbitrary cuts on the momenta of the partons
and of the photons or leptons that are produced in the final
state. Analogous results are available through the FEHiP
code~\cite{Anastasiou:2004xq}, used to produce the plots presented in
Fig.~\ref{fig:gg_to_H_nnlo_exclusive_cuts}, where we can see how the
impact of QCD corrections can vary drastically with different choices
of exclusive cuts~\cite{Anastasiou:2007mz}.

\begin{figure}
\centering
\begin{tabular}{c}
\begin{minipage}{0.8\linewidth}{
\includegraphics[scale=0.5]{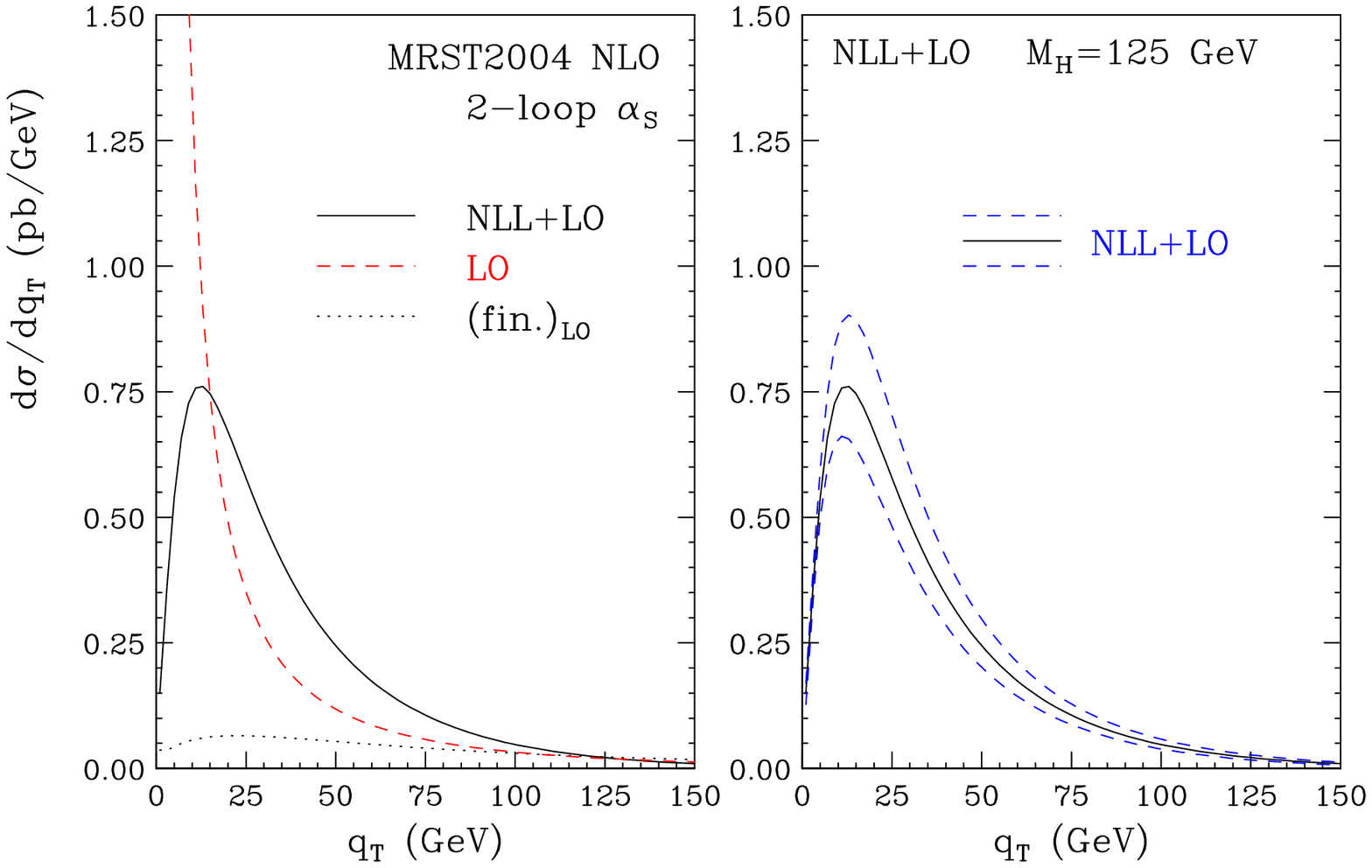}}
\end{minipage} \\
\begin{minipage}{0.8\linewidth}{
\includegraphics[scale=0.5]{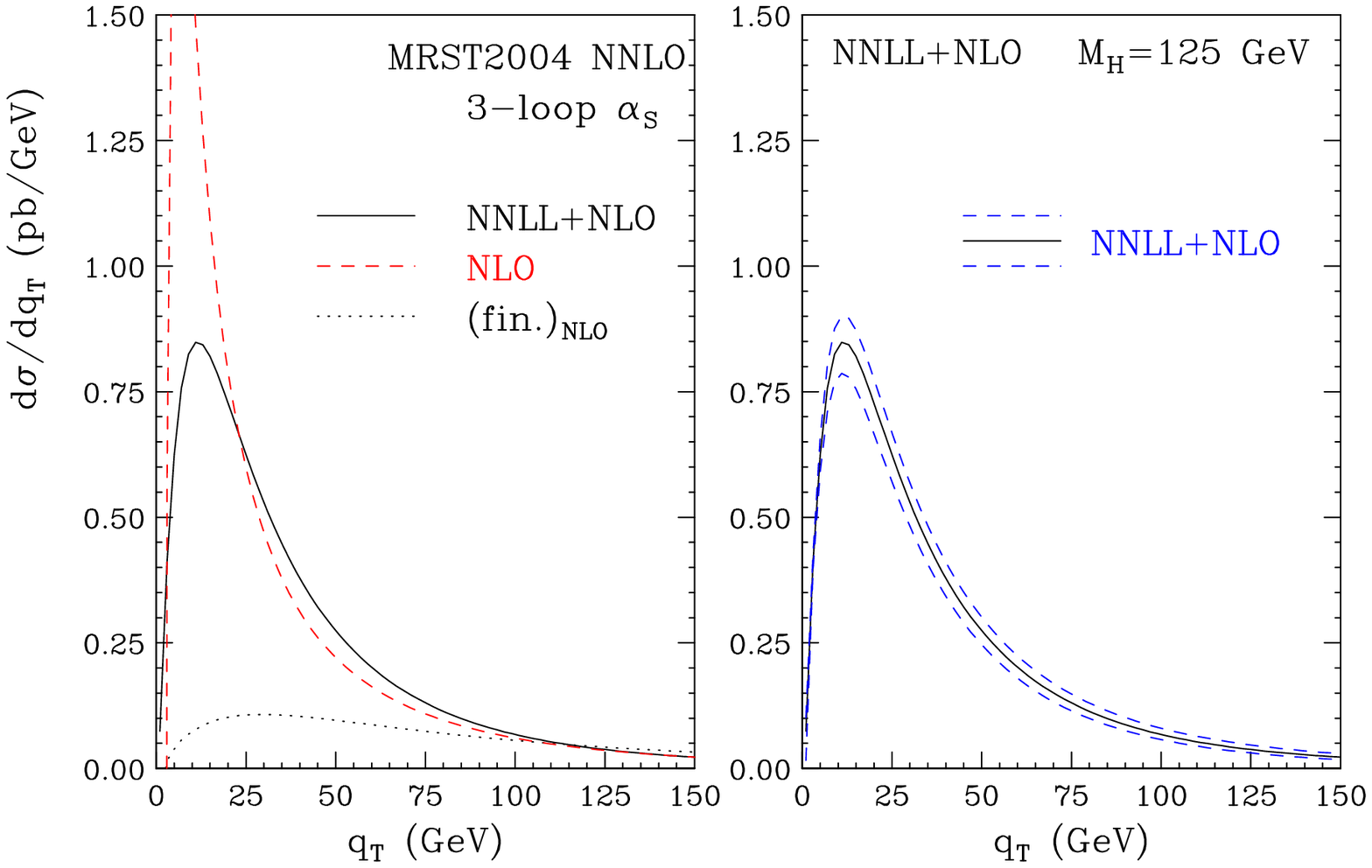}}
\end{minipage}
\end{tabular}
\caption[]{The $q_T$ spectrum  at the LHC with $M_H\!=\!125$~GeV.
The upper plots show: (left) setting $\mu_R\!=\!\mu_F\!=\!Q\!=\!M_H$, the results
at NNL+LO accuracy compared with the LO spectrum and the finite
component of the LO spectrum; (right) the uncertainty band from the
variation of the scales $\mu_R$ and $\mu_F$ at NLL+LO accuracy.
The lower plots show the same at NNLL+NLO accuracy.
From
Ref.~\cite{Bozzi:2005wk}.\label{fig:gg_to_H_nnll_partial_xsc}}
\end{figure}
The NNLO cross section for $gg\rightarrow H$ has been further improved
by Catani et al.~\cite{Catani:2003zt} by resumming up to the
next-to-next-to leading order of soft logarithms. Using the techniques
explained in their papers, they have been able to obtain the
theoretical results shown in Figs.~\ref{fig:gg_to_H_nnll_total_xsc}
and \ref{fig:gg_to_H_nnll_partial_xsc} for the total and differential
cross sections respectively. In particular, we see from
Fig.~\ref{fig:gg_to_H_nnll_total_xsc} that the NNLO and NNLL results
nicely overlap within their uncertainty bands, obtained from the
residual renormalization and factorization scale dependence. The
residual theoretical uncertainty of the NNLO+NNLL results has been
estimated to be 10\% from perturbative origin plus 10\% from the use
of NLO PDF's instead of NNLO PDF's. Moreover, in
Fig.~\ref{fig:gg_to_H_nnll_partial_xsc} we see how the resummation of
NNL crucially modify the shape of the Higgs-boson transverse momentum
distribution at low transverse momentum ($q_T$), where the soft
$\ln(M_H^2/q_T^2)$ are large and change the behavior of the
perturbative expansion in $\alpha_s$~\cite{Bozzi:2005wk}.

\begin{figure}
\begin{tabular}{lr}
\hspace{-0.5truecm}
\begin{minipage}{0.5\linewidth}{
\includegraphics[scale=0.4]{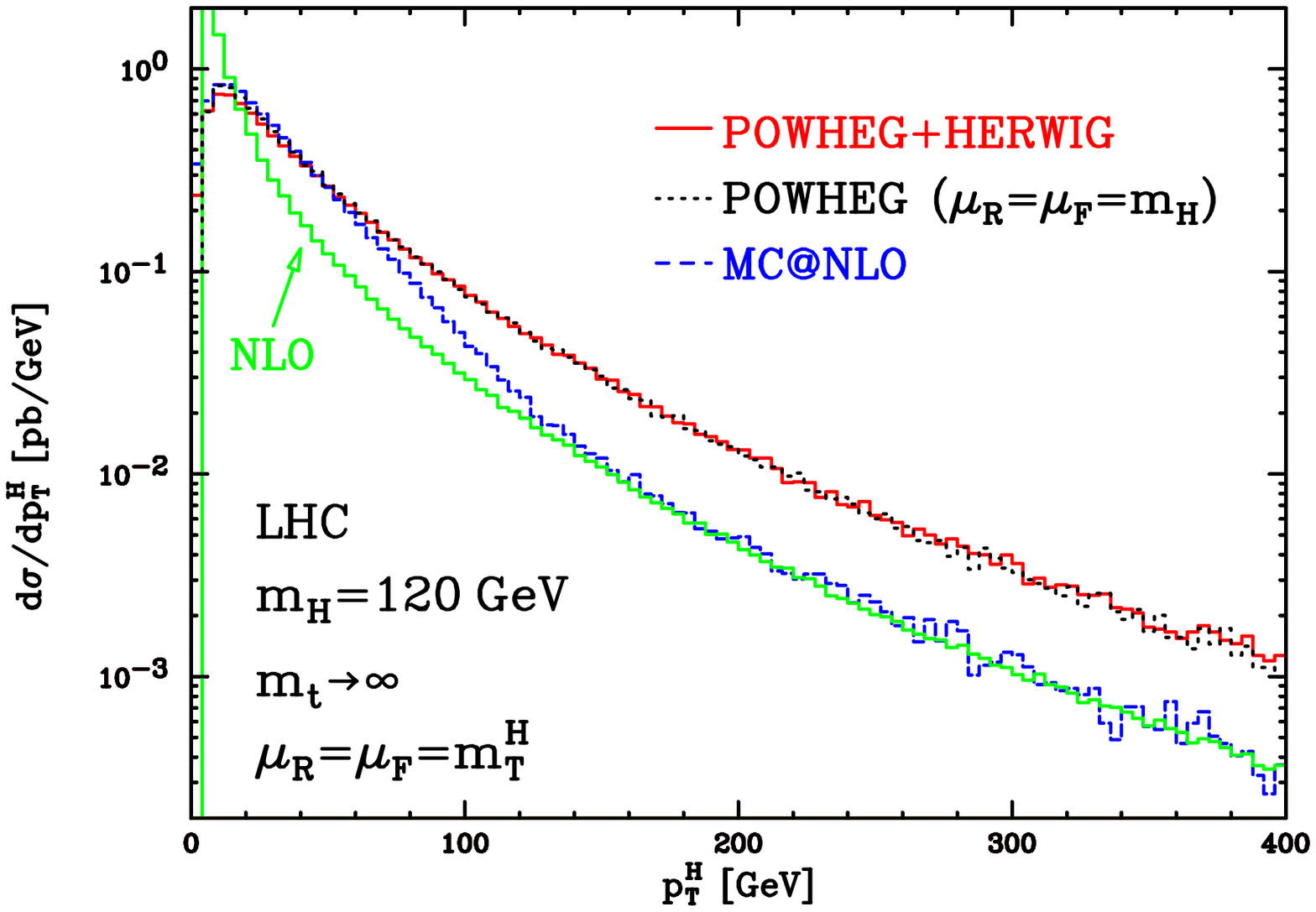}}
\end{minipage} &
\begin{minipage}{0.5\linewidth}{
\includegraphics[scale=0.4]{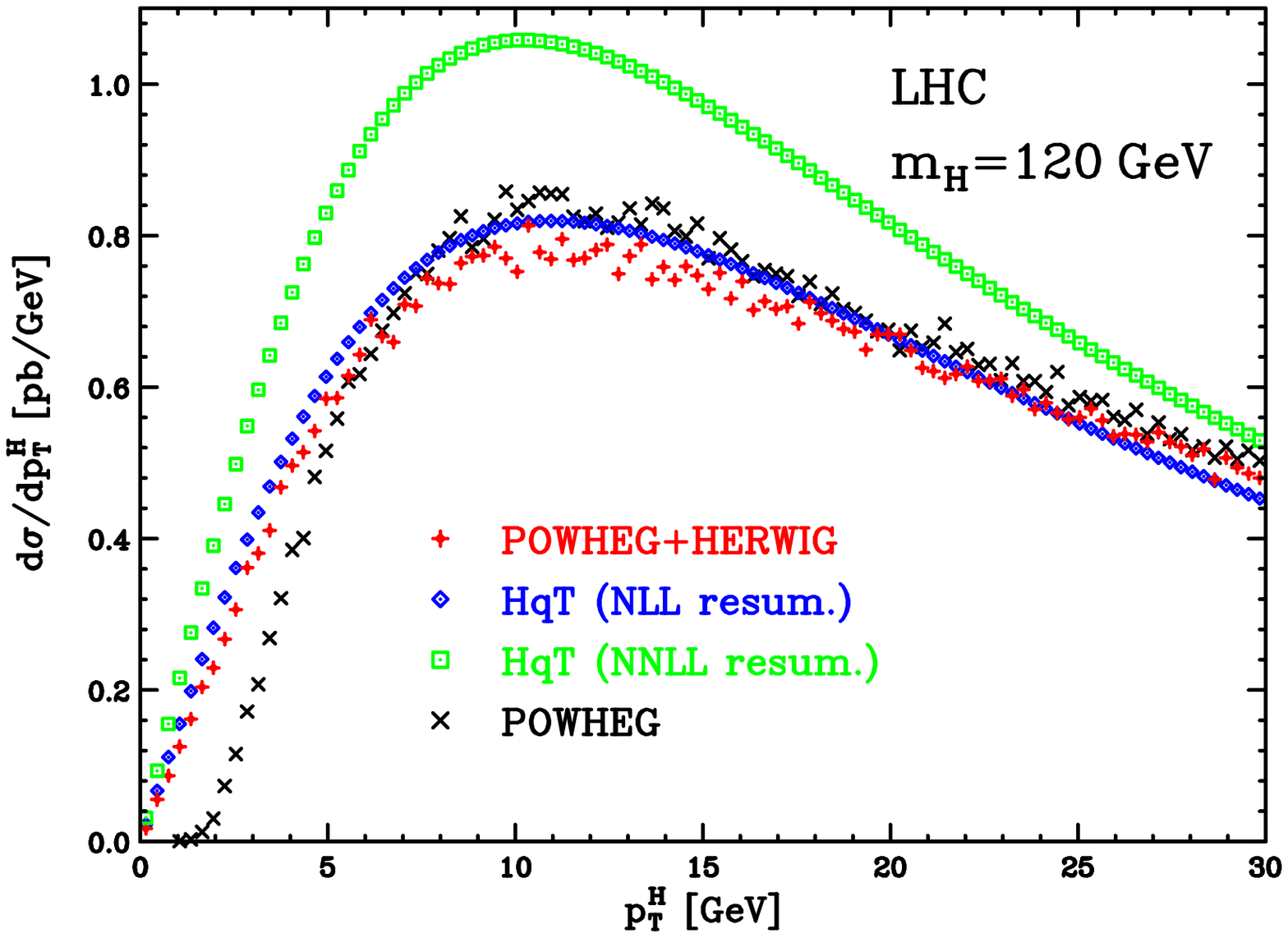}}
\end{minipage} 
\end{tabular}
\caption[]{The HIggs-boson $p_T$ spectrum at the LHC with
  $M_H\!=\!120$~GeV as obtain in the POWHEG and MC@NLO frameworks
  (left) and compared to the NNL and NNLL results.  From
  Ref.~\cite{Alioli:2008tz}.\label{fig:gg_to_H_powheg_mcnlo}}
\end{figure}

Finally, the NLO calculation of $gg\rightarrow H$ has been interfaced
with PS Monte Carlo programs (HERWIG and PYTHIA) using both the MC@NLO
and POWHEG methods~\cite{Alioli:2008tz}. The NNLO calculation cannot
be consistently interfaced with a PS Monte Carlo yet, but the
comparison of the implementation of the NLO calculation into POWHEG
and MC@NLO with Next-to-Leading-Logarithms (NLL) resummed results and
the knowledge of the resummed cross section at the
Next-to-Next-to-Leading-Logarithm (NNLL) level have allowed to provide
an \textit{improved} interface obtained by rescaling the results of
the NLO interface by NNL/NNLL. A sample of results from
Ref.~\cite{Alioli:2008tz} is given in
Fig.~\ref{fig:gg_to_H_powheg_mcnlo}.

\section*{Acknowledgments}
I would like to thank the organizers of TASI 2011 for inviting me to
lecture and for providing such a stimulating atmosphere for both
students and lecturers. This work was supported
in part by the U.S.  Department of Energy under grant
DE-FG02-97ER41022.

\end{document}